# Memory AMP

Lei Liu, *Member, IEEE*, Shunqi Huang, and Brian M. Kurkoski, *Member, IEEE*

***Abstract*—Approximate message passing (AMP) is a low-cost iterative parameter-estimation technique for certain high-dimensional linear systems with non-Gaussian distributions. AMP only applies to independent identically distributed (IID) transform matrices, but may become unreliable (e.g., perform poorly or even diverge) for other matrix ensembles, especially for ill-conditioned ones. To solve this issue, orthogonal/vector AMP (OAMP/VAMP) was proposed for general right-unitarily-invariant matrices. However, the Bayes-optimal OAMP/VAMP (BO-OAMP/VAMP) requires a high-complexity linear minimum mean square error (MMSE) estimator. This prevents OAMP/VAMP from being used in large-scale systems.**

**To address the drawbacks of AMP and BO-OAMP/VAMP, this paper offers a memory AMP (MAMP) framework based on the orthogonality principle, which ensures that estimation errors in MAMP are asymptotically IID Gaussian. To realize the required orthogonality for MAMP, we provide an orthogonalization procedure for the local memory estimators. In addition, we propose a Bayes-optimal MAMP (BO-MAMP), in which a long-memory matched filter is used for interference suppression. The complexity of BO-MAMP is comparable to AMP. To asymptotically characterize the performance of BO-MAMP, a state evolution is derived. The relaxation parameters and damping vector in BO-MAMP are optimized based on state evolution. Most crucially, the state evolution of the optimized BO-MAMP converges to the same fixed point as that of the high-complexity BO-OAMP/VAMP for all right-unitarily-invariant matrices, and achieves the Bayes optimal MSE predicted by the replica method if its state evolution has a unique fixed point. Finally, simulations are provided to verify the theoretical results' validity and accuracy.**

## I. Introduction

Consider the problem of signal reconstruction for a noisy linear system:

$$\boldsymbol{y} = \boldsymbol{A}\boldsymbol{x} + \boldsymbol{n}, \tag{1}$$

where $\boldsymbol{y} \in \mathbb{C}^{M\times 1}$ is a vector of observations, $\boldsymbol{A} \in \mathbb{C}^{M\times N}$ is a transform matrix, $\boldsymbol{x}$ is a vector to be estimated and $\boldsymbol{n} \sim \mathcal{CN}(\boldsymbol{0}, \sigma^2\boldsymbol{I}_M)$ is a vector of Gaussian additive noise samples. The entries of $\boldsymbol{x}$ are independent and identically distributed (IID). In this paper, we consider a large system with $M, N \to \infty$ and a fixed compression ratio $\delta = M/N$. This model covers a wide range of applications, including compressed sensing, multiple-input multiple-output (MIMO), multiple-access system (MAC), random access, deep neural networks, etc. In the special case when $\boldsymbol{x}$ is Gaussian, the optimal solution can be obtained using standard linear minimum mean square error (MMSE) methods. Otherwise, the problem is in general NP-hard [2], [3].

This article has been presented in part at the 2021 IEEE ISIT, Australia, [1]. The source code of this work is publicly available at sites.google.com/site/leihomepage/research.

Lei Liu, Shunqi Huang and Brian M. Kurkoski are with the School of Information Science, Japan Institute of Science and Technology (JAIST), Nomi 923-1292, Japan (e-mail: {leiliu, shunqi.huang, kurkoski}@jaist.ac.jp).

### A. Background

Approximate message passing (AMP), derived from belief-propagation (BP) with Gaussian approximation and the first-order Taylor approximation, has attracted extensive research interest for the problem in (1) [4], [5]. AMP adopts a low-complexity matched filter (MF), so its complexity is as low as $O(MN)$ per iteration if $\boldsymbol{A}$ has no special structure, such as the discrete Fourier transform (DFT) matrix or sparse matrices. Remarkably, the asymptotic performance of AMP can be described by a scalar recursion called state evolution, derived heuristically in [5] and proved rigorously in [4]. State evolution is similar to density evolution [6], except that state evolution was developed for dense graphs while density evolution is only accurate for sparse graphs. State evolution analysis in [7] implies that AMP is Bayes-optimal for zero-mean IID sub-Gaussian sensing matrices when the compression rate is larger than a certain threshold [8]. Spatial coupling [8]–[11] is used for the optimality of AMP for any compression rate. The capacity optimality of AMP was proved in [12] based on matched forward error correction (FEC) coding.

A basic assumption of AMP is that $\boldsymbol{A}$ has IID Gaussian entries [4], [5]. For matrices with correlated entries, AMP may perform poorly or even diverge [13]–[15]. It was discovered in [16], [17] that a variant of AMP based on a unitary transformation, called UTAMP, performs well for difficult (e.g., correlated) matrices $\boldsymbol{A}$. Independently, orthogonal AMP (OAMP) was proposed in [18] (see also closely-related earlier works in [19], [20]) for right-unitarily-invariant $\boldsymbol{A}$ (see Definition 1 in I-E). OAMP involves two local processors: a so-called linear estimator (LE) and a non-linear estimator (NLE) under certain orthogonality constraints, i.e., the input and output estimation errors of each processor are orthogonal. In particular, an MMSE-derived OAMP can be constructed based on the locally optimal MMSE principle. OAMP is related to a variant of the expectation propagation algorithm [21] (called diagonally-restricted expectation consistent inference in [22] or scalar expectation propagation in [23]), as observed in [24], [25]. A closely related algorithm, an MMSE-based vector AMP (VAMP) [25], is equivalent to expectation propagation in its diagonally-restricted form [22]. The accuracy of state evolution for such expectation propagation type algorithms (including VAMP and OAMP) was conjectured in [18] and proved in [24], [25]. Based on state evolution, the replica Bayes optimality of OAMP was derived in [18], [24], [25] when the compression rate is larger than a certain threshold. The capacity optimality of OAMP was proved in [26], [27] based on matched FEC coding. The advantages of AMP-type algorithms over conventional turbo receivers [28], [29] are demonstrated in [12], [27], [30].

The main weakness of Bayes-optimal OAMP/VAMP (BO-OAMP/VAMP) is the high-complexity $O(M^3+M^2N)$ incurred by the linear MMSE (LMMSE) estimator if $\boldsymbol{A}$ has no special structure. Singular-value decomposition (SVD) was used to avoid the high-complexity LMMSE in each iteration [25], but the complexity of the SVD itself is as high as that of the LMMSE estimator. The performance of OAMP/VAMP degrades significantly when the LMMSE estimator is replaced by the low-complexity matched filter [18] used in AMP. This limits the application of OAMP/VAMP to large-scale systems for which LMMSE is too complex.

In summary, the existing replica Bayes-optimal AMP-type algorithms are either limited to IID matrices (e.g., AMP) or need high-complexity LMMSE (e.g., BO-OAMP/VAMP). Hence, a low-complexity replica Bayes-optimal message passing algorithm for right-unitarily-invariant matrices is desired.

Recently, Takeuchi proposed convolutional AMP (CAMP), in which AMP is modified by replacing the Onsager term with a convolution of all preceding messages [31]. Similar to AMP, CAMP consists of an *a-posteriori* NLE and a matched filter with a correction term. The CAMP has low complexity and applies to right-unitarily-invariant matrices. It was proved that the CAMP is replica Bayes-optimal if its state evolution converges to a unique fixed point [31]. However, the CAMP has a relatively low convergence speed and may fail to converge [32] (even if empirical damping is used), particularly for matrices with high condition numbers. A heuristic damping was used to improve the convergence of CAMP. However, the damping used in CAMP is performed on the *a-posteriori* NLE outputs, which breaks orthogonality and the asymptotic Gaussianity of estimation errors [31].

### B. Motivation and Related Works

As mentioned, the potential (e.g., replica) Bayes optimality generally requires locally optimal MMSE estimators. To avoid the high-complexity LMMSE in BO-OAMP/VAMP, a straightforward way is to approach the LMMSE by using inner iterative methods, such as inner Gaussian message passing [33], [34], inner conjugate gradient method [35], [36]. However, these strategies have relatively high complexity since they involve inner iterations. A warm-start technique was proposed to accelerate the convergence of the reset-start conjugate gradient VAMP (CG-VAMP) [36].

Iterative LMMSE receiver for coded MIMO systems [29], [30] provides an important inspiration to avoid the high-complexity inner iteration. High-complexity inner belief-propagation decoding is replaced by a low-complexity memory decoding, where the decoder is initialized by a memory in the previous states and the decoder only updates once at each outer iteration. Therefore, it is natural to ask: can we avoid the high-complexity inner iteration at LE in [33]–[35] using memory? Fortunately, the answer is quite positive. AMP is such an implicit example, where the memory Onsager term assists the matched filter in approaching the locally optimal LMMSE. That is the key point why AMP is Bayes-optimal (in unique fixed-point state evolution) as well as low-complexity. More recently, the concept of long-memory message passing, where all the preceding messages are used for current estimation, was clearly revealed in [31], [37], [38]. Specifically, a long-memory AMP algorithm was constructed in [37] to solve the Thouless-Anderson-Palmer equations for Ising models with general invariant random matrices. The results in [37] were rigorously justified via state evolution in [38].

### C. Contributions

To overcome the difficulties in AMP, OAMP/VAMP and CAMP, we propose a memory AMP (MAMP) framework under the orthogonality principle, which guarantees the asymptotic IID Gaussianity of estimation errors in MAMP. Due to the memory involved in MAMP, stricter orthogonality (compared to the orthogonality of non-memory OAMP/VAMP) is required for MAMP to guarantee the asymptotic IID Gaussianity of estimation errors in MAMP [24], [31]. In detail, the step-by-step orthogonalization between the current input and output estimation errors in non-memory AMP (e.g., OAMP/VAMP) is not sufficient, and instead, the current output estimation error is required to be orthogonal to all input estimation errors. We present an orthogonalization procedure for the local memory estimators to realize the required orthogonality for MAMP.

We propose a Bayes-optimal MAMP (BO-MAMP) using a low-complexity long-memory matched filter for interference suppression. Using the asymptotic IID Gaussianity, a covariance-matrix state evolution is established for BO-MAMP. Based on state evolution, relaxation parameters and damping vectors, preserving the orthogonality (e.g., the asymptotic Gaussianity of estimation errors), are analytically optimized to improve and guarantee the convergence of BO-MAMP. BO-MAMP has the following two properties.

- The complexity of BO-MAMP is comparable to AMP and much lower than BO-OAMP/VAMP.
- The state evolution of the optimized BO-MAMP converges to the same fixed point as that of the high-complexity BO-OAMP/VAMP for all right-unitarily-invariant matrices. Hence, the optimized BO-MAMP achieves the Bayes-optimal MSE predicted by the replica method if its state evolution has a unique fixed point.

Part of the results in this paper has been published in [1]. In this paper, we additionally provide the derivation of state evolution, detailed proofs and more numerical results.

### D. Comparisons with Existing Works

Table I gives an overview of the key properties of the past and present AMP-type algorithms including AMP, BO-OAMP/VAMP, MF-OAMP/VAMP, CAMP and BO-MAMP, which are summarized as follows.

- AMP is potentially Bayes-optimal, and has low complexity as it uses a memory MF-LE. However, it is limited to IID matrices, and has a low convergence speed since only a single memory in last iteration is used [5].
- CAMP is potentially Bayes-optimal, applies to right-unitarily-invariant matrices, and has low complexity as

TABLE I
AN OVERVIEW OF THE AMP-TYPE ALGORITHMS

| Algorithms | Matrix prerequisite | Linear Estimator | Optimality | State-evolution convergence | Relative speed |
|---|---|---|---|---|---|
| AMP [5] | **IID** | Memory MF (low complexity) | Bayes-optimal (potential) | Converges | 4 |
| BO-OAMP/VAMP [18], [25] | Right unitarily invariant | **Non-memory LMMSE** (**high complexity**) | Bayes-optimal (potential) | Converges | 1 (best) |
| MF-OAMP/VAMP [18], [25] | Right unitarily invariant | Non-memory MF (low complexity) | **Sub-optimal** in **ill-conditioned matrices** | Converges | 5 (worst) |
| CAMP [31] | Right unitarily invariant | Long-memory MF (low complexity) | Bayes-optimal (potential) | **Fails to converge in high condition numbers** | 3 |
| BO-MAMP (proposed) | Right unitarily invariant | Long-memory MF (low complexity) | Bayes-optimal (potential) | Converges | 2 |

Notes: The properties in bold indicate weakness. The relative speed means the number of iterations to achieve the same MSE performance for right-unitarily-invariant matrices (except IID Gaussian and identical-singular-value matrices in IV-E). "MF" means matched filter, and "potential" means that the algorithm achieves the Bayes-optimal MSE predicted by the replica method if its state evolution has a unique fixed point. For zero-mean IID Gaussian matrices, the convergence speeds of AMP, CAMP and BO-MAMP are almost the same. For identical-singular-value matrices, MF-OAMP/VAMP, BO-OAMP/VAMP and BO-MAMP are equivalent.

it uses an MF-LE. Since long memory is considered, its convergence speed is higher than AMP and MF-OAMP/VAMP for general non-IID and ill-conditioned matrices. However, it fails to converge for matrices with high condition numbers [31], [32].

- MF-OAMP/VAMP applies to right-unitarily-invariant matrices, and has low complexity as it uses a non-memory MF-LE. However, it is sub-optimal, i.e., the performance is poor for ill-conditioned matrices [18].
- BO-OAMP/VAMP is potentially Bayes-optimal and applies to right-unitarily-invariant matrices. It has the highest convergence speed as locally optimal LE (i.e., LMMSE) is used, which costs high complexity [18], [25].
- BO-MAMP proposed in this paper is potentially Bayes-optimal, applies to right-unitarily-invariant matrices, and has low complexity since it uses an MF-LE. BO-MAMP solves the convergence problem of CAMP. We prove that the state evolution of the optimized BO-MAMP converges to the same fixed point as BO-OAMP/VAMP. Furthermore, for general ill-conditioned matrices, the convergence speed of BO-MAMP is higher than AMP, CAMP and MF-OAMP/VAMP, but relatively lower than BO-OAMP/VAMP.

More specifically, distinct from the AMP, CAMP and long-memory AMP in [37], [38] that consist of an *a-posteriori* NLE and a matched filter with a correction term, the proposed MAMP consists of an orthogonal NLE and an orthogonal long-memory matched filter, which is similar to the structure of OAMP/VAMP. Apart from that, distinct from AMP and CAMP that perform a heuristic scalar damping on the *a-posteriori* NLE outputs and break the orthogonality and Gaussianity, an analytically optimized vector damping for the orthogonal NLE outputs is derived for BO-MAMP to guarantee and improve the convergence as well as preserve the orthogonality and Gaussianity.

Recently, Skuratovs and Davies proposed a warm-start CG-VAMP (WS-CG-VAMP) to accelerate the convergence of CG-VAMP in [36]. However, the state evolution of WS-CG-VAMP was not provided in the first preprint version of [36] on arXiv. After we posted the preprint of this work (which provided a rigorous state evolution for BO-MAMP) [39], a rigorous state evolution was derived for WS-CG-VAMP in the latest version of [36]. In addition, the convergence and replica Bayes optimality of the state evolution for WS-CG-VAMP have not yet been rigorously proved, while the convergence and replica Bayes optimality of the state evolution for the proposed BO-MAMP are rigorously proved in this paper.

More recently, the damping optimization technique in this paper has been used to analyze the convergence of BO-OAMP/VAMP in [40], [41] from a sufficient statistic perspective. Motivated by the damping optimization in this paper and its statistical interpretation in [40], [41], a sufficient statistic memory AMP (SS-MAMP) is proposed in [58], [59] to ensure the convergence of any MAMP-type algorithms in principle.

### E. Notation and Definitions

Boldface lowercase letters represent vectors and boldface uppercase symbols denote matrices. $\mathrm{E}\{\cdot\}$ is the expectation operation over all random variables involved in the brackets, except when otherwise specified, e.g., $\mathrm{E}\{a\}$ is the expectation of $a$. $\mathrm{Var}\{a\}$ is $\mathrm{E}\left\{\|a-\mathrm{E}\{a\}\|^2\right\}$, $\mathrm{sign}(a)$ is the sign of $a$. We say that $x = x_{\rm Re} + ix_{\rm Im}$ is circularly-symmetric complex Gaussian if $x_{\rm Re}$ and $x_{\rm Im}$ are two independent Gaussian distributed random variables with $\mathrm{E}\{x_{\rm Re}\} = \mathrm{E}\{x_{\rm Im}\} = 0$ and $\mathrm{Var}\{x_{\rm Re}\} = \mathrm{Var}\{x_{\rm Im}\}$. We define $\mathrm{Var}\{x\} \equiv \mathrm{Var}\{x_{\rm Re}\} + \mathrm{Var}\{x_{\rm Im}\}$. $\boldsymbol{I}$ is the identity matrix of appropriate size, $\boldsymbol{0}$ the zero matrix/vector, $\boldsymbol{1}$ the all one matrix/vector, $\boldsymbol{A}^*$ the conjugate of $\boldsymbol{A}$, $\boldsymbol{A}^{\rm T}$ the transpose of $\boldsymbol{A}$, $\boldsymbol{A}^{\rm H}$ the conjugate transpose of $\boldsymbol{A}$, $\|\boldsymbol{a}\|$ the $\ell_2$-norm of $\boldsymbol{a}$, and $\mathcal{CN}(\boldsymbol{\mu},\boldsymbol{\Sigma})$ the circularly-symmetric complex Gaussian distribution with mean $\boldsymbol{\mu}$ and covariance $\boldsymbol{\Sigma}$. For an $N$-by-$N$ matrix $\boldsymbol{A} = [a_{i,j}]_{N\times N}$, $\mathrm{diag}\{\boldsymbol{A}\} \equiv [a_{1,1},\dots,a_{N,N}]^{\rm T}$ denotes the diagonal vector, $\det(\boldsymbol{A})$ the determinant of $\boldsymbol{A}$, and $\mathrm{tr}\{\boldsymbol{A}\}$ the trace of $\boldsymbol{A}$. A matrix is said to be column-wise IID if the elements in each of its columns are IID and row-wise joint-Gaussian if the elements in each of its rows are jointly Gaussian. The notation $X \sim Y$ means that $X$ follows the same distribution as $Y$. $\min\{\mathcal{S}\}$ and $\max\{\mathcal{S}\}$ respectively takes the minimum and maximum value in set $\mathcal{S}$. The notation $\overset{\text{a.s.}}{=}$ denotes almost sure equivalence, $\binom{n}{k} = n/(k!(n-k)!)$ denotes $n$ choose $k$.

Throughout this paper, unless stated otherwise, we will assume that the length of a vector is $N$ and $\boldsymbol{x}$ is normalized, i.e., $\frac{1}{N}\mathrm{E}\{\|\boldsymbol{x}\|^2\} = 1$. For simplicity, in this paper, unless stated otherwise, BO-OAMP/VAMP refers to the MMSE variant where the LE uses LMMSE and NLE uses MMSE.

*Definition 1 (Haar Matrix):* Let $\mathcal{U}_n$ be the space of all $n\times n$ unitary matrices. A unitary random matrix $\boldsymbol{U} \in \mathcal{U}_n$ is called a Haar matrix if $\boldsymbol{U}$ is uniformly distributed on $\mathcal{U}_n$.

### F. Paper Outline

This paper is organized as follows. Section II gives the preliminaries including problem formulation and the review of OAMP/VAMP. Section III proposes a memory AMP under an orthogonality principle. A Bayes-optimal memory AMP (BO-MAMP) is developed in Section IV. Parameter optimization for BO-MAMP is given in Section V. Numerical results are shown in Section VI.

## II. Preliminaries

In this section, we first give the problem formulation and assumptions. Then, we introduce a non-memory iterative process and review OAMP/VAMP and its properties.

### A. Problem Formulation

Fig. 1(a) illustrates the system in (1) with two constraints:

$$\Gamma: \quad \boldsymbol{y} = \boldsymbol{A}\boldsymbol{x} + \boldsymbol{n}, \tag{2a}$$

$$\Phi: \quad x_i \sim P_x, \ \ \forall i. \tag{2b}$$

Our aim is to use the AMP-type iterative approach in Fig. 1(b) to find an MMSE estimate of $\boldsymbol{x}$. That is, its MSE converges to [42]

$$\mathrm{mmse}\{\boldsymbol{x}|\boldsymbol{y},\boldsymbol{A},\Gamma,\Phi\} \equiv \tfrac{1}{N}\mathrm{E}\{\|\hat{\boldsymbol{x}}_{\rm post} - \boldsymbol{x}\|^2\}, \tag{3}$$

where $\hat{\boldsymbol{x}}_{\rm post} = \mathrm{E}\{\boldsymbol{x}|\boldsymbol{y},\boldsymbol{A},\Gamma,\Phi\}$ is the *a-posteriori* mean of $\boldsymbol{x}$.

*Definition 2 (Bayes Optimality):* An iterative approach is said to be Bayes optimal if its MSE converges to the MMSE of the system in (1).

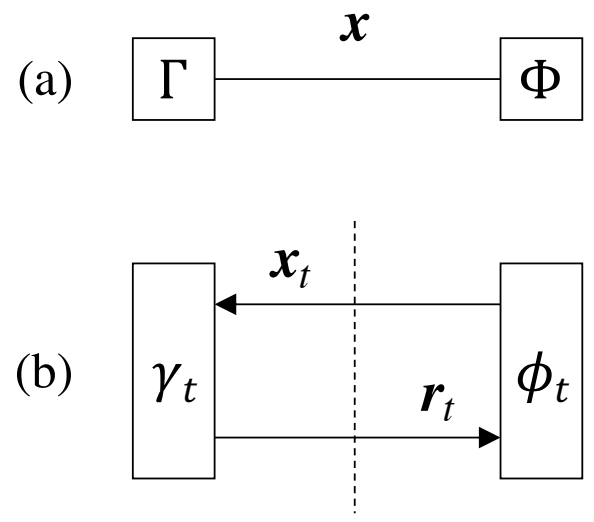

Fig. 1. Graphic illustrations for (a) a system model with two constraints $\Phi$ and $\Gamma$, and (b) a non-memory iterative process involving two local processors $\gamma_t$ and $\phi_t$.

### B. Assumptions

*Assumption 1:* The entries of $\boldsymbol{x}$ are IID with zero mean. The variance of $\boldsymbol{x}$ is normalized, i.e., $\frac{1}{N}\mathrm{E}\{\|\boldsymbol{x}\|^2\}$=1, and the $(2+\epsilon)$th moments of $\boldsymbol{x}$ are finite for some $\epsilon$.

*Assumption 2:* Let the singular value decomposition of $\boldsymbol{A}$ be $\boldsymbol{A} = \boldsymbol{U}\boldsymbol{\Sigma}\boldsymbol{V}^{\rm H}$, where $\boldsymbol{U} \in \mathbb{C}^{M\times M}$ and $\boldsymbol{V} \in \mathbb{C}^{N\times N}$ are unitary matrices, and $\boldsymbol{\Sigma}$ is a rectangular diagonal matrix. We assume that $M,N \to \infty$ with a fixed $\delta = M/N$, and $\boldsymbol{A}$ is known and is right-unitarily-invariant, i.e., $\boldsymbol{U}\boldsymbol{\Sigma}$ and $\boldsymbol{V}$ are independent, and $\boldsymbol{V}$ is Haar distributed (or equivalently, isotropically random). Furthermore, the empirical eigenvalue distribution of $\boldsymbol{A}\boldsymbol{A}^{\rm H}$ converges almost surely to a compactly supported deterministic distribution with unit first moment in the large system limit, i.e., $\lim_{N\to\infty}\frac{1}{N}\mathrm{tr}\{\boldsymbol{A}\boldsymbol{A}^{\rm H}\} \overset{\text{a.s.}}{=} 1$ [31]. Let $\lambda_t = \frac{1}{N}\mathrm{tr}\{(\boldsymbol{A}\boldsymbol{A}^{\rm H})^t\}$ and $\lambda^\dagger \equiv [\lambda_{\max} + \lambda_{\min}]/2$, where $\lambda_{\min}$ and $\lambda_{\max}$ denote the minimal and maximal eigenvalues of $\boldsymbol{A}\boldsymbol{A}^{\rm H}$, respectively. Without loss of generality, we assume that $\{\lambda_{\min},\lambda_{\max}\}$ and $\{\lambda_t, \forall t \le 2\mathcal{T}\}$ are known, where $\mathcal{T}$ is the maximum number of iterations.

For specific random matrices such as IID Gaussian, Wigner and Wishart matrices, $\{\lambda_{\min},\lambda_{\max},\lambda_t\}$ can be calculated from the eigenvalue distribution $P_\lambda$ [43]. Otherwise, the following approximations can be used.

- $\{\lambda_t\}$ can be approximated as follows. The complexity is as low as $O(MN)$ per iteration.

  *Proposition 1:* The $\{\lambda_t\}$ can be approximated by

$$\lambda_t \overset{\text{a.s.}}{=} \lim_{N\to\infty} \|\boldsymbol{s}_t\|^2, \quad \text{for } t = 1,2,\dots \tag{4a}$$

  where $\boldsymbol{s}_t$ is given by the following recursion: Starting with $t = 1$ and $\boldsymbol{s}_0 \sim \mathcal{CN}(\boldsymbol{0}_{N\times 1}, \frac{1}{N}\boldsymbol{I}_{N\times N})$,

$$\boldsymbol{s}_t = \begin{cases} \boldsymbol{A}\boldsymbol{s}_{t-1}, & \text{if } t \text{ is odd} \\ \boldsymbol{A}^{\rm H}\boldsymbol{s}_{t-1}, & \text{if } t \text{ is even} \end{cases}. \tag{4b}$$

  *Proof:* Following (4b), we can expand $\boldsymbol{s}_t$ to

$$\boldsymbol{s}_t = \begin{cases} \boldsymbol{A}(\boldsymbol{A}^{\rm H}\boldsymbol{A})^{\frac{t-1}{2}}\boldsymbol{s}_0, & \text{if } t = 1,3,5,\cdots \\ (\boldsymbol{A}^{\rm H}\boldsymbol{A})^{\frac{t}{2}}\boldsymbol{s}_0, & \text{if } t = 2,4,6,\cdots \end{cases}. \tag{5}$$

  Since $\boldsymbol{s}_0 \sim \mathcal{CN}(\boldsymbol{0}_{N\times 1}, \frac{1}{N}\boldsymbol{I}_{N\times N})$, we have

$$\lim_{N\to\infty}\|\boldsymbol{s}_t\|^2 \overset{\text{a.s.}}{=} \tfrac{1}{N}\mathrm{tr}\{(\boldsymbol{A}^{\rm H}\boldsymbol{A})^t\} = \lambda_t. \tag{6}$$

  Thus, we complete the proof of Proposition 1. ■
- $\lambda_{\min}$ and $\lambda_{\max}$ can be replaced respectively by a lower bound $\lambda_{\min}^{\rm low}$ of the minimal eigenvalue and an upper bound $\lambda_{\max}^{\rm up}$ of the maximal eigenvalue of $\boldsymbol{A}\boldsymbol{A}^{\rm H}$. Theorem 2 in Subsection V-E needs $\{\lambda_{\min},\lambda_{\max}\}$ to satisfy $\rho(\theta_t\boldsymbol{B}) < 1$, which does not change by using the bounds $\{\lambda_{\min}^{\rm low},\lambda_{\max}^{\rm up}\}$ (see Remark 1 in V-A). Therefore, Theorem 2 still holds with the bounds $\{\lambda_{\min}^{\rm low},\lambda_{\max}^{\rm up}\}$. Loose bounds may decrease the convergence speed of BO-MAMP. Since $\lambda_{\min} \ge 0$ and $\lambda_{\max}^{\tau} \le N\lambda_\tau$, we can set

$$\lambda_{\min}^{\rm low} = 0, \qquad \lambda_{\max}^{\rm up} = (N\lambda_\tau)^{1/\tau}. \tag{7a}$$

  The $\lambda_{\max}^{\rm up}$ is tighter for larger $\tau$. In practice, we set $\tau = 2\mathcal{T}$, where $\mathcal{T}$ is the maximum number of iterations.

As is shown in Fig. 13, the performance of BO-MAMP with approximate $\{\lambda_t\}$ in Proposition 1 and bounds $\lambda_{\min}^{\text{low}} = 0$ and $\lambda_{\max}^{\text{up}} = (N\lambda_\tau)^{1/\tau}$ are almost the same as that with the exact values of $\{\lambda_t, \lambda_{\min}, \lambda_{\max}\}$.

### C. Non-memory Iterative Process and Orthogonality

*Non-memory Iterative Process:* Fig. 1(b) illustrates a non-memory iterative process[1] consisting of a linear estimator (LE) and a non-linear estimator (NLE): Starting with $t = 1$,

$$\text{LE}: \quad \boldsymbol{r}_t = \gamma_t(\boldsymbol{x}_t) = \boldsymbol{Q}_t \boldsymbol{y} + \boldsymbol{\mathcal{P}}_t \boldsymbol{x}_t, \tag{8a}$$

$$\text{NLE}: \quad \boldsymbol{x}_{t+1} = \phi_t(\boldsymbol{r}_t), \tag{8b}$$

where $\gamma_t(\cdot)$ and $\phi_t(\cdot)$ process the two constraints $\Gamma$ and $\Phi$ separately, and $\boldsymbol{Q}_t\boldsymbol{A}$ and $\boldsymbol{\mathcal{P}}_t$ are polynomials in $\boldsymbol{A}^{\text{H}}\boldsymbol{A}$. Without loss of generality, we assume that the norms of $\boldsymbol{Q}_t$ and $\boldsymbol{\mathcal{P}}_t$ are finite. For any $\boldsymbol{x}_{t,1}, \boldsymbol{x}_{t,2} \in \mathbb{C}^N$,

$$\|\gamma_t(\boldsymbol{x}_{t,1}) - \gamma_t(\boldsymbol{x}_{t,2})\| = \|\boldsymbol{\mathcal{P}}_t(\boldsymbol{x}_{t,1} - \boldsymbol{x}_{t,2})\| \tag{9a}$$

$$\leq \|\boldsymbol{\mathcal{P}}_t\|\|\boldsymbol{x}_{t,1} - \boldsymbol{x}_{t,2}\|, \tag{9b}$$

where $\|\boldsymbol{\mathcal{P}}_t\|$ is finite. Hence, $\gamma_t(\cdot)$ is Lipschitz-continuous.

We call (8) a non-memory iterative process since both $\gamma_t(\cdot)$ and $\phi_t(\cdot)$ are memoryless, depending only on their current inputs $\boldsymbol{x}_t$ and $\boldsymbol{r}_t$ respectively. Let

$$\boldsymbol{r}_t = \boldsymbol{x} + \boldsymbol{g}_t, \tag{10a}$$

$$\boldsymbol{x}_t = \boldsymbol{x} + \boldsymbol{f}_t, \tag{10b}$$

where $\boldsymbol{g}_t$ and $\boldsymbol{f}_t$ indicate the estimation errors with zero means and variances:

$$v_t^\gamma = \tfrac{1}{N}\text{E}\{\|\boldsymbol{g}_t\|^2\}, \qquad v_t^\phi = \tfrac{1}{N}\text{E}\{\|\boldsymbol{f}_t\|^2\}. \tag{10c}$$

The asymptotic IID Gaussian property of a non-memory iterative process was conjectured in [18] and proved in [24], [25] based on the following error orthogonality. For more details, refer to Section V-D and Section V-E in [24] for the proof of asymptotically IID Gaussian properties (see A1 and B1 in Theorem 4 in [24]) using the orthogonality (see A2 and B2 in Theorem 4 in [24]), which does not rely on the explicit constructions of $\gamma_t(\cdot)$ and $\phi_t(\cdot)$.

*Lemma 1 (Orthogonality and Asymptotic IID Gaussianity):* Suppose that Assumptions 1-2 and the following orthogonality holds: for all $t \geq 1$,

$$\lim_{N\to\infty} \tfrac{1}{N}\boldsymbol{x}^{\text{H}}\boldsymbol{g}_t \overset{\text{a.s.}}{=} 0, \tag{11a}$$

$$\lim_{N\to\infty} \tfrac{1}{N}\boldsymbol{f}_t^{\text{H}}\boldsymbol{g}_t \overset{\text{a.s.}}{=} 0, \tag{11b}$$

$$\lim_{N\to\infty} \tfrac{1}{N}\boldsymbol{g}_t^{\text{H}}\boldsymbol{f}_{t+1} \overset{\text{a.s.}}{=} 0. \tag{11c}$$

Then, for the $\{\gamma_t(\cdot)\}$ in (8a) and separable-and-Lipschitz-continuous [44] $\{\phi_t(\cdot)\}$, we have: $\forall t \geq 1$,

$$\lim_{N\to\infty} v_t^\gamma \overset{\text{a.s.}}{=} \tfrac{1}{N}\text{E}\big\{\|\gamma_t(\boldsymbol{x} + \boldsymbol{\eta}_t^\phi) - \boldsymbol{x}\|^2\big\}, \tag{12a}$$

$$\lim_{N\to\infty} v_{t+1}^\phi \overset{\text{a.s.}}{=} \tfrac{1}{N}\text{E}\big\{\|\phi_t(\boldsymbol{x} + \boldsymbol{\eta}_t^\gamma) - \boldsymbol{x}\|^2\big\}, \tag{12b}$$

where $\boldsymbol{\eta}_t^\phi \sim \mathcal{CN}(\boldsymbol{0}, v_t^\phi \boldsymbol{I})$ and $\boldsymbol{\eta}_t^\gamma \sim \mathcal{CN}(\boldsymbol{0}, v_t^\gamma \boldsymbol{I})$ are independent of $\boldsymbol{x}$.

Lemma 1 is a special case of Lemma 4 for the memory iterative process in III-A. In fact, the sufficient condition of IID Gaussian property for an iterative process is the "full orthogonality" in (20), i.e., the current output estimation error is orthogonal to all preceding input estimation errors in each estimation. For a non-memory iterative process, this condition is relaxed to the "step-by-step orthogonality" in (11), i.e., the current input and output estimation errors are orthogonal in each estimation, following the generalized Stein's lemma [31], [45] below. Note that in general, "step-by-step orthogonality" in (11) is much looser than "full orthogonality" in (20), except for the non-memory iterative process discussed above.

*Lemma 2 (Generalized Stein's Lemma):* Let $\boldsymbol{z} = [z_1 \cdots z_t]^{\text{T}} \sim \mathcal{CN}(\boldsymbol{0}, \boldsymbol{\Sigma}_z)$. Any Lipschitz-continuous function $\psi: \mathbb{C} \to \mathbb{C}$ satisfies: For $1 \leq i \leq t$ and $1 \leq j \leq t$,

$$\text{E}\{z_j^* \psi(z_i)\} = \frac{\text{E}\{z_j^* z_i\}}{\text{E}\{\|z_i\|^2\}}\text{E}\{z_i^* \psi(z_i)\}. \tag{13a}$$

Furthermore, any differentiable[2] and Lipschitz-continuous function $\phi: \mathbb{C}^t \to \mathbb{C}$ satisfies: For $1 \leq i \leq t$,

$$\text{E}\{z_i^* \phi(\boldsymbol{z})\} = \textstyle\sum_{j=1}^t \text{E}\{z_i^* z_j\}\text{E}\{\partial\phi(\boldsymbol{z})/\partial z_j\}. \tag{13b}$$

The generalized Stein's lemma has been widely used in the design of orthogonal iterative processes (e.g., OAMP/VAMP and expectation propagation algorithm).

### D. Overview of BO-OAMP/VAMP

The following BO-OAMP/VAMP [18], [25] is a non-memory iterative process that solves the problem in (2) with right-unitarily-invariant matrix.

*BO-OAMP/VAMP [18], [25]:* Let $\rho_t = \sigma^2/v_t^\phi$, $\hat{\phi}_t(\cdot)$ be a separable and Lipschitz-continuous estimator, and define $\hat{\gamma}_t(\cdot)$ as an estimate of $\boldsymbol{x}$ at $t$-th iteration:

$$\hat{\gamma}_t(\boldsymbol{x}_t) \equiv \boldsymbol{A}^{\text{H}}(\rho_t \boldsymbol{I} + \boldsymbol{A}\boldsymbol{A}^{\text{H}})^{-1}(\boldsymbol{y} - \boldsymbol{A}\boldsymbol{x}_t). \tag{14}$$

A BO-OAMP/VAMP is then defined as: Starting with $t = 1$, $v_1^\phi = 1$ and $\boldsymbol{x}_1 = \boldsymbol{0}$,

$$\text{LE}: \quad \boldsymbol{r}_t = \gamma_t(\boldsymbol{x}_t) \equiv \tfrac{1}{\epsilon_t^\gamma}\hat{\gamma}_t(\boldsymbol{x}_t) + \boldsymbol{x}_t, \tag{15a}$$

$$\text{NLE}: \quad \boldsymbol{x}_{t+1} = \phi_t(\boldsymbol{r}_t) \equiv \tfrac{1}{\epsilon_{t+1}^\phi}\big[\hat{\phi}_t(\boldsymbol{r}_t) + (\epsilon_{t+1}^\phi - 1)\boldsymbol{r}_t\big], \tag{15b}$$

where

$$\epsilon_t^\gamma = \tfrac{1}{N}\text{tr}\big\{\boldsymbol{A}^{\text{H}}(\rho_t \boldsymbol{I} + \boldsymbol{A}\boldsymbol{A}^{\text{H}})^{-1}\boldsymbol{A}\big\}, \tag{15c}$$

$$v_t^\gamma = \gamma_{\text{SE}}(v_t^\phi) \equiv v_t^\phi[(\epsilon_t^\gamma)^{-1} - 1], \tag{15d}$$

$$\epsilon_{t+1}^\phi = 1 - \tfrac{1}{N v_t^\gamma}\|\hat{\phi}_t(\boldsymbol{x} + \sqrt{v_t^\gamma}\boldsymbol{\eta}) - \boldsymbol{x}\|^2, \tag{15e}$$

$$v_{t+1}^\phi = \phi_{\text{SE}}(v_t^\gamma) \equiv v_t^\gamma[(\epsilon_{t+1}^\phi)^{-1} - 1], \tag{15f}$$

[1]The $\boldsymbol{V}$ transformed error recursion of a non-memory iterative process in (8) coincides with that in [24]. (See also Point 2 in Subsection III-D for more details)

[2]For a complex number $z = x + \text{i}y$, the complex derivative is defined as $\partial/\partial z = 1/2\,(\partial/\partial x - \text{i}\cdot\partial/\partial y)$. For a complex function $\phi$: $\mathbb{C} \to \mathbb{C}$, we write $(\partial/\partial z)(\text{Re}[\phi] + \text{i}\cdot\text{Im}[\phi])$ as $\partial\phi/\partial z$.

and $\boldsymbol{\eta} \sim \mathcal{CN}(\mathbf{0}, \boldsymbol{I})$ is independent of $\boldsymbol{x}$. The final estimate is $\hat{\phi}_t(\boldsymbol{r}_t)$.

Using Taylor series expansion, $\left(\rho_t \boldsymbol{I} + \boldsymbol{A}\boldsymbol{A}^{\rm H}\right)^{-1}$ in (14) can be rewritten to a polynomial in $\boldsymbol{A}^{\rm H}\boldsymbol{A}$. Therefore, the BO-OAMP/VAMP LE in (15a) is a special case of the NIM-LE in (8a). Without loss of generality, throughout this paper, we assume that the $\hat{\phi}_t(\cdot)$ in NLE is a symbol-by-symbol MMSE estimator given by

$$\hat{\phi}_t(\boldsymbol{r}_t) \equiv {\rm E}\{\boldsymbol{x}|\boldsymbol{r}_t, \Phi\}. \tag{16}$$

*Assumption 3:* Each posterior variance ${\rm E}\{|x_n|^2|r_{t,n}, \Phi_n\} - |\hat{\phi}_t(r_{t,n})|^2, \forall n$ is almost surely bounded, where $x_n, r_{t,n}$ and $\Phi_n$ are the $n$-th entry/constraint of $\boldsymbol{x}, \boldsymbol{r}_t$ and $\Phi$, respectively [24].

The MMSE estimator $\hat{\phi}_t(\cdot)$ in (16) is Lipschitz-continuous for $\boldsymbol{x}$ under Assumption 1, Assumption 3 and an equivalent Gaussian observation[3] $\boldsymbol{r}_t$ (see [24, Lemma 2]). It was proved in [24] that BO-OAMP/VAMP satisfies the orthogonality in (11). Hence, the IID Gaussian property in (12) holds for BO-OAMP/VAMP, which results in the following state evolution.

*Proposition 2 (State Evolution):* Suppose that Assumptions 1-3 hold. The iterative performance of BO-OAMP/VAMP can be tracked by state evolution: Starting with $t = 1$ and $v_1^{\phi} = 1$,

$$\text{LE}: \quad v_t^{\gamma} = \gamma_{\rm SE}(v_t^{\phi}), \tag{17a}$$

$$\text{NLE}: \quad v_{t+1}^{\phi} = \phi_{\rm SE}(v_t^{\gamma}), \tag{17b}$$

where $\gamma_{\rm SE}(\cdot)$ and $\phi_{\rm SE}(\cdot)$ are defined in (15d) and (15f), respectively.

The compression limit was analyzed by replica method for right-unitarily-invariant matrices in [47], [48]. Based on state evolution, it was proved in [18], [25] that the fixed points of BO-OAMP/VAMP are consistent with the replica predicted MMSE. Recently, the replica method is proved in [49] for a sub-class of right-unitarily-invariant matrices $\boldsymbol{A} = \boldsymbol{U}\boldsymbol{W}$, where $\boldsymbol{W}$ are IID Gaussian, $\boldsymbol{U}$ is a product of a finite number of independent matrices, each with IID matrix-elements that are either bounded or standard Gaussian, and $\boldsymbol{U}$ and $\boldsymbol{W}$ are independent. Therefore, we have the following lemma.

*Lemma 3 (Potential Bayes Optimality):* Suppose that Assumptions 1-3 hold and the state evolution of BO-OAMP/VAMP has a unique fixed point. Then, BO-OAMP/VAMP achieves the Bayes optimal MSE predicted by the replica method.

In general, the NLE in BO-OAMP/VAMP is a symbol-by-symbol estimator, whose time complexity is as low as $\mathcal{O}(N)$. The complexity of BO-OAMP/VAMP is dominated by LMMSE-LE, which costs $\mathcal{O}(M^2N+M^3)$ time complexity per iteration for matrix multiplication and matrix inversion. Therefore, to reduce the complexity, it is desired to design a low-complexity potentially Bayes-optimal LE for the message passing algorithm.

[3] In OAMP/VAMP and MAMP, $\boldsymbol{r}_t$ can be treated as Gaussian observation of $\boldsymbol{x}$, which can be guaranteed by the orthogonality in (11) or (20) (see Lemma 1 and Lemma 4).

## III. Memory AMP

In this section, we propose a memory AMP (MAMP) under an orthogonal principle. We construct a kind of local orthogonal memory estimators to realize the required orthogonality for MAMP.

### A. Memory AMP

*Definition 3 (Memory Iterative Process):* A memory iterative process[4] consists of a memory linear estimator (LE) and a memory non-linear estimator (NLE): Starting with $t = 1$,

$$\text{Memory LE}: \boldsymbol{r}_t = \gamma_t\left(\boldsymbol{x}_1, \cdots, \boldsymbol{x}_t\right) = \boldsymbol{Q}_t\boldsymbol{y} + \textstyle\sum_{i=1}^{t}\boldsymbol{\mathcal{P}}_{t,i}\boldsymbol{x}_i, \tag{18a}$$

$$\text{Memory NLE}: \boldsymbol{x}_{t+1} = \phi_t\left(\boldsymbol{r}_1, \cdots, \boldsymbol{r}_t\right), \tag{18b}$$

where $\boldsymbol{Q}_t\boldsymbol{A}$ and $\boldsymbol{\mathcal{P}}_{t,i}$ are polynomials in $\boldsymbol{A}^{\rm H}\boldsymbol{A}$. Without loss of generality, we assume that the norms of $\boldsymbol{Q}_t$ and $\{\boldsymbol{\mathcal{P}}_{t,i}\}$ are finite. Let $\boldsymbol{x}_t^{(1)} \equiv [\boldsymbol{x}_{1,1}^{\rm T}, \cdots, \boldsymbol{x}_{t,1}^{\rm T}]^{\rm T}$, $\boldsymbol{x}_t^{(2)} \equiv [\boldsymbol{x}_{1,2}^{\rm T}, \cdots, \boldsymbol{x}_{t,2}^{\rm T}]^{\rm T}$ and $\boldsymbol{\mathcal{P}}_t \equiv {\rm diag}\{\boldsymbol{\mathcal{P}}_{t,1}, \cdots, \boldsymbol{\mathcal{P}}_{t,t}\}$ (a block diagonal matrix). For any $\boldsymbol{x}_t^{(1)}, \boldsymbol{x}_t^{(2)} \in \mathbb{C}^{Nt}$,

$$\|\gamma_t(\boldsymbol{x}_t^{(1)}) - \gamma_t(\boldsymbol{x}_t^{(2)})\| = \|\boldsymbol{\mathcal{P}}_t(\boldsymbol{x}_t^{(1)} - \boldsymbol{x}_t^{(2)})\| \tag{19a}$$

$$\leq \|\boldsymbol{\mathcal{P}}_t\|\|\boldsymbol{x}_t^{(1)} - \boldsymbol{x}_t^{(2)}\|, \tag{19b}$$

where $\|\boldsymbol{\mathcal{P}}_t\|$ is finite since $\{\|\boldsymbol{\mathcal{P}}_{t,i}\|, \forall i\}$ are finite. Hence, $\gamma_t(\cdot)$ in (18a) is Lipschitz-continuous.

We call (18) a memory iterative process since $\gamma_t(\cdot)$ and $\phi_t(\cdot)$ contain memories $\{\boldsymbol{x}_i, 1 \leq i < t\}$ and $\{\boldsymbol{r}_i, 1 \leq i < t\}$, respectively. Fig. 2 give a graphical illustration of the memory iterative process. The memory iterative process is degraded to the classic non-memory iterative process if $\gamma_t\left(\boldsymbol{x}_1, \cdots, \boldsymbol{x}_t\right) = \gamma_t\left(\boldsymbol{x}_t\right)$ and $\phi_t\left(\boldsymbol{r}_1, \cdots, \boldsymbol{r}_t\right) = \phi_t\left(\boldsymbol{r}_t\right)$.

*Definition 4 (Memory AMP):* The memory iterative process in III-A is said to be a memory AMP (MAMP) if the following orthogonal constraints hold for $1 \leq t' \leq t$:

$$\lim_{N\to\infty} \tfrac{1}{N}\boldsymbol{x}^{\rm H}\boldsymbol{g}_t \overset{\rm a.s.}{=} 0, \tag{20a}$$

$$\lim_{N\to\infty} \tfrac{1}{N}\boldsymbol{f}_{t'}^{\rm H}\boldsymbol{g}_t \overset{\rm a.s.}{=} 0, \tag{20b}$$

$$\lim_{N\to\infty} \tfrac{1}{N}\boldsymbol{g}_{t'}^{\rm H}\boldsymbol{f}_{t+1} \overset{\rm a.s.}{=} 0. \tag{20c}$$

Specifically, $\gamma_t(\cdot)$ and $\phi_t(\cdot)$ can be partial memory estimators such as

$$\gamma_t\left(\boldsymbol{x}_1, \cdots, \boldsymbol{x}_t\right) = \gamma_t(\boldsymbol{x}_{\tau_{\gamma_t}}, \cdots, \boldsymbol{x}_t), \quad 1 \leq \tau_{\gamma_t} \leq t, \tag{21a}$$

$$\phi_t\left(\boldsymbol{x}_1, \cdots, \boldsymbol{x}_t\right) = \phi_t(\boldsymbol{r}_{\tau_{\phi_t}}, \cdots, \boldsymbol{r}_t), \quad 1 \leq \tau_{\phi_t} \leq t. \tag{21b}$$

[4] The $\boldsymbol{V}$-transformed error recursion of the memory iterative process in (18) coincides with that in [31] (See Point 2 in Subsection III-D).

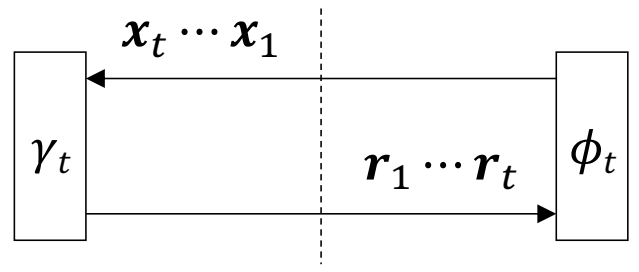

Fig. 2. Graphic illustration for a memory iterative process.

In this case, based on the generalized Stein's lemma in Lemma 2, the full orthogonality in (20b) and (20c) can be reduced to the following partial orthogonality.

$$\lim_{N\to\infty} \tfrac{1}{N}\boldsymbol{f}_{t'}^{\rm H}\boldsymbol{g}_t \overset{\rm a.s.}{=} 0, \quad \tau_{\gamma_t} \le t' \le t, \tag{22a}$$

$$\lim_{N\to\infty} \tfrac{1}{N}\boldsymbol{g}_{t'}^{\rm H}\boldsymbol{f}_{t+1} \overset{\rm a.s.}{=} 0, \quad \tau_{\phi_t} \le t' \le t. \tag{22b}$$

When $\tau_{\gamma_t} = \tau_{\phi_t} = t$, the MAMP degenerates to the non-memory process discussed in Section II.

Note that in general, the full orthogonalization in (20) is stricter than the step-by-step orthogonalization in (11), except for the non-memory iterative process in Subsection II-C. For example, the step-by-step orthogonalization in (11) in a non-memory iterative process (e.g., OAMP/VAMP) is not sufficient to guarantee the asymptotic IID Gaussianity for MAMP. Thus, a stricter orthogonality is required, i.e., the error of $\gamma_t(\cdot)$ is required to be orthogonal to all input estimation errors. This is the main difference between MAMP and the non-memory AMP-type algorithms (e.g., OAMP/VAMP). In III-B and III-C, we will show how to construct a memory LE and a memory NLE to satisfy the orthogonality in (20).

Define the covariances as follows:

$$v_{t,t'}^{\gamma} = (v_{t',t}^{\gamma})^* \equiv \tfrac{1}{N}{\rm E}\{\boldsymbol{g}_t^{\rm H}\boldsymbol{g}_{t'}\}, \tag{23a}$$

$$v_{t,t'}^{\phi} = (v_{t',t}^{\phi})^* \equiv \tfrac{1}{N}{\rm E}\{\boldsymbol{f}_t^{\rm H}\boldsymbol{f}_{t'}\}. \tag{23b}$$

The following lemma is proved in [31], [46] (see "Proof of Property (A1)" in Appendix A-D in [31]). It is important for the state evolution of MAMP.

*Lemma 4 (Orthogonality and Asymptotic IID Gaussianity):* Suppose that Assumptions 1-2 hold. For MAMP with the orthogonality in (20), the $\{\gamma_t(\cdot)\}$ in (18a) and separable-and-Lipschitz-continuous [44] $\{\phi_t(\cdot)\}$, we have $\forall 1\le t'\le t$,

$$\lim_{N\to\infty} v_{t,t'}^{\gamma} \overset{\rm a.s.}{=} \tfrac{1}{N}{\rm E}\Big\{\big[\gamma_t(\boldsymbol{x}+\boldsymbol{\eta}_1^{\phi},\dots,\boldsymbol{x}+\boldsymbol{\eta}_t^{\phi})-\boldsymbol{x}\big]^{\rm H} \big[\gamma_{t'}(\boldsymbol{x}+\boldsymbol{\eta}_1^{\phi},\dots,\boldsymbol{x}+\boldsymbol{\eta}_{t'}^{\phi})-\boldsymbol{x}\big]\Big\}, \tag{24a}$$

$$\lim_{N\to\infty} v_{t+1,t'+1}^{\phi} \overset{\rm a.s.}{=} \tfrac{1}{N}{\rm E}\Big\{\big[\phi_t(\boldsymbol{x}+\boldsymbol{\eta}_1^{\gamma},\dots,\boldsymbol{x}+\boldsymbol{\eta}_t^{\gamma})-\boldsymbol{x}\big]^{\rm H} \big[\phi_{t'}(\boldsymbol{x}+\boldsymbol{\eta}_1^{\gamma},\dots,\boldsymbol{x}+\boldsymbol{\eta}_{t'}^{\gamma})-\boldsymbol{x}\big]\Big\}, \tag{24b}$$

where $[\boldsymbol{\eta}_1^{\gamma}\dots\boldsymbol{\eta}_t^{\gamma}]$ and $[\boldsymbol{\eta}_1^{\phi}\dots\boldsymbol{\eta}_t^{\phi}]$ are column-wise IID Gaussian, row-wise joint-Gaussian and independent of $\boldsymbol{x}$. In addition, $\boldsymbol{\eta}_t^{\phi}\sim\mathcal{CN}(\boldsymbol{0},v_{t,t}^{\phi}\boldsymbol{I})$ with ${\rm E}\{\boldsymbol{\eta}_t^{\phi}(\boldsymbol{\eta}_{t'}^{\phi})^{\rm H}\} = v_{t',t}^{\phi}\boldsymbol{I}$ and $\boldsymbol{\eta}_t^{\gamma}\sim\mathcal{CN}(\boldsymbol{0},v_{t,t}^{\gamma}\boldsymbol{I})$ with ${\rm E}\{\boldsymbol{\eta}_t^{\gamma}(\boldsymbol{\eta}_{t'}^{\gamma})^{\rm H}\}=v_{t',t}^{\gamma}\boldsymbol{I}$.

### B. Orthogonal Memory LE

We construct an orthogonal memory LE $\gamma_t(\cdot)$ to satisfy the orthogonality in (20a) and (20b).

*Definition 5 (Orthogonal Memory LE):* A general orthogonal memory LE is defined as

$$\gamma_t(\boldsymbol{x}_1,\cdots,\boldsymbol{x}_t) = \boldsymbol{Q}_t\boldsymbol{y} + \textstyle\sum_{i=1}^t \boldsymbol{\mathcal{P}}_{t,i}\boldsymbol{x}_i, \tag{25a}$$

where $\boldsymbol{Q}_t\boldsymbol{A}$ and $\boldsymbol{\mathcal{P}}_{t,i}$ are polynomials in $\boldsymbol{A}^{\rm H}\boldsymbol{A}$, and

$$\tfrac{1}{N}{\rm tr}\big\{\boldsymbol{Q}_t\boldsymbol{A}\big\} = 1, \tag{25b}$$

$${\rm tr}\big\{\boldsymbol{\mathcal{P}}_{t,i}\big\} = 0, \quad i=1,\dots,t. \tag{25c}$$

Recall $\boldsymbol{g}_t = \gamma_t(\boldsymbol{x}_1,\cdots,\boldsymbol{x}_t)-\boldsymbol{x}$ and $\boldsymbol{f}_i = \boldsymbol{x}_i-\boldsymbol{x}$, $\forall i$, and assume[5] that $\boldsymbol{F}_t = [\boldsymbol{f}_1\cdots\boldsymbol{f}_t]$ is column-wise IID with zero mean and independent of $\boldsymbol{n}$. Then, the memory LE in (25) satisfies the orthogonalization:

$$\lim_{N\to\infty} \tfrac{1}{N}\boldsymbol{x}^{\rm H}\boldsymbol{g}_t \overset{\rm a.s.}{=} 0, \tag{26a}$$

$$\lim_{N\to\infty} \tfrac{1}{N}\boldsymbol{f}_{t'}^{\rm H}\boldsymbol{g}_t \overset{\rm a.s.}{=} 0, \quad \forall 1\le t'\le t. \tag{26b}$$

*Proof:* First, we rewrite (25) to

$$\gamma_t(\boldsymbol{x}_1,\cdots,\boldsymbol{x}_t) = \boldsymbol{Q}_t(\boldsymbol{A}\boldsymbol{x}+\boldsymbol{n}) + \textstyle\sum_{i=1}^t \boldsymbol{\mathcal{P}}_{t,i}(\boldsymbol{x}+\boldsymbol{f}_i). \tag{27}$$

Then, we have

$$\boldsymbol{g}_t = \underbrace{(\boldsymbol{Q}_t\boldsymbol{A}-\boldsymbol{I}+\textstyle\sum_{i=1}^t \boldsymbol{\mathcal{P}}_{t,i})}_{\boldsymbol{L}_t}\boldsymbol{x} + \textstyle\sum_{i=1}^t \boldsymbol{\mathcal{P}}_{t,i}\boldsymbol{f}_i + \boldsymbol{Q}_t\boldsymbol{n}. \tag{28}$$

For $1\le t'\le t$,

$$\lim_{N\to\infty} \tfrac{1}{N}\boldsymbol{x}^{\rm H}\boldsymbol{g}_t \overset{\rm a.s.}{=} \tfrac{1}{N}\Big[{\rm E}\Big\{{\rm tr}\big\{\boldsymbol{x}\boldsymbol{x}^{\rm H}\boldsymbol{L}_t\big\}\Big\} + \sum_{i=1}^t {\rm E}\Big\{{\rm tr}\big\{\boldsymbol{f}_i\boldsymbol{x}^{\rm H}\boldsymbol{\mathcal{P}}_{t,i}\big\}\Big\} + {\rm E}\Big\{{\rm tr}\big\{\boldsymbol{n}\boldsymbol{x}^{\rm H}\boldsymbol{Q}_t\big\}\Big\}\Big] \tag{29a}$$

$$= 0, \tag{29b}$$

$$\lim_{N\to\infty} \tfrac{1}{N}\boldsymbol{f}_{t'}^{\rm H}\boldsymbol{g}_t \overset{\rm a.s.}{=} \tfrac{1}{N}\Big[{\rm E}\Big\{{\rm tr}\big\{\boldsymbol{x}\boldsymbol{f}_{t'}^{\rm H}\boldsymbol{L}_t\big\}\Big\} + \sum_{i=1}^t {\rm E}\Big\{{\rm tr}\big\{\boldsymbol{f}_i\boldsymbol{f}_{t'}^{\rm H}\boldsymbol{\mathcal{P}}_{t,i}\big\} + {\rm E}\Big\{{\rm tr}\big\{\boldsymbol{n}\boldsymbol{f}_{t'}^{\rm H}\boldsymbol{Q}_t\big\}\Big\}\Big] \tag{29c}$$

$$= 0, \tag{29d}$$

where (29a) and (29c) follow that ${\rm E}\{\boldsymbol{x}\boldsymbol{x}^{\rm H}\} = \boldsymbol{I}$, ${\rm E}\{\boldsymbol{n}\boldsymbol{x}^{\rm H}\} = {\rm E}\{\boldsymbol{n}\boldsymbol{f}_{t'}^{\rm H}\} = \boldsymbol{0}$, ${\rm E}\{\boldsymbol{f}_i\boldsymbol{x}^{\rm H}\} \overset{\rm a.s.}{=} \frac{1}{N}(\boldsymbol{x}^{\rm H}\boldsymbol{f}_i)\boldsymbol{I}$, ${\rm E}\{\boldsymbol{x}\boldsymbol{f}_{t'}^{\rm H}\} \overset{\rm a.s.}{=} \frac{1}{N}(\boldsymbol{f}_{t'}^{\rm H}\boldsymbol{x})\boldsymbol{I}$ and ${\rm E}\{\boldsymbol{f}_i\boldsymbol{f}_{t'}^{\rm H}\} = v_{t',i}^{\phi}\boldsymbol{I}$, and (29b) and (29d) follow ${\rm tr}\big\{\boldsymbol{Q}_t\boldsymbol{A}-\boldsymbol{I}\big\} = {\rm tr}\big\{\boldsymbol{\mathcal{P}}_{t,i}\big\} = 0$ (see (25)). ■

The orthogonal memory LE in (25) is a generalization of the de-correlated LE in OAMP [18]. The lemma below constructs an orthogonal memory LE based on a general memory LE.

*Lemma 5 (Orthogonal Memory LE Construction):* Given a general memory LE (not necessary orthogonal)

$$\hat{\gamma}_t(\boldsymbol{x}_1,\cdots,\boldsymbol{x}_t) = \boldsymbol{Q}_t\boldsymbol{y} + \textstyle\sum_{i=1}^t \boldsymbol{P}_{t,i}\boldsymbol{x}_i, \tag{30a}$$

where $\boldsymbol{Q}_t\boldsymbol{A}$ and $\boldsymbol{P}_{t,i}$ are polynomials in $\boldsymbol{A}^{\rm H}\boldsymbol{A}$, we can construct an orthogonal memory LE by

$$\gamma_t(\boldsymbol{x}_1,\cdots,\boldsymbol{x}_t) = \tfrac{1}{\varepsilon_t^{\gamma}}\big(\hat{\gamma}_t(\boldsymbol{x}_1,\cdots,\boldsymbol{x}_t) - [\boldsymbol{x}_1\cdots\boldsymbol{x}_t]\boldsymbol{p}_t\big), \tag{30b}$$

where

$$\varepsilon_t^{\gamma} = \tfrac{1}{N}{\rm tr}\big\{\boldsymbol{Q}_t\boldsymbol{A}\big\}, \quad \boldsymbol{p}_t = \big[\tfrac{1}{N}{\rm tr}\{\boldsymbol{P}_{t,1}\}\cdots\tfrac{1}{N}{\rm tr}\{\boldsymbol{P}_{t,t}\}\big]^{\rm T}. \tag{30c}$$

It is easy to verify that $\gamma_t(\cdot)$ in (30) satisfies (25), i.e., it is an orthogonal memory LE.

### C. Orthogonal Memory NLE

We construct an orthogonal memory NLE $\phi_t(\cdot)$ to satisfy the orthogonality in (20c).

[5]The assumption that $\boldsymbol{F}$ is column-wise IID with zero mean and independent of $\boldsymbol{n}$ is guaranteed by the orthogonal memory NLE given in III-C by induction [31].

*Definition 6 (Orthogonal Memory NLE):* Recall $\boldsymbol{f}_{t+1} = \phi_t(\boldsymbol{r}_1,\cdots,\boldsymbol{r}_t) - \boldsymbol{x}$ and $\boldsymbol{g}_i = \boldsymbol{r}_i - \boldsymbol{x}$, $\forall i$. An memory NLE $\phi_t(\boldsymbol{r}_1,\cdots,\boldsymbol{r}_t)$ is called orthogonal memory NLE if

$$\lim_{N\to\infty} \tfrac{1}{N}\boldsymbol{f}_{t+1}^{\rm H}\boldsymbol{g}_{t'} \overset{\rm a.s.}{=} 0, \quad \forall 1\le t'\le t. \tag{31}$$

The orthogonal memory NLE in (31) is a generalization of the divergence-free NLE in OAMP [18]. The lemma below constructs an orthogonal memory NLE based on an arbitrary memory NLE.

*Lemma 6:* Assume[6] that $\boldsymbol{G}_t = [\boldsymbol{g}_1\cdots\boldsymbol{g}_t]$ are column-wise IID Gaussian and row-wise joint Gaussian with zero mean and $\lim_{N\to\infty}\frac{1}{N}\boldsymbol{G}_t^{\rm H}\boldsymbol{x} \overset{\rm a.s.}{=} \boldsymbol{0}$. Given an arbitrary separable-and-Lipschitz-continuous $\hat{\phi}_t(\boldsymbol{r}_1,\cdots,\boldsymbol{r}_t)$ (not necessary orthogonal), we can construct an orthogonal memory NLE by

$$\phi_t(\boldsymbol{r}_1,\cdots,\boldsymbol{r}_t) = \tfrac{1}{\varepsilon_t^\phi}\big(\hat{\phi}_t(\boldsymbol{r}_1,\cdots,\boldsymbol{r}_t) - [\boldsymbol{r}_1\cdots\boldsymbol{r}_t]\boldsymbol{u}_t\big), \tag{32a}$$

where

$$\boldsymbol{u}_t = (\boldsymbol{G}_t^{\rm H}\boldsymbol{G}_t)^{-1}\boldsymbol{G}_t^{\rm H}\hat{\phi}_t(\boldsymbol{r}_1,\cdots,\boldsymbol{r}_t), \tag{32b}$$

and $\varepsilon_t^\phi$ is an arbitrary constant.

*Proof:* We prove that $\phi_t(\cdot)$ in (32) satisfies the orthogonalization in (31) as follows.

$$\begin{aligned}
&\lim_{N\to\infty}\tfrac{1}{N}\boldsymbol{G}_t^{\rm H}\boldsymbol{f}_{t+1}\\
&= \lim_{N\to\infty}\tfrac{1}{N}\boldsymbol{G}_t^{\rm H}\big[\phi_t(\boldsymbol{r}_1,\cdots,\boldsymbol{r}_t) - \boldsymbol{x}\big] &(33a)\\
&\overset{\rm a.s.}{=} \lim_{N\to\infty}\tfrac{1}{N}\boldsymbol{G}_t^{\rm H}\phi_t(\boldsymbol{r}_1,\cdots,\boldsymbol{r}_t) &(33b)\\
&= \lim_{N\to\infty}\tfrac{1}{N\varepsilon_t^\phi}\big[\boldsymbol{G}_t^{\rm H}\big(\hat{\phi}_t(\boldsymbol{r}_1,\cdots,\boldsymbol{r}_t) - [\boldsymbol{r}_1\cdots\boldsymbol{r}_t]\boldsymbol{u}_t\big)\big] &(33c)\\
&\overset{\rm a.s.}{=} \tfrac{1}{N\varepsilon_t^\phi}\big(\boldsymbol{G}_t^{\rm H}\hat{\phi}(\boldsymbol{r}_1,\cdots,\boldsymbol{r}_t) - \boldsymbol{G}_t^{\rm H}\boldsymbol{G}_t\boldsymbol{u}_t\big) &(33d)\\
&= \boldsymbol{0}, &(33e)
\end{aligned}$$

where (33b) and (33d) follow $\lim_{N\to\infty}\frac{1}{N}\boldsymbol{G}_t^{\rm H}\boldsymbol{x} \overset{\rm a.s.}{=} \boldsymbol{0}$, (33c) follows (32a), and (33e) follows (32b). ■

In general, $\varepsilon_t^\phi$ in (32a) is determined by minimizing the MSE of $\phi_t(\cdot)$.

*Lemma 7:* Assume that $\boldsymbol{G}_t = [\boldsymbol{g}_1\cdots\boldsymbol{g}_t]$ are column-wise IID Gaussian and row-wise joint Gaussian with zero mean and $\lim_{N\to\infty}\frac{1}{N}\boldsymbol{G}_t^{\rm H}\boldsymbol{x} \overset{\rm a.s.}{=} \boldsymbol{0}$. For an arbitrary separable-and-Lipschitz-continuous and differentiable $\hat{\phi}_t(\boldsymbol{r}_1,\cdots,\boldsymbol{r}_t)$, using generalized Stein's Lemma [45] (see Lemma 2), we have

$$\lim_{N\to\infty}\tfrac{1}{N}\boldsymbol{G}_t^{\rm H}\hat{\phi}_t \overset{\rm a.s.}{=} \tfrac{1}{N}\boldsymbol{G}_t^{\rm H}\boldsymbol{G}_t\big[{\rm E}\{\tfrac{\partial\hat{\phi}_t}{\partial r_1}\}\cdots{\rm E}\{\tfrac{\partial\hat{\phi}_t}{\partial r_t}\}\big]^{\rm T}. \tag{34}$$

Then, $\boldsymbol{u}_t$ in (32b) can be rewritten to

$$\boldsymbol{u}_t = \big[{\rm E}\{\tfrac{\partial\hat{\phi}_t}{\partial r_1}\}\cdots{\rm E}\{\tfrac{\partial\hat{\phi}_t}{\partial r_t}\}\big]^{\rm T}. \tag{35}$$

Fig. 3 gives a graphic illustration of MAMP with orthogonal $\gamma_t$ constructed in III-B and orthogonal $\phi_t$ constructed in III-C.

### *D. Connection to Related Frameworks and Algorithms*

*1) Connection to the Non-Memory Iterative Process in II-C:* In (18), let $\mathcal{P}_{t,i} = \boldsymbol{0}, \forall i<t$ and $\phi_t(\boldsymbol{r}_1,\cdots,\boldsymbol{r}_t) = \phi_t(\boldsymbol{r}_t)$. Then the MIP is reduced to the non-memory iterative process in (8). In this case, the full orthogonality in (20) for MAMP can be reduced to the step-by-step orthogonality in (11) using the generalized Stein's lemma in Lemma 2. As a result, Lemma 4 is reduced to Lemma 1.

*2) Connection to the Unified Framework:* MAMP basically coincides with the unified framework in [31], [46].

- Assumptions 1-2 in this paper correspond to Assumptions 1-2 in [31], respectively. Assumption 3 in [31] holds in this paper, since we consider an AWGN vector $\boldsymbol{n}\sim\mathcal{CN}(\boldsymbol{0},\sigma^2\boldsymbol{I})$. In addition, the Lipschitz-continuous $\gamma_t(\cdot)$ in (18a) and separable-and-Lipschitz-continuous $\phi_t(\cdot)$ satisfy the Lipschitz continuous conditions in Assumption 4 in [31]. In addition, the nonlinear condition in Assumption 4 in [31] is un-necessary in this paper, since we consider a noisy recovery for un-coded non-trivial IID $\boldsymbol{x}$, in which error-free recovery is impossible, i.e., the norms of error vectors are non-zero in MAMP.
- The unified framework in [31], [46] is defined on a general error model under $\boldsymbol{V}$ transform (see (9)-(12) in [31]), while the MAMP in this paper is straightforwardly defined on the estimation recursion. Recall $\boldsymbol{A}=\boldsymbol{U\Sigma V}^{\rm H}$, and let $\boldsymbol{y}'=\boldsymbol{U}^{\rm H}\boldsymbol{y}$, $\boldsymbol{n}'=\boldsymbol{U}^{\rm H}\boldsymbol{n}$ and $\boldsymbol{x}'=\boldsymbol{V}^{\rm H}\boldsymbol{x}$. Thus, $\boldsymbol{y}'=\boldsymbol{\Sigma x}'+\boldsymbol{n}'$. Then, the memory iterative process can be rewritten to

$$\begin{aligned}
\text{Memory LE}: \ &\boldsymbol{x}'_t = \boldsymbol{V}^{\rm H}\boldsymbol{x}_t, &(36a)\\
&\boldsymbol{r}'_t = \boldsymbol{\Sigma}_{\boldsymbol{Q}_t\boldsymbol{A}}\boldsymbol{y}' + \textstyle\sum_{i=1}^t \boldsymbol{\Sigma}_{\mathcal{P}_{t,i}}\boldsymbol{x}'_i, &(36b)\\
&\boldsymbol{r}_t = \boldsymbol{V}\boldsymbol{r}'_t, &(36c)\\
\text{NLE}: \ &\boldsymbol{x}_{t+1} = \phi_t(\boldsymbol{r}_1,\ldots,\boldsymbol{r}_t), &(36d)
\end{aligned}$$

  where $\boldsymbol{\Sigma}_{\boldsymbol{Q}_t\boldsymbol{A}}$ and $\boldsymbol{\Sigma}_{\mathcal{P}_{t,i}}$ are the diagonal matrices corresponding to $\boldsymbol{Q}_t\boldsymbol{A}$ and $\mathcal{P}_{t,i}$, respectively. Based on (36), we then obtain an equivalent error recursion under $\boldsymbol{V}$ transform as that in [31].
- The unified framework in [31], [46] used two derivative correction terms to realize the orthogonality in (20), which results in the asymptotic error Gaussianity in Lemma 4. The MAMP in this paper is straightforwardly defined on the orthogonality in (20). Since the derivative correction is a sufficient (not necessary) condition of the orthogonality, the MAMP is more general than that in [31], [46] (e.g., for non-differentiable processors [50]).

*3) Connection to AMP and CAMP:* Both AMP and CAMP

[6]The assumption that $\lim_{N\to\infty}\frac{1}{N}\boldsymbol{G}_t^{\rm H}\boldsymbol{x}\overset{\rm a.s.}{=}\boldsymbol{0}$ and $\boldsymbol{G}_t$ are column-wise IID Gaussian and row-wise joint Gaussian with zero mean are guaranteed by the orthogonal memory LE given in III-B by induction [31].

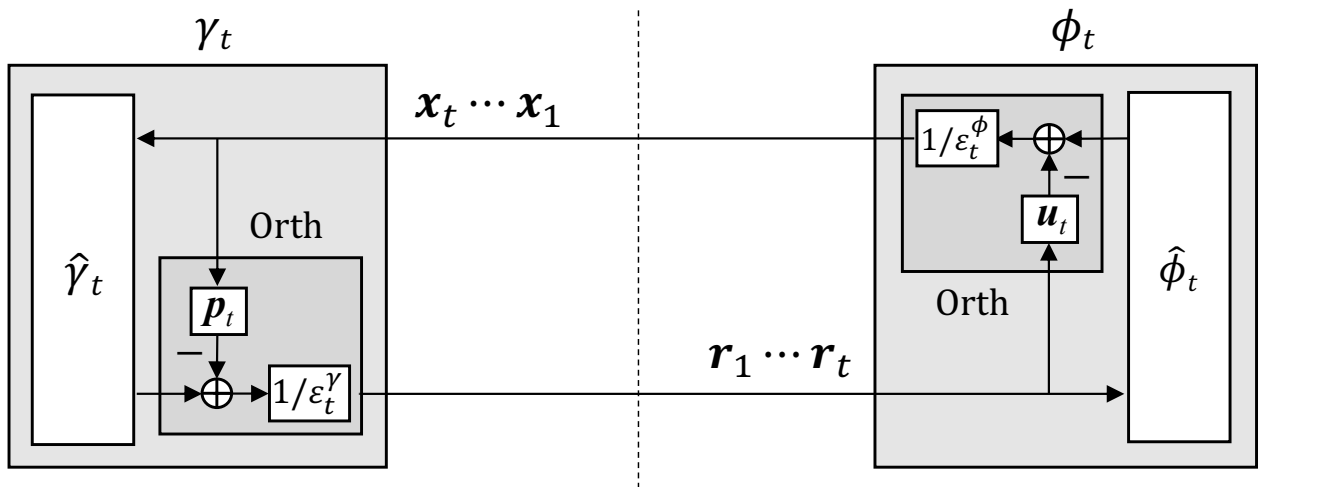

Fig. 3. Graphic illustration of MAMP involving two local orthogonal processors $\gamma_t$ (for $\Gamma$) and $\phi_t$ (for $\Phi$), which are realized by orthogonalization for $\hat{\gamma}_t$ (see (30)) and $\hat{\phi}_t$ (see (32)).

are instances of the unified framework in [31], [46]. As mentioned above, MAMP basically coincides with the unified framework in [31], [46]. Hence, AMP and CAMP can be treated as instances of MAMP.

*4) Connection to OAMP/VAMP:* An OAMP/VAMP is an instance of MAMP as shown below.

- Let $\boldsymbol{P}_{t,t} = -\boldsymbol{Q}_t\boldsymbol{A}$ and $\boldsymbol{P}_{t,t'} = \boldsymbol{0}, \forall t' < t$. Then, the orthogonal memory LE in (30b) is degraded to the decorrelated LE in OAMP.
- Let $\boldsymbol{u}_t = \big[0, \cdots, 0, \mathrm{E}\{\frac{d\hat{\phi}_t}{dr_t}\}\big]$. In this case, the orthogonal memory NLE in (32) is degraded to the divergence-free NLE in OAMP.

Furthermore, BO-OAMP/VAMP is an instance of MAMP as shown below.

- Let $\boldsymbol{Q}_t = \boldsymbol{A}^{\rm H}(\rho_t\boldsymbol{I} + \boldsymbol{A}^{\rm H}\boldsymbol{A})^{-1}$, $\boldsymbol{P}_{t,t} = -\boldsymbol{Q}_t\boldsymbol{A}$ and $\boldsymbol{P}_{t,t'} = \boldsymbol{0}, \forall t' < t$. Then, the orthogonal memory LE in (30b) is degraded to the LMMSE-LE in (15a) of BO-OAMP/VAMP.
- Let $\varepsilon_t^{\phi} = 1 - \mathrm{E}\{\frac{d\hat{\phi}_t}{dr_t}\}$, $\hat{\phi}_t$ be an MMSE estimator given by $\hat{\phi}_t(\boldsymbol{r}_t) \equiv \mathrm{E}\{\boldsymbol{x}|\boldsymbol{r}_t, \Phi\}$. Then, $\boldsymbol{u}_t = \big[0, \cdots, 0, \mathrm{E}\{\frac{d\hat{\phi}_t}{dr_t}\}\big]$ and $\mathrm{E}\{\frac{d\hat{\phi}_t}{dr_t}\} \overset{\text{a.s.}}{=} \lim\limits_{N\to\infty} \frac{\|\hat{\phi}_t(\boldsymbol{r}_t)-\boldsymbol{x}\|^2}{\|\boldsymbol{r}_t-\boldsymbol{x}\|^2}$. In this case, the orthogonal memory NLE in (32) is degraded to the MMSE-NLE in (15b) of BO-OAMP/VAMP.

## IV. Bayes-optimal Memory AMP

In this section, we construct a Bayes-optimal MAMP (BO-MAMP) using a long-memory MF-LE, and provide some properties of BO-MAMP such as orthogonality and asymptotic IID Gaussianity, based on which we derive the state evolution of BO-MAMP. Interestingly, with the assistance of memory, we prove that the state evolution of the optimized BO-MAMP converges to the same fixed point as that of BO-OAMP/VAMP even if a low-complexity matched filter is used.

### A. Bayes-Optimal Memory AMP (BO-MAMP)

The following is a BO-MAMP algorithm. The convergence of BO-MAMP is optimized with relaxation parameters $\{\theta_t\}$, weights $\{\xi_t\}$ and damping vectors $\{\boldsymbol{\zeta}_t\}$.

*Bayes-Optimal MAMP (BO-MAMP):* Let $\boldsymbol{B} = \lambda^{\dagger}\boldsymbol{I} - \boldsymbol{A}\boldsymbol{A}^{\rm H}$. Consider a long-memory linear estimation:

$$\hat{\boldsymbol{r}}_t = \theta_t\boldsymbol{B}\hat{\boldsymbol{r}}_{t-1} + \xi_t(\boldsymbol{y} - \boldsymbol{A}\boldsymbol{x}_t). \tag{37}$$

A BO-MAMP is: Starting with $t = 1$ and $\boldsymbol{x}_1 = \hat{\boldsymbol{r}}_0 = \boldsymbol{0}$,

$$\text{Memory LE}: \quad \boldsymbol{r}_t = \gamma_t(\boldsymbol{x}_1, \cdots, \boldsymbol{x}_t) \equiv \tfrac{1}{\varepsilon_t^{\gamma}}\big(\boldsymbol{A}^{\rm H}\hat{\boldsymbol{r}}_t - \textstyle\sum_{i=1}^{t} p_{t,i}\boldsymbol{x}_i\big), \tag{38a}$$

$$\text{NLE}: \quad \boldsymbol{x}_{t+1} = \bar{\phi}_t(\boldsymbol{r}_t) = [\boldsymbol{x}_1, \dots, \boldsymbol{x}_t, \phi_t(\boldsymbol{r}_t)] \cdot \boldsymbol{\zeta}_{t+1}, \tag{38b}$$

where $\boldsymbol{\zeta}_{t+1} = [\zeta_{t+1,1}, \dots, \zeta_{t+1,t+1}]^{\rm T}$, and $\phi_t(\cdot)$ is the same as that in BO-OAMP/VAMP (see (15b)).

See Algorithm 1 in Appendix H for more details of BO-MAMP. The following are some intuitive interpretations of the parameters in BO-MAMP. We will optimize these parameters in Section V.

- In a memory LE, all preceding messages are utilized in $\sum_{i=1}^{t} p_{t,i}\boldsymbol{x}_i$ to guarantee the orthogonality in (20). In NLE, the preceding messages $\{\boldsymbol{x}_1, \dots, \boldsymbol{x}_t\}$ are used for damping to guarantee and improve the convergence of BO-MAMP.
- The relaxation parameter $\theta_t = (\lambda^{\dagger} + \rho_t)^{-1}$, optimized in Subsection V-A, minimizes the spectral radius of $\theta_t\boldsymbol{B}$ to improve the convergence speed of BO-MAMP and ensure that BO-MAMP has a potentially Bayes-optimal fixed point if converges.
- The weight $\xi_t$, optimized in Subsection V-B, adjusts the contribution of $\boldsymbol{x}_t$ to the estimate $\boldsymbol{r}_t$. Higher weight $\xi_t$ means that the contribution of $\boldsymbol{x}_t$ to $\boldsymbol{r}_t$ increases. As a result, the optimized $\xi_t$ accelerates the convergence of BO-MAMP. However, the choice of $\xi_t$ does not affect the convergence and fixed point (i.e., potential Bayes optimality) of the state evolution for BO-MAMP. See Theorem 2 and its proof in Appendix G.
- The normalization parameters $\{\varepsilon_t^{\gamma}\}$ and the orthogonalization parameters $\{p_{t,i}\}$, given in (40), guarantee the orthogonality in (20) (see Theorem 1).
- The damping vector $\boldsymbol{\zeta}_t = [\zeta_{t,1}, \cdots, \zeta_{t,t}]^{\rm T}$, optimized in Subsection V-C (see (61)), is under the constraint of $\sum_{i=1}^{t} \zeta_{t,i} = 1$. We optimize the linear combination of all the messages $\{\boldsymbol{x}_t\}$ to guarantee and improve the convergence of BO-MAMP. In particular, no damping is applied for $\boldsymbol{\zeta}_t = [0, \cdots, 0, 1]^{\rm T}$. In practice, we consider a maximum damping length $L$, i.e., the number of non-zero entries in $\boldsymbol{\zeta}_t$. In general, we set $L = 3$ or 2.

We call (38) memory AMP as it involves a process with memory $\{\boldsymbol{x}_i, i < t\}$ at LE which is distinct from the non-memory LE in OAMP/VAMP. Intuitively, $\boldsymbol{A}^{\rm H}\hat{\boldsymbol{r}}_t$ is an estimate of $\boldsymbol{x}$ that converges to the LMMSE estimate in BO-OAMP/VAMP. As can be seen, matrix-vector multiplications instead of matrix inverse are involved in each iteration. Thus, the time complexity of BO-MAMP is as low as $O(MN)$ per iteration, which is comparable to AMP.

### B. Orthogonality and Asymptotic IID Gaussianity

For $t \ge 0$, we define[7]

$$\boldsymbol{W}_t \equiv \boldsymbol{A}^{\rm H}\boldsymbol{B}^t\boldsymbol{A}, \tag{39a}$$

$$b_t \equiv \tfrac{1}{N}\mathrm{tr}\{\boldsymbol{B}^t\} = \textstyle\sum_{i=0}^{t}\binom{t}{i}(-1)^i(\lambda^{\dagger})^{t-i}\lambda_i, \tag{39b}$$

$$w_t \equiv \tfrac{1}{N}\mathrm{tr}\{\boldsymbol{W}_t\} = \lambda^{\dagger}b_t - b_{t+1}, \tag{39c}$$

$$\bar{w}_{i,j} \equiv \lambda^{\dagger}w_{i+j} - w_{i+j+1} - w_iw_j, \quad i \ge 0, j \ge 0. \tag{39d}$$

For $1 \le i \le t$,

$$\vartheta_{t,i} \equiv \begin{cases} \xi_t, & i = t \\ \xi_i \prod_{\tau=i+1}^{t}\theta_\tau, & i < t \end{cases}, \tag{40a}$$

[7]If the eigenvalue distribution $p_\lambda$ is available, $b_t$ can be calculated by $b_t = \int_{\lambda_{\min}}^{\lambda_{\max}}(\lambda^{\dagger} - \lambda)^t p_\lambda\, d\lambda$.

$$p_{t,i} \equiv -\vartheta_{t,i} w_{t-i}, \tag{40b}$$

$$\varepsilon_t^{\gamma} \equiv -\textstyle\sum_{i=1}^t p_{t,i}. \tag{40c}$$

Furthermore, $\vartheta_{t,i} = 1$ if $i > t$.

*Proposition 3:* The $\{\boldsymbol{r}_t\}$ in (38) and the corresponding errors can be expanded to

$$\boldsymbol{r}_t = \tfrac{1}{\varepsilon_t^{\gamma}}\big(\boldsymbol{Q}_t \boldsymbol{y} + \textstyle\sum_{i=1}^t \boldsymbol{H}_{t,i}\boldsymbol{x}_i\big), \tag{41a}$$

$$\boldsymbol{g}_t = \tfrac{1}{\varepsilon_t^{\gamma}}\big(\boldsymbol{Q}_t \boldsymbol{n} + \textstyle\sum_{i=1}^t \boldsymbol{H}_{t,i}\boldsymbol{f}_i\big), \tag{41b}$$

where $\boldsymbol{g}_t$ and $\boldsymbol{f}_i$ are respectively the estimation errors of $\boldsymbol{r}_t$ and $\boldsymbol{x}_i$ (see (10)), and

$$\boldsymbol{Q}_t \equiv \textstyle\sum_{i=1}^t \vartheta_{t,i} \boldsymbol{A}^{\rm H}\boldsymbol{B}^{t-i}, \tag{41c}$$

$$\boldsymbol{H}_{t,i} \equiv \vartheta_{t,i}(w_{t-i}\boldsymbol{I} - \boldsymbol{W}_{t-i}). \tag{41d}$$

*Proof:* By expanding (38a) one by one, we can obtain (41a) directly. Following (41a), the errors of a memory LE are given by

$$\boldsymbol{g}_t = \tfrac{1}{\varepsilon_t^{\gamma}}\big(\boldsymbol{Q}_t \boldsymbol{y} + \textstyle\sum_{i=1}^t \boldsymbol{H}_{t,i}\boldsymbol{x}_i\big) - \boldsymbol{x} \tag{42a}$$

$$= \tfrac{1}{\varepsilon_t^{\gamma}}\big[\boldsymbol{Q}_t(\boldsymbol{A}\boldsymbol{x} + \boldsymbol{n}) + \textstyle\sum_{i=1}^t \boldsymbol{H}_{t,i}(\boldsymbol{x} + \boldsymbol{f}_i)\big] - \boldsymbol{x} \tag{42b}$$

$$= \tfrac{1}{\varepsilon_t^{\gamma}}\Big[\big(\boldsymbol{Q}_t \boldsymbol{n} + \sum_{i=1}^{t} \boldsymbol{H}_{t,i}\boldsymbol{f}_i\big) + \big(\sum_{i=1}^{t} \boldsymbol{H}_{t,i} + \boldsymbol{Q}_t\boldsymbol{A} - \varepsilon_t^{\gamma}\boldsymbol{I}\big)\boldsymbol{x}\Big] \tag{42c}$$

$$= \tfrac{1}{\varepsilon_t^{\gamma}}\big(\boldsymbol{Q}_t \boldsymbol{n} + \textstyle\sum_{i=1}^t \boldsymbol{H}_{t,i}\boldsymbol{f}_i\big), \tag{42d}$$

where (42b) is from (10), and (42d) from (39)-(41). ■

The theorem below follows Lemma 4 and Proposition 3.

*Theorem 1 (Orthogonality and Asymptotic IID Gaussianity):* Suppose that Assumptions 1-3 hold. Then, the orthogonality in (20) holds for BO-MAMP. Furthermore, the limits in (24) hold for BO-MAMP.

*Proof:* See Appendix A. ■

Using Theorem 1, we can track the MSE of BO-MAMP using the state evolution discussed in the following subsection.

### C. State Evolution

Using the IID Gaussian property in Theorem 1, we can establish a state evolution equation for the dynamics of the MSE of BO-MAMP. The main challenge is the correlation between the long-memory inputs of the memory LE. It requires a covariance-matrix state evolution to track the dynamics of MSE, which is different from AMP/OAMP/VAMP whose MSE can be tracked by a scalar state evolution.

Define the covariance vectors and covariance matrices as:

$$\boldsymbol{v}_t^{\gamma} \equiv [v_{t,1}^{\gamma}, \cdots, v_{t,t}^{\gamma}]^{\rm T}, \qquad \boldsymbol{V}_t^{\gamma} \equiv [v_{i,j}^{\gamma}]_{t\times t}, \tag{43a}$$

$$\boldsymbol{v}_t^{\bar{\phi}} \equiv [v_{t,1}^{\bar{\phi}}, \cdots, v_{t,t}^{\bar{\phi}}]^{\rm T}, \qquad \boldsymbol{V}_t^{\bar{\phi}} \equiv [v_{i,j}^{\bar{\phi}}]_{t\times t}, \tag{43b}$$

where $v_{i,j}^{\gamma}$ and $v_{i,j}^{\bar{\phi}}$ is defined in (23). Furthermore, define the error covariance matrix of $\{\boldsymbol{x}_1, \cdots, \boldsymbol{x}_t, \phi_t(\boldsymbol{r}_t)\}$ (inputs of $\bar{\phi}_t$) as

$$\boldsymbol{\mathcal{V}}_{t+1} \equiv \begin{bmatrix} \boldsymbol{V}_t^{\bar{\phi}} & \begin{matrix} v_{1,t+1}^{\phi} \\ \vdots \end{matrix} \\ v_{t+1,1}^{\phi} \;\cdots & v_{t+1,t+1}^{\phi} \end{bmatrix}_{(t+1)\times(t+1)}, \tag{44a}$$

where

$$v_{t+1,t'}^{\phi} \equiv \tfrac{1}{N}{\rm E}\big\{[\phi_t(\boldsymbol{r}_t) - \boldsymbol{x}]^{\rm H}\boldsymbol{f}_{t'}\big\}. \tag{44b}$$

*Proposition 4 (State Evolution):* Suppose that Assumptions 1-3 hold. The error covariance matrices of BO-MAMP can be tracked by state evolution: Starting with $v_{1,1}^{\bar{\phi}} = 1$,

$$\text{Memory LE}: \quad \boldsymbol{v}_t^{\gamma} = \gamma_{\rm SE}(\boldsymbol{V}_t^{\bar{\phi}}), \tag{45a}$$

$$\text{NLE}: \quad \boldsymbol{v}_{t+1}^{\bar{\phi}} = \bar{\phi}_{\rm SE}(\boldsymbol{V}_t^{\gamma}), \tag{45b}$$

where $\{\boldsymbol{v}_t^{\gamma}, \boldsymbol{v}_t^{\bar{\phi}}, \boldsymbol{V}_t^{\gamma}, \boldsymbol{V}_t^{\bar{\phi}}\}$ are defined in (43), and $\gamma_{\rm SE}(\cdot)$ and $\bar{\phi}_{\rm SE}(\cdot)$ are given as follows.

- From (23a), Proposition 3 and Theorem 1, $\gamma_{\rm SE}(\cdot)$ is given by: For $1 \le t' \le t$,

$$v_{t,t'}^{\gamma} = \tfrac{1}{\varepsilon_t^{\gamma}\varepsilon_{t'}^{\gamma}}\textstyle\sum_{i=1}^t\sum_{j=1}^{t'}\vartheta_{t,i}\vartheta_{t',j}[\sigma^2 w_{t+t'-i-j} + v_{i,j}^{\bar{\phi}}\bar{w}_{t-i,t'-j}], \tag{46}$$

  where $\{w_i, \bar{w}_{i,j}\}$ are defined in (39) and $\{\varepsilon_t^{\gamma}, \vartheta_{t,i}\}$ in (40).

  *Proof:* See Appendix B. ■
- From (38b), (23b) and Theorem 1, $\bar{\phi}_{\rm SE}(\cdot)$ is given by

$$(\boldsymbol{v}_{t+1}^{\bar{\phi}})^{\rm T} = \boldsymbol{\zeta}_{t+1}^{\rm H}\boldsymbol{\mathcal{V}}_{t+1}\left[\boldsymbol{I}_{(t+1)\times t} \;\; \boldsymbol{\zeta}_{t+1}\right], \tag{47a}$$

  where $\boldsymbol{\mathcal{V}}_{t+1}$ is defined in (44), $\boldsymbol{\zeta}_{t+1} = [\zeta_{t+1,1}, \cdots, \zeta_{t+1,t+1}]^{\rm T}$ is a damping vector, $\boldsymbol{I}_{(t+1)\times t}$ is a $(t+1)\times t$ sub-matrix of $\boldsymbol{I}_{(t+1)\times(t+1)}$ excluding the last column, and

$$\boldsymbol{v}_{t+1}^{\phi} \equiv {\rm E}\big\{[\phi_t(x+\eta_t) - x]^*[\tilde{\boldsymbol{x}}_{t+1} - x\boldsymbol{1}]\big\}, \tag{47b}$$

  with $\tilde{\boldsymbol{x}}_{t+1} = [x_1 \cdots x_t \;\; \phi_t(x+\eta_t)]^{\rm T}$ and $[\eta_1 \cdots \eta_t] \sim \mathcal{CN}(\boldsymbol{0}, \boldsymbol{V}_t^{\gamma})$ being independent of $x \sim P_x$. The expectation in (47b) can be evaluated by Monte Carlo method. See Algorithm 2 in Appendix H and Appendix C for more details.

In practice, $\{v_{t,t'}^{\phi}\}$ can be simply estimated using the following proposition.

*Proposition 5:* Define $\boldsymbol{z}_t \equiv \boldsymbol{y} - \boldsymbol{A}\,\phi_{t-1}(\boldsymbol{r}_{t-1})$ and $\bar{\boldsymbol{z}}_{t'} \equiv \boldsymbol{y} - \boldsymbol{A}\boldsymbol{x}_{t'}$. Then, $\{v_{t,t'}^{\phi}\}$ can be approximated by: For $1 \le t' < t$,

$$v_{t,t}^{\phi} \overset{\rm a.s.}{=} \lim_{N\to\infty} [\tfrac{1}{N}\boldsymbol{z}_t^{\rm H}\boldsymbol{z}_t - \delta\sigma^2]/w_0, \tag{48a}$$

$$v_{t,t'}^{\phi} = (v_{t',t}^{\phi})^* \overset{\rm a.s.}{=} \lim_{N\to\infty} [\tfrac{1}{N}\boldsymbol{z}_t^{\rm H}\bar{\boldsymbol{z}}_{t'} - \delta\sigma^2]/w_0. \tag{48b}$$

*Proof:* See Appendix D. ■

To reduce the time complexity, we calculate $\{\bar{\boldsymbol{z}}_t\}$ by $\bar{\boldsymbol{z}}_t = [\bar{\boldsymbol{z}}_1 \ldots \bar{\boldsymbol{z}}_{t-1} \;\; \boldsymbol{z}_t]\boldsymbol{\zeta}_t$, and store $\{\bar{\boldsymbol{z}}_t\}$ to avoid the unnecessarily repetitive calculation in each iteration. The calculation of $\boldsymbol{z}_t$ costs $\mathcal{O}(MN)$ time complexity.

After we posted the preprint of this work [39], a rigorous state evolution was derived for WS-CG-VAMP in the latest version of [36]. The expanded output of WS-CG in Theorem 3 in [36] corresponds to the expanded output of memory LE in (41) in this paper, and the output covariance of WS-CG in Theorem 4 in [36] corresponds to the output covariance of memory LE in (46) in this paper. Moreover, the output covariance of NLE in (59) in [36] is the same as (48) in this paper, but this work wasn't cited in the latest version of [36].

TABLE II
TIME AND SPACE COMPLEXITY COMPARISON BETWEEN AMP, BO-OAMP/VAMP, CAMP AND BO-MAMP

| Algorithms | Time complexity | Space complexity |
|---|---|---|
| AMP [5] | $O(MN\mathcal{T})$ | $O(MN)$ |
| BO-OAMP/VAMP (SVD) [18], [25] | $O(M^2N+MN\mathcal{T})$ | $O(N^2+MN)$ |
| BO-OAMP/VAMP (matrix inverse) [18], [25] | $O\big((M^2N+M^3)\mathcal{T}\big)$ | $O(MN)$ |
| CAMP [31] | $O(MN\mathcal{T}+M\mathcal{T}^2+\mathcal{T}^4)$ | $O(MN+M\mathcal{T}+\mathcal{T}^2)$ |
| BO-MAMP (proposed) | $O(MN\mathcal{T}+N\mathcal{T}^2+\mathcal{T}^3+L^3\mathcal{T})$ | $O(MN+N\mathcal{T}+\mathcal{T}^2)$ |

### *D. Complexity Comparison*

We assume that $\boldsymbol{A}$ has no special structures, such as DFT or sparse matrices. Let $\mathcal{T}$ be the number of iterations. The time and space complexity of BO-MAMP is given below.

- BO-MAMP costs $O(MN\mathcal{T})$ time complexity for matrix-vector multiplications $\{\boldsymbol{A}\boldsymbol{A}^{\rm H}\hat{\boldsymbol{r}}_t\}$ and $\{\boldsymbol{A}\boldsymbol{x}_t\}$, which is dominant for the case $\mathcal{T} \ll N$, $O\big((N+M)\mathcal{T}^2\big)$ for $\{\sum_{i=1}^t p_{t,i}\boldsymbol{x}_i\}$ and $\{\boldsymbol{z}_t^{\rm H}\bar{\boldsymbol{z}}_{t'}, 1 \leq t' \leq t\}$, $O(\mathcal{T}^3)$ for calculating $\{v_{t,t}^\gamma\}$, and $O(L^3\mathcal{T})$ for calculating $\{\zeta_t\}$ (see V-C), which is negligible for $L \ll \mathcal{T} \ll N$, where $L$ is damping length.
- BO-MAMP needs $O(MN)$ space complexity to store $\boldsymbol{A}$, which is dominant for the case $\mathcal{T} \ll N$, $O\big((M+N)\mathcal{T}\big)$ space for $\{\boldsymbol{x}_t\}$ and $\{\boldsymbol{z}_t\}$, and $O\big(\mathcal{T}^2\big)$ for $\{\boldsymbol{V}_t^\phi\}$.

Table II compares the time and space complexity of BO-MAMP, BO-OAMP/VAMP, AMP and CAMP. BO-MAMP and CAMP have similar time and space complexity. For BO-OAMP/VAMP that uses matrix inverse LMMSE, the time complexity is as high as $O\big((M^2N+M^3)\mathcal{T}\big)$ due to the matrix multiplication and matrix inverse in each iteration. For SVD-BO-OAMP/VAMP, the SVD of $\boldsymbol{A}$ requires time complexity of $O\big(M^2N\big)$ unless $\boldsymbol{A}$ has a special structure that can use efficient SVD. Hence, BO-OAMP/VAMP has higher complexity than AMP, CAMP and BO-MAMP, while BO-MAMP and CAMP have comparable complexity to AMP for $\mathcal{T} \ll N$.

### *E. BO-MAMP for Identical-Singular-Value Matrices*

*1) Over-Loaded Systems:* Consider the over-loaded case $N \geq M$ and the rectangular diagonal matrix of the SVD of $\boldsymbol{A}$ is $\boldsymbol{\Sigma} = \sqrt{\frac{N}{M}}[\boldsymbol{I}_M \; \boldsymbol{0}_{M\times(N-M)}]$. Therefore, we have $\lambda^\dagger = 1$ and $\boldsymbol{B} = \boldsymbol{0}$. In this case, the performance of BO-MAMP is independent of $\xi_t$. Thus, we can set $\xi_t = \varepsilon_t^\gamma = 1$. In addition, for $1 \leq i \leq t$,

$$w_t = \begin{cases} 1, & t = 0 \\ 0, & t > 1 \end{cases}, \tag{49a}$$

$$p_{t,i} = \begin{cases} -1, & i = t \\ 0, & \text{otherwise} \end{cases}. \tag{49b}$$

Then, BO-MAMP can be simplified to: Starting with $t = 1$ and $\boldsymbol{x}_1 = \boldsymbol{0}$,

$$\boldsymbol{r}_t = \gamma_t(\boldsymbol{x}_t) \equiv \boldsymbol{A}^{\rm H}(\boldsymbol{y} - \boldsymbol{A}\boldsymbol{x}_t) + \boldsymbol{x}_t, \tag{50a}$$

$$\boldsymbol{x}_{t+1} = \phi_t(\boldsymbol{r}_t). \tag{50b}$$

*2) Under-Loaded Systems:* Consider the under-loaded case $N \leq M$ and the rectangular diagonal matrix of the SVD of $\boldsymbol{A}$ is $\boldsymbol{\Sigma} = [\boldsymbol{I}_N \; \boldsymbol{0}_{N\times(M-N)}]^{\rm T}$. Therefore, we have $\lambda^\dagger = 1$, $\boldsymbol{B} = \boldsymbol{0}$ and $\boldsymbol{A}^{\rm H}\boldsymbol{A} = \boldsymbol{I}$. Following (50), BO-MAMP can be further simplified to:

$$\boldsymbol{r} = \boldsymbol{A}^{\rm H}\boldsymbol{y}, \tag{51a}$$

$$\hat{\boldsymbol{x}} = \hat{\phi}(\boldsymbol{r}), \tag{51b}$$

where $\hat{\phi}(\cdot)$ is given in (16). In this case, one step LE and one step NLE is sufficient, and the iteration between LE and NLE is un-necessary.

Comparing (50) and (51) with the BO-OAMP/VAMP in (15), we obtain the following.

*Proposition 6:* BO-MAMP is equivalent to BO-OAMP/VAMP (MF-OAMP/VAMP) when matrix $\boldsymbol{A}$ has identical singular values.

## V. OPTIMIZATION, CONVERGENCE AND POTENTIAL BAYES OPTIMALITY OF BO-MAMP

In this section, we optimize $\{\theta_t, \xi_t, \zeta_{t+1}\}$ step-by-step for each iteration assuming that the parameters $\{\theta_{t'}, \xi_{t'}, \zeta_{t'+1}, t' < t\}$ in previous iterations are fixed. More specifically, we first optimize $\theta_t$. Given $\theta_t$, we optimize $\xi_t$. Then, given $\theta_t$ and $\xi_t$, we optimize $\zeta_{t+1}$.

### *A. Optimization of $\theta_t$*

Proposition 7 gives a Taylor series expansion for matrix inverse.

*Proposition 7 (Taylor Series Expansion):* Assume that the matrix $\boldsymbol{I} - \boldsymbol{C}$ is invertible and $\rho(\boldsymbol{C}) < 1$, where $\rho(\boldsymbol{C})$ is the spectral radius of $\boldsymbol{C}$. Then,

$$\lim_{t\to\infty} \textstyle\sum_{i=0}^t \boldsymbol{C}^i = (\boldsymbol{I} - \boldsymbol{C})^{-1}. \tag{52}$$

The convergence condition and convergence speed mainly depend on the spectral radius of $\theta_t\boldsymbol{B}$. Next, we optimize $\theta_t$ to minimize the spectral radius of $\theta_t\boldsymbol{B}$.

Consider the $t$-th iteration, the goal of a memory LE is to approach the matrix inverse $\big(\rho_t\boldsymbol{I} + \boldsymbol{A}\boldsymbol{A}^{\rm H}\big)^{-1}$, where $\rho_t = \sigma^2/v_{t,t}^{\phi}$. Following Proposition 7, we need to construct an iterative matrix $\tilde{\boldsymbol{C}}_t = \boldsymbol{I} - (\rho_t\boldsymbol{I} + \boldsymbol{A}\boldsymbol{A}^{\rm H})$, which may not satisfy the convergence condition $\rho(\tilde{\boldsymbol{C}}_t) < 1$, where $\rho(\tilde{\boldsymbol{C}}_t)$ is the spectral radius of $\tilde{\boldsymbol{C}}_t$. To solve this problem, we introduce a relaxation parameter $\theta_t$ and construct $\boldsymbol{C}_t = \boldsymbol{I} - \theta_t(\rho_t\boldsymbol{I} + \boldsymbol{A}\boldsymbol{A}^{\rm H})$ for the matrix inverse $[\theta_t(\rho_t\boldsymbol{I} + \boldsymbol{A}\boldsymbol{A}^{\rm H})]^{-1}$. The spectral radius of $\boldsymbol{C}_t$ is then minimized by [33], [34]

$$\theta_t = (\lambda^\dagger + \rho_t)^{-1}, \tag{53}$$

where $\lambda^\dagger = [\lambda_{\max} + \lambda_{\min}]/2$ and $\rho_t = \sigma^2/v_{t,t}^{\bar{\phi}}$. As a result,

$$\rho(\theta_t\boldsymbol{B}) = \rho(\boldsymbol{C}_t) \tag{54a}$$

$$= \frac{\lambda_{\max} - \lambda_{\min}}{\lambda_{\max} + \lambda_{\min} + 2\rho_t} \tag{54b}$$

$$< 1. \tag{54c}$$

Therefore, the convergence condition holds with the optimized $\theta_t$ in (53). Furthermore, the convergence speed of BO-MAMP decreases with $\rho(\theta_t\boldsymbol{B})$, i.e., the smaller $\rho(\theta_t\boldsymbol{B})$ is, the faster convergence speed it has. That is, the $\theta_t$ in (53) optimizes the convergence speed of BO-MAMP.

*Remark 1:* When $\lambda_{\min}$ and $\lambda_{\max}$ are unavailable, we can replace $\lambda_{\min}$ with a lower bound $\lambda_{\min}^{\rm low}$ and $\lambda_{\max}$ with an upper bound $\lambda_{\max}^{\rm up}$. By doing so, the condition $\rho(\theta_t\boldsymbol{B})<1$ still holds. In detail, since $\theta_t\boldsymbol{B}=\boldsymbol{I}-\theta_t(\rho_t\boldsymbol{I}+\boldsymbol{A}\boldsymbol{A}^{\rm H})$, we have $\lambda_{\theta_t\boldsymbol{B}}=1-\theta_t(\rho_t+\lambda_{\boldsymbol{A}\boldsymbol{A}^{\rm H}})$. Since $0\le\lambda_{\min}\le\lambda_{\boldsymbol{A}\boldsymbol{A}^{\rm H}}\le\lambda_{\max}$, we have $\rho(\lambda_{\theta_t\boldsymbol{B}})<1$ if and only if

$$0<\theta_t<\frac{2}{\rho_t+\lambda_{\max}}. \tag{55}$$

Let $\lambda_{\rm approx}^{\dagger}=(\lambda_{\min}^{\rm low}+\lambda_{\max}^{\rm up})/2$. Then,

$$\begin{aligned}\theta_t^{\rm approx}&\equiv\frac{1}{\rho_t+\lambda_{\rm approx}^{\dagger}} &(56a)\\ &=\frac{2}{2\rho_t+\lambda_{\min}^{\rm low}+\lambda_{\max}^{\rm up}} &(56b)\\ &<\frac{2}{\rho_t+\lambda_{\max}}, &(56c)\end{aligned}$$

where (56c) follows $\rho_t>0$, $\lambda_{\min}^{\rm low}\ge 0$ and $\lambda_{\max}^{\rm up}\ge\lambda_{\max}>0$.

Theorem 2 needs $\{\lambda_{\min},\lambda_{\max}\}$ to satisfy $\rho(\theta_t\boldsymbol{B})<1$, which does not change by using the bounds $\{\lambda_{\min}^{\rm low},\lambda_{\max}^{\rm up}\}$. That is, Theorem 2 still holds with the bounds $\{\lambda_{\min}^{\rm low},\lambda_{\max}^{\rm up}\}$. Loose bounds may decrease the convergence speed of BO-MAMP since the spectral radius of $\theta_t\boldsymbol{B}$ is not minimized. In Fig. 13, the performance of BO-MAMP with bounds $\lambda_{\min}^{\rm low}=0$ and $\lambda_{\max}^{\rm up}=(N\lambda_\tau)^{1/\tau}$ are almost the same as that with the exact values of $\{\lambda_{\min},\lambda_{\max}\}$.

*Remark 2:* The value of $\lambda_t$ increases exponentially with $t$, while $\vartheta_{t,i}$ decreases exponentially with $t-i$. To avoid the overflow, we can store $\{\vartheta_{t,i}\lambda_{t-i}\}$ or $\{\vartheta_{t,i}w_{t-i}\}$ for BO-MAMP.

### B. Optimization of $\xi_t$

Given $\theta_t$, we optimize $\xi_t$ to minimize the MSE $v_{t,t}^{\gamma}$ of $\boldsymbol{r}_t$ in BO-MAMP. From (46), we have

$$\begin{aligned}&v_{t,t}^{\gamma}(\xi_t)\\ &=\frac{1}{(\varepsilon_t^{\gamma})^2}\sum_{i=1}^{t}\sum_{j=1}^{t}\vartheta_{t,i}\vartheta_{t,j}\big(\sigma^2 w_{2t-i-j}+v_{i,j}^{\bar\phi}\bar w_{t-i,t-j}\big) &(57a)\\ &=\frac{c_{t,1}\xi_t^2-2c_{t,2}\xi_t+c_{t,3}}{w_0^2(\xi_t+c_{t,0})^2}, &(57b)\end{aligned}$$

where

$$\begin{aligned}c_{t,0}&=-\sum_{i=1}^{t-1}p_{t,i}/w_0, &(58a)\\ c_{t,1}&=\sigma^2 w_0+v_{t,t}^{\bar\phi}\bar w_{0,0}, &(58b)\\ c_{t,2}&=-\sum_{i=1}^{t-1}\vartheta_{t,i}(\sigma^2 w_{t-i}+{\rm Re}(v_{t,i}^{\bar\phi})\bar w_{0,t-i}), &(58c)\\ c_{t,3}&=\sum_{i=1}^{t-1}\sum_{j=1}^{t-1}\vartheta_{t,i}\vartheta_{t,j}\big(\sigma^2 w_{2t-i-j}+v_{i,j}^{\bar\phi}\bar w_{t-i,t-j}\big), &(58d)\end{aligned}$$

$v_{i,j}^{\bar\phi}$ is defined in (23b), $\{\varepsilon_t^{\gamma},p_{t,i},\vartheta_{t,i},w_i,\bar w_{i,j}\}$ are defined in (39) and (40), and ${\rm Re}(\cdot)$ denotes real value operation. Proposition 8 gives an optimized $\xi_t$.

*Proposition 8:* For a fixed $\theta_t$, a $\xi_t$ that minimizes the MSE $v_{t,t}^{\gamma}$ of $\boldsymbol{r}_t$ in BO-MAMP is given by: $\xi_1^{\rm opt}=1$ and for $t\ge 2$,

$$\xi_t^{\rm opt}=\begin{cases}\dfrac{c_{t,2}c_{t,0}+c_{t,3}}{c_{t,1}c_{t,0}+c_{t,2}}, & \text{if } c_{t,1}c_{t,0}+c_{t,2}\neq 0\\ +\infty, & \text{otherwise}\end{cases}, \tag{59}$$

where $\{c_{t,0},c_{t,1},c_{t,2},c_{t,3}\}$ are given in (58).

*Proof:* See Appendix E. ■

When $c_{t,1}c_{t,0}+c_{t,2}$ is equal to or close to zero, to avoid overflow in simulation, we can set $\xi_t^{\rm opt}={\rm sign}\{\xi_t^{\rm opt}\}\cdot C_{\max}$ if $|\xi_t^{\rm opt}|>C_{\max}$, where $C_{\max}$ is a sufficiently large number.

### C. Optimization of $\zeta_{t+1}$

For simplicity, we let $\mathcal{I}_t$ be an index set for the non-trivial (or effective) memories, and its complementary set $\bar{\mathcal{I}}_t$ be the index set of the trivial memories $\bar{\mathcal{I}}_t=\{\tau:\boldsymbol{x}_\tau=\boldsymbol{x}_{\tau'},\tau'<\tau\le t\}$. Hence, we have $\mathcal{I}_t\cap\bar{\mathcal{I}}_t=\emptyset$ and $\mathcal{I}_t\cup\bar{\mathcal{I}}_t=\{1,\dots,t\}$. For example, consider $t=5$ and a sequence $[\boldsymbol{x}_1,\dots,\boldsymbol{x}_5]$ with $\boldsymbol{x}_2=\boldsymbol{x}_3$ and $\boldsymbol{x}_4=\boldsymbol{x}_5$. We have $\mathcal{I}_5=\{1,2,4\}$ and $\bar{\mathcal{I}}_5=\{3,5\}$.

We do not allow the trivial memories join the damping process since they do not improve the MSE performance, i.e.,

$$\zeta_{t+1,i}=0,\quad i\in\bar{\mathcal{I}}_t. \tag{60}$$

Therefore, we only need to optimize $\zeta_{t+1}^{\mathcal{I}}=[\zeta_{t+1,i},i\notin\bar{\mathcal{I}}_t]$ for the effective inputs, i.e., $\phi_t(\boldsymbol{r}_t)$ and non-trivial memories $\{\boldsymbol{x}_i,i\in\mathcal{I}_t\}$, whose covariance matrix is denoted as $\boldsymbol{\mathcal{V}}_{t+1}^{\mathcal{I}}$. Lemma 8 gives an optimized damping vector $\zeta_{t+1}^{\mathcal{I}}$.

*Lemma 8 (Optimal Damping):* For a fixed $\theta_t$ and $\xi_t$, a $\zeta_{t+1}^{\mathcal{I}}$ that minimizes the MSE $v_{t+1,t+1}^{\bar\phi}$ of $\boldsymbol{x}_{t+1}$ in BO-MAMP is given by

$$\zeta_{t+1}^{\mathcal{I}}=\begin{cases}\dfrac{[\boldsymbol{\mathcal{V}}_{t+1}^{\mathcal{I}}]^{-1}\boldsymbol{1}}{\boldsymbol{1}^{\rm T}[\boldsymbol{\mathcal{V}}_{t+1}^{\mathcal{I}}]^{-1}\boldsymbol{1}}, & \text{if } \boldsymbol{\mathcal{V}}_{t+1}^{\mathcal{I}} \text{ is invertible}\\ [0,\cdots,1,0]^{\rm T}, & \text{otherwise}\end{cases}, \tag{61}$$

where $[\boldsymbol{\mathcal{V}}_{t+1}^{\mathcal{I}}]^{-1}$ is the inverse of $\boldsymbol{\mathcal{V}}_{t+1}^{\mathcal{I}}$. Furthermore, if $\boldsymbol{\mathcal{V}}_{t+1}^{\mathcal{I}}$ is invertible, we have $v_{t+1,t+1}^{\bar\phi}<v_{t,t}^{\bar\phi}$.

*Proof:* See Appendix F. ■

Following (47) and (61), $\boldsymbol{v}_{t+1}^{\bar\phi}$ can be rewritten to

$$\boldsymbol{v}_{t+1}^{\bar\phi}=\begin{cases}\frac{1}{\boldsymbol{1}^{\rm T}[\boldsymbol{\mathcal{V}}_{t+1}^{\mathcal{I}}]^{-1}\boldsymbol{1}}, & \text{if } \boldsymbol{\mathcal{V}}_{t+1}^{\mathcal{I}} \text{ is invertible}\\ v_{t,t}^{\bar\phi}\boldsymbol{1}, & \text{otherwise}\end{cases}. \tag{62}$$

That is $v_{t+1,t'}^{\bar\phi}=v_{t+1,t+1}^{\bar\phi},\forall 1\le t'\le t$. Furthermore, the MSE of current iteration (i.e., $\boldsymbol{x}_{t+1}$) with optimized damping is not worse than that of the previous iterations (e.g., $\zeta_{t+1}=[0,\cdots,1,0]^{\rm T}$). Therefore, the MSE of BO-MAMP with optimized $\{\zeta_t\}$ is monotonically decreasing in the iterations, i.e., $\{v_{t,t}^{\bar\phi}\}$ is a monotonically decreasing sequence. Besides, $\{v_{t,t}^{\bar\phi}\}$ has a lower bound 0. Thus, sequence $\{v_{t,t}^{\bar\phi}\}$ converges to a certain value, i.e., $\lim_{t\to\infty}v_{t,t}^{\bar\phi}\to v_*^{\bar\phi}$. In other words, the optimized $\zeta_{t+1}$ guarantees the convergence of BO-MAMP as well as improving the convergence speed of BO-MAMP. Therefore, we have the following lemma.

*Lemma 9 (Convergence of $\{\boldsymbol{V}_t^{\bar{\phi}}\}$):* For the BO-MAMP with optimized $\{\zeta_t\}$ in (61), the covariance matrix $\{\boldsymbol{V}_t^{\bar{\phi}}\}$ has the following properties.

- Covariance matrix $\boldsymbol{V}_t^{\bar{\phi}}$ is L-banded, i.e., the elements in each "L band" of the matrix are the same:

$$v^{\bar{\phi}}_{t',\tau} = v^{\bar{\phi}}_{\tau,t'} = v^{\bar{\phi}}_{t',t'}, \quad \forall 1 \leq \tau \leq t',\ \forall t' \leq t. \tag{63}$$

  That is,

$$\boldsymbol{V}_t^{\bar{\phi}} = \begin{pmatrix} v^{\bar{\phi}}_{1,1} & v^{\bar{\phi}}_{2,2} & \cdots & v^{\bar{\phi}}_{t,t} \\ v^{\bar{\phi}}_{2,2} & v^{\bar{\phi}}_{2,2} & \cdots & v^{\bar{\phi}}_{t,t} \\ \vdots & \vdots & \ddots & \vdots \\ v^{\bar{\phi}}_{t,t} & v^{\bar{\phi}}_{t,t} & \cdots & v^{\bar{\phi}}_{t,t} \end{pmatrix}. \tag{64}$$

- Sequence $\{v^{\bar{\phi}}_{t,t}\}$ is monotonically decreasing and converges to a certain value, i.e.,

$$v^{\bar{\phi}}_{t,t} \leq v^{\bar{\phi}}_{t',t'},\ \forall t' \leq t, \quad \text{and} \quad \lim_{t\to\infty} v^{\bar{\phi}}_{t,t} \to v^{\bar{\phi}}_{*}. \tag{65}$$

Therefore, the convergence of $\{\boldsymbol{V}_t^{\bar{\phi}}\}$ is guaranteed in the optimized BO-MAMP.

*Remark 3:* In Lemma 8, if $\mathcal{V}^{\mathcal{I}}_{t+1}$ is singular, we have $\boldsymbol{x}_{t+1} = \boldsymbol{x}_t$, which will lead to the damping covariance matrix in the next iterations be singular. To avoid this, we will not allow $\boldsymbol{x}_{t+1}$ join the damping in the next iterations if $\boldsymbol{x}_{t+1} = \boldsymbol{x}_t$. By doing so, we can always ensure that the covariance matrix $\boldsymbol{V}_t^{\bar{\phi}_{\mathcal{I}}}$ of the non-trivial memories in each damping is invertible.

*Proof:* We prove Remark 3 by induction as follows.

- For $t = 1$, we have $\boldsymbol{V}_1^{\bar{\phi}_{\mathcal{I}}} = v^{\bar{\phi}}_{1,1} = 1$, which is invertible.
- Assume that $\boldsymbol{V}_t^{\bar{\phi}_{\mathcal{I}}}$ is invertible, we next prove $\boldsymbol{V}_{t+1}^{\bar{\phi}_{\mathcal{I}}}$ is invertible.
  - If $\mathcal{V}^{\mathcal{I}}_{t+1}$ is singular, $\boldsymbol{x}_{t+1}$ will not join the damping in the next iterations. In this case, $\boldsymbol{V}_{t+1}^{\bar{\phi}_{\mathcal{I}}} = \boldsymbol{V}_t^{\bar{\phi}_{\mathcal{I}}}$, which is invertible since $\boldsymbol{V}_t^{\bar{\phi}_{\mathcal{I}}}$ is invertible.
  - If $\mathcal{V}^{\mathcal{I}}_{t+1}$ is invertible, from Lemma 8 and Lemma 9,

$$v^{\bar{\phi}}_{t+1,t+1} < v^{\bar{\phi}}_{t',t'}, \quad \forall t' \leq t. \tag{66}$$

    Furthermore, since $\boldsymbol{V}_t^{\bar{\phi}_{\mathcal{I}}}$ is invertible, we have

$$v^{\bar{\phi}}_{i,i} < v^{\bar{\phi}}_{j,j}, \quad \forall j < i, i \in \mathcal{I}_t, j \in \mathcal{I}_t. \tag{67}$$

    Otherwise there must exists one $i \in \mathcal{I}_t$ that $v^{\bar{\phi}}_{i,i} = v^{\bar{\phi}}_{i+1,i+1}$ (see (65)). Then, $\det\big(\boldsymbol{V}_t^{\bar{\phi}_{\mathcal{I}}}\big) = 0$ (see Lemma 8 in [58], [59]), which contradicts the fact that $\boldsymbol{V}_t^{\bar{\phi}_{\mathcal{I}}}$ is invertible. Following (66) and (67), we have

$$v^{\bar{\phi}}_{i,i} < v^{\bar{\phi}}_{j,j}, \quad \forall j < i, i \in \mathcal{I}_{t+1}, j \in \mathcal{I}_{t+1}. \tag{68}$$

    Then, following Lemma 8 in [58], [59], we have $\det\big(\boldsymbol{V}_{t+1}^{\bar{\phi}_{\mathcal{I}}}\big) > 0$, i.e., $\boldsymbol{V}_{t+1}^{\bar{\phi}_{\mathcal{I}}}$ is invertible.

Therefore, we complete the proof of Remark 3. ■

*Lemma 10:* Following Remark 3, we can always assume that the memory covariance matrix $\boldsymbol{V}_t^{\bar{\phi}_{\mathcal{I}}}$ in each damping is invertible. Then, from Lemma 8 and (62), we have

$$(\boldsymbol{V}_t^{\bar{\phi}_{\mathcal{I}}})^{-1}\mathbf{1} = \big[0, \ldots, 0, 1/v^{\bar{\phi}}_{t,t}\big]^{\mathrm{T}}, \tag{69a}$$

$$\mathbf{1}^{\mathrm{T}}(\boldsymbol{V}_t^{\bar{\phi}_{\mathcal{I}}})^{-1}\mathbf{1} = 1/v^{\bar{\phi}}_{t,t}. \tag{69b}$$

*Proof:* Since $\boldsymbol{V}_t^{\bar{\phi}_{\mathcal{I}}}$ is invertible and denotes the covariance matrix of the optimally damped estimates, following Lemma 8 and (62), we have

$$v^{\bar{\phi}}_{t,t} = [\mathbf{1}^{\mathrm{T}}(\boldsymbol{V}_t^{\bar{\phi}_{\mathcal{I}}})^{-1}\mathbf{1}]^{-1}, \tag{70a}$$

$$\zeta_t^{\mathcal{I}} = (\boldsymbol{V}_t^{\bar{\phi}_{\mathcal{I}}})^{-1}\mathbf{1}[\mathbf{1}^{\mathrm{T}}(\boldsymbol{V}_t^{\bar{\phi}_{\mathcal{I}}})^{-1}\mathbf{1}]^{-1} = [0, \cdots, 0, 1]^{\mathrm{T}}, \tag{70b}$$

where (70b) follows the fact that the optimal damping of $[\boldsymbol{x}_1 \ldots \boldsymbol{x}_t]$ is $\boldsymbol{x}_t$. Eq. (69) straightforwardly follows (70). Therefore, we complete the proof of Lemma 10. ■

The calculation of $\{\zeta_t^{\mathcal{I}}\}$ in (61) costs time complexity[8] of $\mathcal{O}(\mathcal{T}^4)$ due to the matrix inverse $\{(\mathcal{V}_t^{\mathcal{I}})^{-1}\}$, where $\mathcal{T}$ is the number of iterations. In practice, to reduce the computational complexity, instead of doing the full damping, the maximum damping length is generally set as $L$ (see Subsection V-D and the simulation results in Section VI), which is much less than $\mathcal{T}$. In this case, the time complexity is reduced to $\mathcal{O}(L^3\mathcal{T})$. In general, the best choice of $L$ is less than or equal to 3 (see the simulation results in Section VI). Therefore, $\mathcal{O}(L^3\mathcal{T})$ is negligible in the complexity of BO-MAMP as $L \ll \mathcal{T} \ll M$.

### D. Revision for Fixed Maximum Damping Length

Let $L$ be the maximum damping length of BO-MAMP. Then, the following should be revised. The NLE of BO-MAMP in (38b) is reduced to

$$\boldsymbol{x}_{t+1} = \bar{\phi}_t(\boldsymbol{r}_t) = [\boldsymbol{x}_{t+2-l_{t+1}}, \ldots, \boldsymbol{x}_t, \phi_t(\boldsymbol{r}_t)] \cdot \zeta_{t+1}, \tag{71}$$

where $l_{t+1} = \min\{L, t+1\}$ and $\zeta_{t+1} = [\zeta_{t+1,1}, \ldots, \zeta_{t+1,l_{t+1}}]^{\mathrm{T}}$. The $l_{t+1} \times l_{t+1}$ covariance matrix $\mathcal{V}_{t+1}$ in (44a) is changed to

$$\mathcal{V}_{t+1} \equiv \begin{bmatrix} v^{\bar{\phi}}_{t-l_{t+1}+2,t-l_{t+1}+2} & \cdots & v^{\bar{\phi}}_{t-l_{t+1}+2,t} & v^{\phi}_{t-l_{t+1}+2,t+1} \\ \vdots & \ddots & \vdots & \vdots \\ v^{\bar{\phi}}_{t,t-l_{t+1}+2} & \cdots & v^{\bar{\phi}}_{t,t} & v^{\phi}_{t,t+1} \\ v^{\phi}_{t+1,t-l_{t+1}+2} & \cdots & v^{\phi}_{t+1,t} & v^{\phi}_{t+1,t+1} \end{bmatrix}. \tag{72}$$

The NLE state evolution function $\bar{\phi}_{\mathrm{SE}}(\cdot)$ in (47a) is changed to

$$v^{\bar{\phi}}_{t+1,t'} = \begin{cases} \zeta^*_{t+1,l_{t+1}} v^{\phi}_{t+1,t'} + \sum_{i=1}^{l_{t+1}-1} \zeta^*_{t+1,i} v^{\bar{\phi}}_{t-l_{t+1}+1+i,t'}, & 1 \leq t' \leq t \\ \zeta^{\mathrm{H}}_{t+1} \mathcal{V}_{t+1} \zeta_{t+1}, & t' = t+1 \end{cases}. \tag{73}$$

Eqn. (62) is changed to

$$v^{\bar{\phi}}_{t+1,\tau} = \frac{1}{\mathbf{1}^{\mathrm{T}}[\mathcal{V}^{\mathcal{I}}_{t+1}]^{-1}\mathbf{1}}, \quad t+2-l_{t+1} \leq \tau \leq t+1. \tag{74}$$

### E. Convergence and Potential Bayes Optimality of BO-MAMP

The following theorem gives the convergence and potential Bayes optimality of BO-MAMP.

[8]The time complexity of calculating $\{[\mathcal{V}^{\mathcal{I}}_{t+1}]^{-1}\}$ can be reduced to $\mathcal{O}(\mathcal{T}^3)$ if we calculate it based on $\{(\boldsymbol{V}_t^{\bar{\phi}})^{-1}\}$ instead of calculating it straightforwardly.

*Theorem 2 (Convergence and Potential Bayes optimality):* Suppose that Assumptions 1-3 hold. For any[9] non-zero $\xi_*$ and $\{\xi_t = \xi_*, \forall t\}$, the state evolution of BO-MAMP with optimized $\{\theta_t, \zeta_t\}$ (see Section V) converges to the same fixed point as that of BO-OAMP/VAMP. That is, the optimized BO-MAMP achieves the Bayes optimal MSE predicted by the replica method if its state evolution has a unique fixed point.

*Proof:* See Appendix G. The following are the key points. Damping guarantees the convergence of state evolution of the optimized BO-MAMP (see Lemma 9). The fixed points of the optimized BO-MAMP are the same as those of BO-OAMP/VAMP. Therefore, the optimized BO-MAMP converges to the same fixed point as that of BO-OAMP/VAMP. Finally, following with the potential Bayes optimality of BO-OAMP/VAMP (see Lemma 3), the optimized BO-MAMP achieves the Bayes optimal MSE predicted by the replica method if its state evolution has a unique fixed point. ■

## VI. Simulation Results

We study a compressed sensing problem where $\boldsymbol{x}$ follows a symbol-wise Bernoulli-Gaussian distribution, i.e. $\forall i$,

$$x_i \sim \begin{cases} 0, & \text{probability} = 1-\mu \\ \mathcal{N}(0,\mu^{-1}), & \text{probability} = \mu \end{cases}. \tag{75}$$

The variance of $x_i$ is normalized to 1. The signal-to-noise ratio (SNR) is defined as SNR $= 1/\sigma^2$.

Let the SVD of $\boldsymbol{A}$ be $\boldsymbol{A} = \boldsymbol{U}\boldsymbol{\Sigma}\boldsymbol{V}^{\mathbf{H}}$. The system model in (1) is rewritten as:

$$\boldsymbol{y} = \boldsymbol{U}\boldsymbol{\Sigma}\boldsymbol{V}^{\mathrm{H}}\boldsymbol{x} + \boldsymbol{n}. \tag{76}$$

Note that $\boldsymbol{U}^{\mathrm{H}}\boldsymbol{n}$ has the same distribution as $\boldsymbol{n}$. Thus, we can assume $\boldsymbol{U} = \boldsymbol{I}$ without loss of generality. To reduce the calculation complexity of BO-OAMP/VAMP, unless otherwise specified, we approximate a large random unitary matrix by $\boldsymbol{V}^{\mathrm{H}} = \boldsymbol{\Pi}\boldsymbol{F}$, where $\boldsymbol{\Pi}$ is a random permutation matrix and $\boldsymbol{F}$ is a DFT matrix. Note that all the algorithms involved here admit fast implementation for this matrix model. The entries $\{d_i\}$ of diagonal matrix $\boldsymbol{\Sigma}$ are generated as: $d_i/d_{i+1} = \kappa^{1/J}$ for $i = 1,\ldots,J-1$ and $\sum_{i=1}^{J} d_i^2 = N$, where $J = \min\{M,N\}$. Here, $\kappa \geq 1$ controls the condition number of $\boldsymbol{A}$. Note that BO-MAMP does not require the SVD structure of $\boldsymbol{A}$. BO-MAMP only needs the right-unitarily invariance of $\boldsymbol{A}$.

### *A. Influence of Relaxation Parameters and Damping*

Fig. 4 shows MSE as a function of iterations and the influence of the relaxation parameters $\{\lambda^\dagger, \theta_t, \xi_t\}$ and damping. As can be seen, without damping (e.g., $L = 1$) the convergence of BO-MAMP is not guaranteed, and the optimization of $\{\lambda^\dagger, \theta_t, \xi_t\}$ has significant improvement in the MSE of BO-MAMP. The following observations can be made from Fig. 4.

(i) Damping guarantees the convergence of BO-MAMP.
(ii) The relaxation parameters $\{\lambda^\dagger, \theta_t, \xi_t\}$ do not change the convergence and fixed point of BO-MAMP, but they can be optimized to improve the convergence speed.

[9]Theorem 2 shows that the choice of $\xi_*$ does not affect the state-evolution fixed point (i.e., potential Bayes optimality) of BO-MAMP. However, as is shown in V-B and Fig. 4, $\{\xi_t\}$ can be optimized to improve the convergence speed of BO-MAMP.

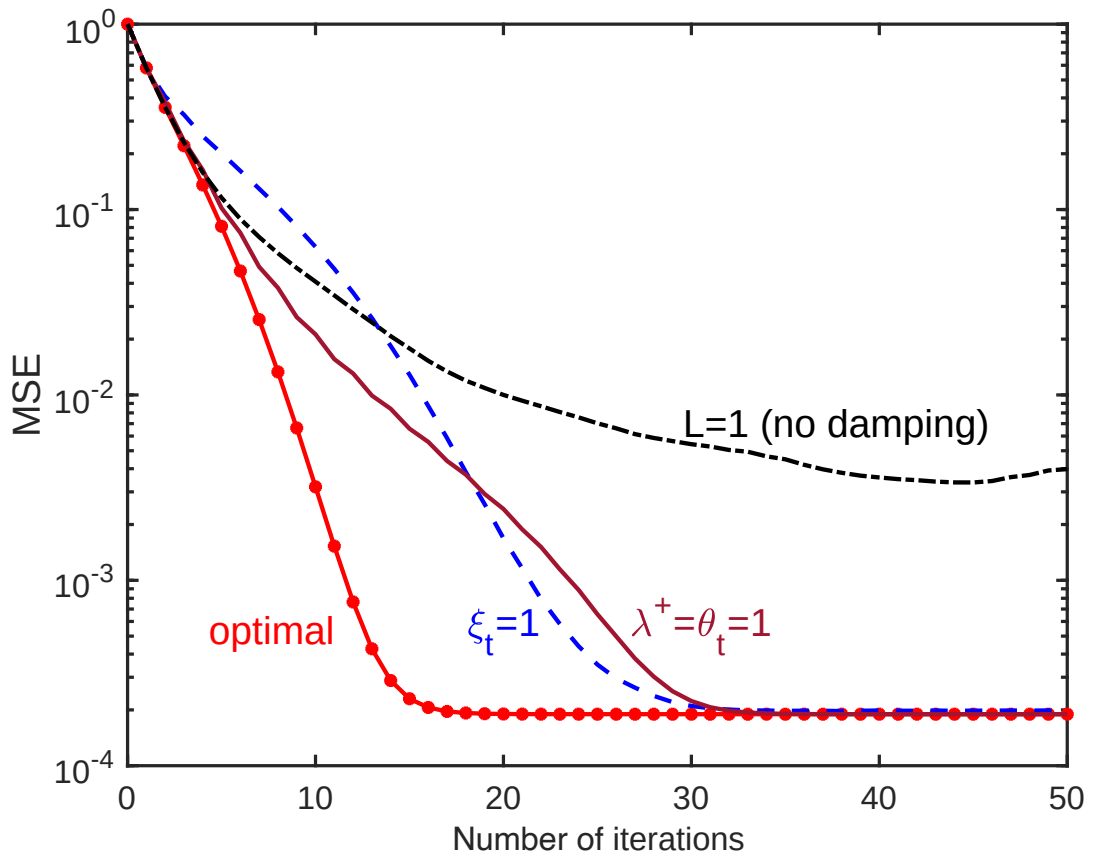

Fig. 4. MSE versus the number of iterations for BO-MAMP with different parameters $\{\lambda^\dagger, \theta_t, \xi_t, L\}$. $M = 4096$, $N = 8192$, $\mu = 0.1$, $\kappa = 10$ and SNR = 30 dB. "optimal" denotes BO-MAMP with optimized $\{\lambda^\dagger, \theta_t, \xi_t\}$ and $L = 3$ (damping length). The other curves denote BO-MAMP with the same parameters as the "optimal" except the one marked on each curve.

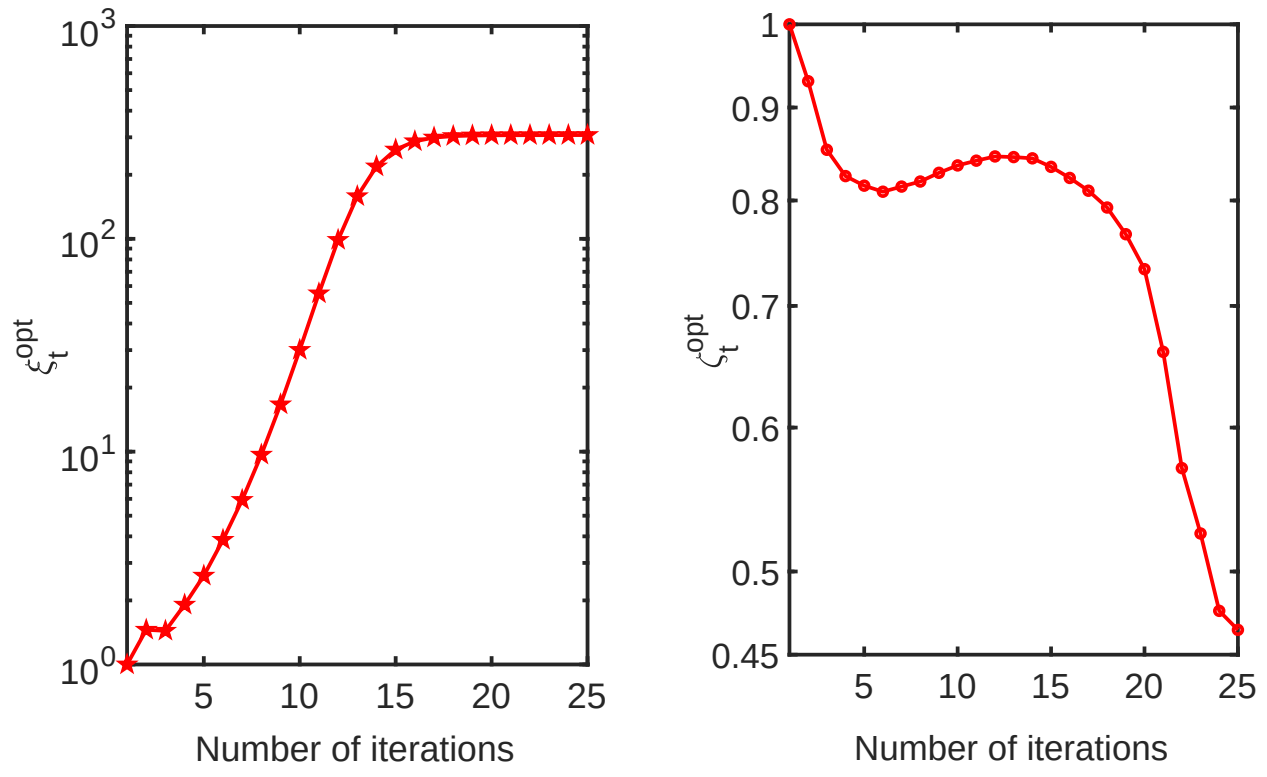

Fig. 5. $\xi_t^{\rm opt}$ and $\zeta_t^{\rm opt}$ versus the number of iterations for BO-MAMP. $M = 4096$, $N = 8192$, $\mu = 0.1$, $\kappa = 10$ and SNR = 30 dB. Left sub-figure: $L = 3$; Right sub-figure: $L = 2$.

The left subfigure of Fig. 5 shows the optimal relaxation parameter $\xi_t^{\rm opt}$ versus the number of iterations for BO-MAMP. As can be seen, $\xi_t^{\rm opt}$ is monotone increasing with the number of iterations. Since $\xi_t^{\rm opt}$ is the weight of the current input of a memory LE, it means that as the number of iterations increases, the current input becomes more and more important, while the previous inputs (memory terms) become less and less important. In addition, when BO-MAMP converges (e.g., $t \geq 16$, see the "optimal" curve in Fig. 4), the sequence $\{\xi_t^{\rm opt}\}$ also converges to a fixed value. We currently can not explain why the curve is not smooth at the point $t = 3$.

The right subfigure of Fig. 5 shows the optimal damping parameter $\{\zeta_t^{\rm opt}\}$ versus the number of iterations for BO-MAMP. As can be seen, before BO-MAMP converges (e.g., $t \leq 17$, see the curve $L = 2$ in Fig. 9(a)), the values of $\{\zeta_t^{\rm opt}\}$ are around 0.8. After BO-MAMP converges (e.g., $t \geq 18$), $\zeta_t^{\rm opt}$ is monotone decreasing with the number of iterations, but in this case the value of $\zeta_t$ does not impact the performance of BO-MAMP significantly (since it has converged).

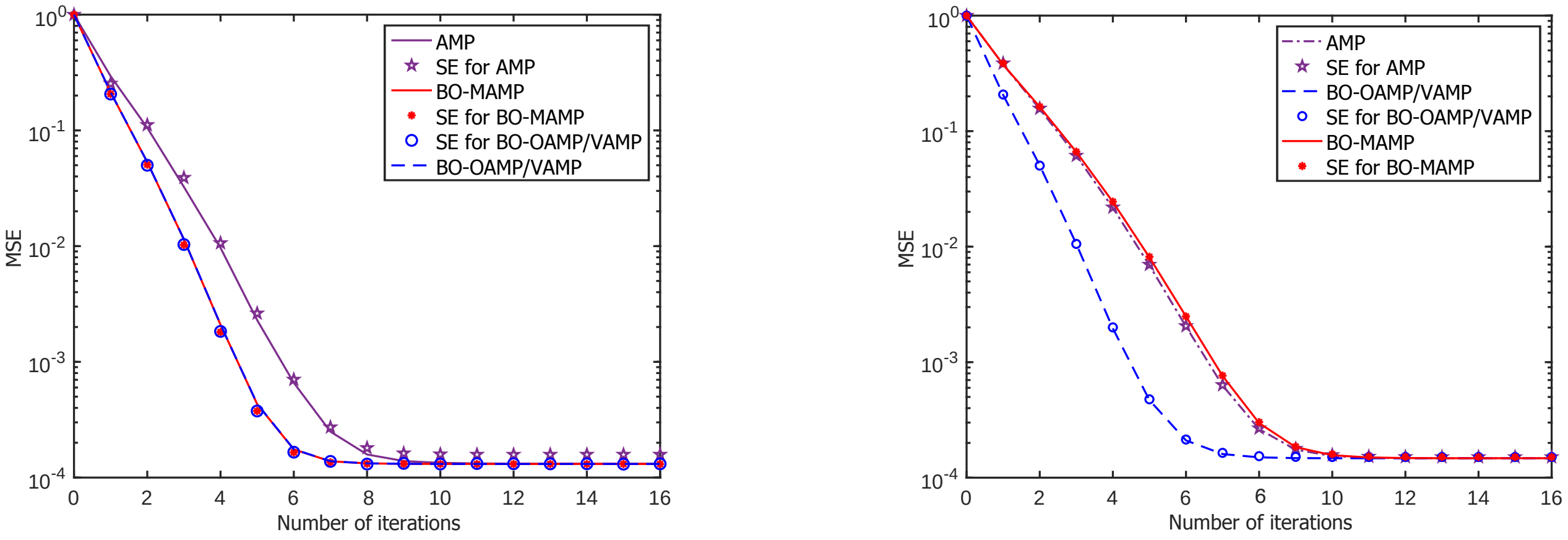

(a) Well-conditioned matrix ($\kappa = 1$) (b) Zero-mean IID Gaussian matrix

Fig. 6. MSE versus the number of iterations for AMP, BO-OAMP/VAMP and BO-MAMP. $M = 8192$, $N = 16384$, $\mu = 0.1$ and SNR = 30 dB. SE denotes state evolution. (a) Well-conditioned $\boldsymbol{A}$ with $\kappa = 1$ and $L = 1$ (no damping); (b) zero-mean IID Gaussian $\boldsymbol{A}$ with $\{\mathrm{E}\{a_{ij}^2\} = 1/M\}$ and $L = 3$.

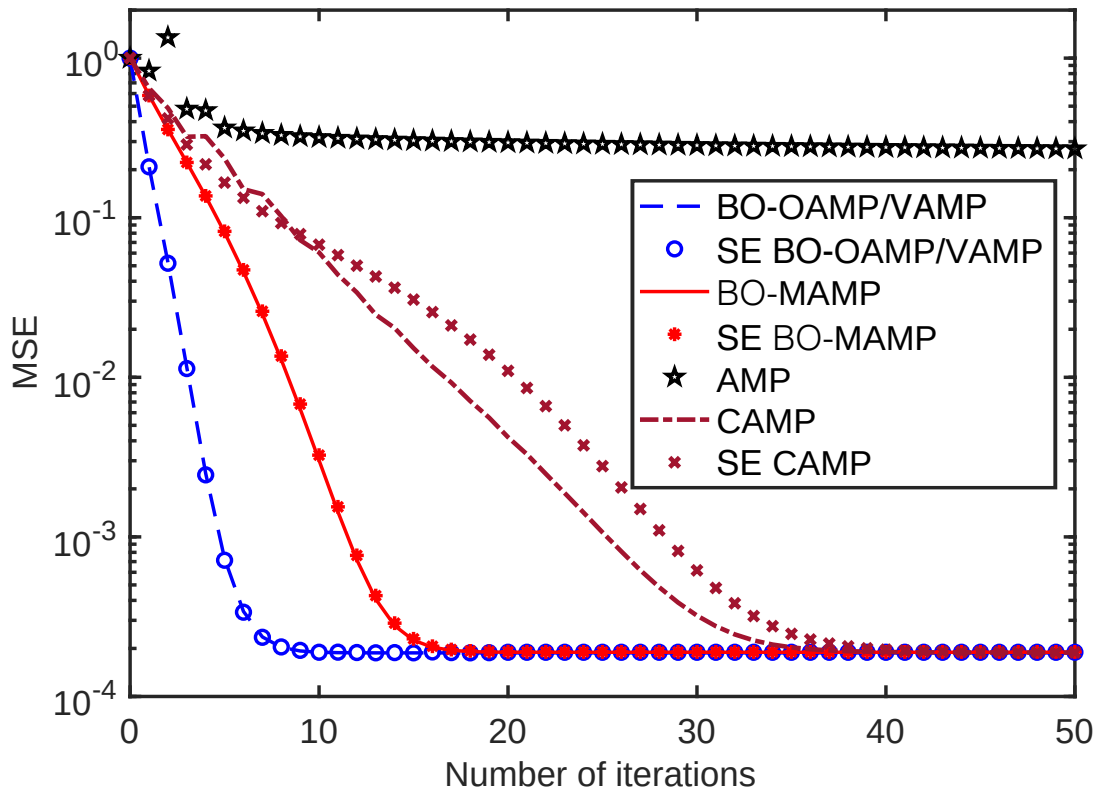

Fig. 7. MSE versus the number of iterations for AMP, CAMP, BO-OAMP/VAMP and BO-MAMP. $M = 8192$, $N = 16384$, $\mu = 0.1$, $\kappa = 10$, $L = 3$ and SNR = 30 dB. SE denotes state evolution. The AMP and CAMP curves are from Fig. 2 in [31].

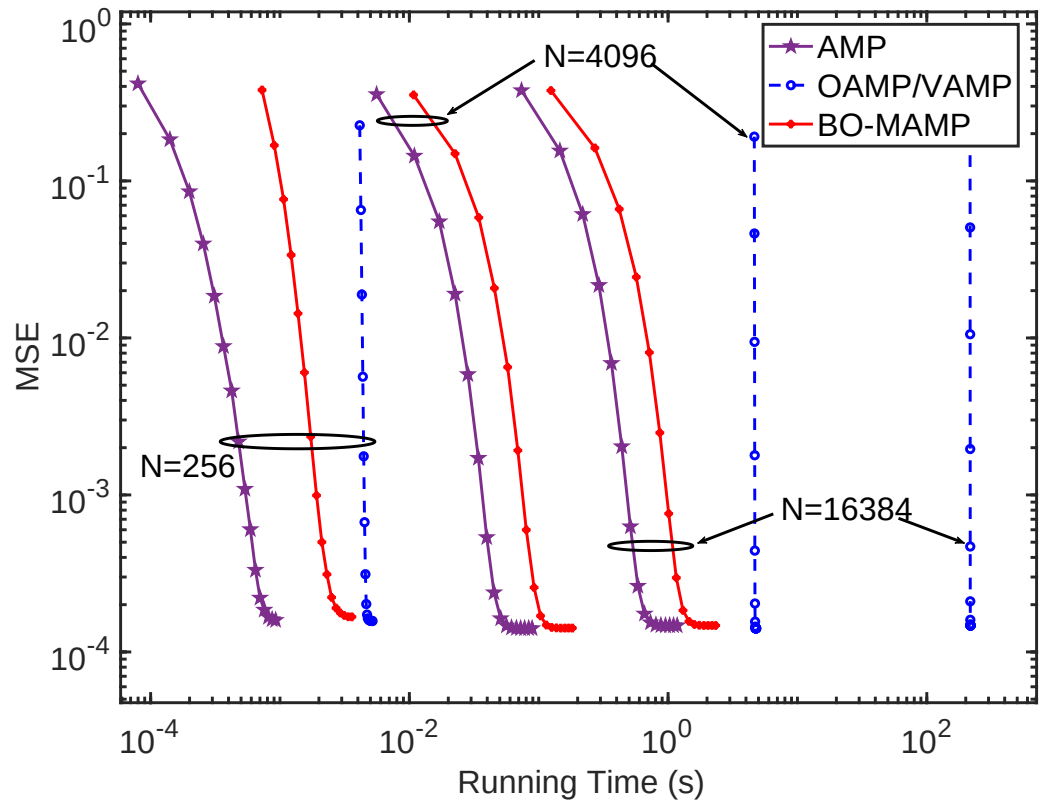

Fig. 8. Running time comparison between AMP, BO-OAMP/VAMP (SVD) and BO-MAMP for zero-mean IID Gaussian matrices with $N \in \{256, 4096, 16384\}$, $\{\mathrm{E}\{a_{ij}^2\} = 1/M\}$, $\delta = 0.5$, $\mu = 0.1$, $L = 3$ and SNR = 30 dB.

## *B. Comparison with AMP and CAMP*

Fig. 6 shows MSE versus the number of iterations for AMP, BO-OAMP/VAMP and BO-MAMP with (a) well-conditioned $\boldsymbol{A}$ (e.g., $\kappa = 1$) and (b) zero-mean IID Gaussian $\boldsymbol{A}$. As can be seen, for well-conditioned $\boldsymbol{A}$ with $\kappa = 1$, the MSEs of BO-OAMP/VAMP and BO-MAMP overlap, which is consistent with Proposition 6, i.e., BO-OAMP/VAMP and BO-MAMP are equivalent for matrices with identical eigenvalues. BO-MAMP and OAMP/VAMP converge faster than AMP. For zero-mean IID Gaussian $\boldsymbol{A}$, the MSEs of AMP and BO-MAMP overlap, i.e., they have almost the same convergence speed, and BO-OAMP/VAMP converges faster than AMP and BO-MAMP. Furthermore, the state evolution of BO-MAMP matches well with the simulated MSE in both cases.

Fig. 7 shows MSE versus the number of iterations for AMP, CAMP, BO-OAMP/VAMP and BO-MAMP. To improve the convergence, both AMP and CAMP are damped. As can be seen, for an ill-conditioned matrix with $\kappa = 10$, the MSE performance of AMP is poor. CAMP converges to the same performance as that of BO-OAMP/VAMP. However, the state evolution of CAMP is inaccurate since the damping is made on the *a-posteriori* NLE outputs, which breaks the orthogonality and Gaussianity of the estimation errors. The proposed BO-MAMP in this paper converges faster than CAMP to BO-OAMP/VAMP. Furthermore, the state evolution of BO-MAMP is accurate since the damping is made on the *orthogonal* NLE outputs, which preserves the orthogonality and Gaussianity of the estimation errors.

Fig. 8 shows MSE versus the MATLAB program running time of AMP, BO-OAMP/VAMP (SVD) and BO-MAMP for zero-mean IID Gaussian matrix $\boldsymbol{A}$ with $N \in \{256, 4096, 16384\}$ and $\delta = 0.5$. As can be seen, the running time of BO-MAMP is lower than BO-OAMP/VAMP. The time complexity improvement of BO-MAMP is more significant for larger matrix sizes. The running time of MAMP is relatively higher than (about double of) AMP. The reason is that, for zero-mean IID Gaussian $\boldsymbol{A}$, the convergence speed of AMP and BO-MAMP are almost the same, but the matrix-vector multiplications per iteration in BO-MAMP is double of AMP.

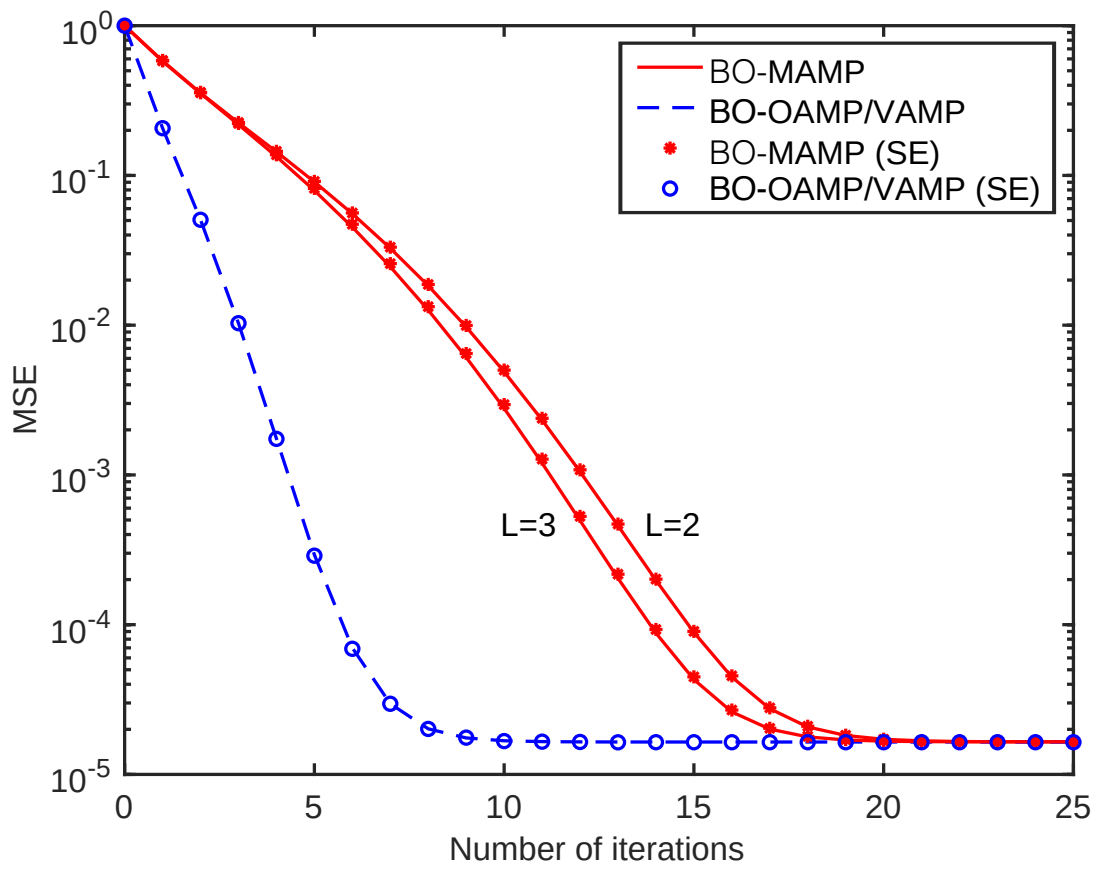

(a) $\kappa = 10$

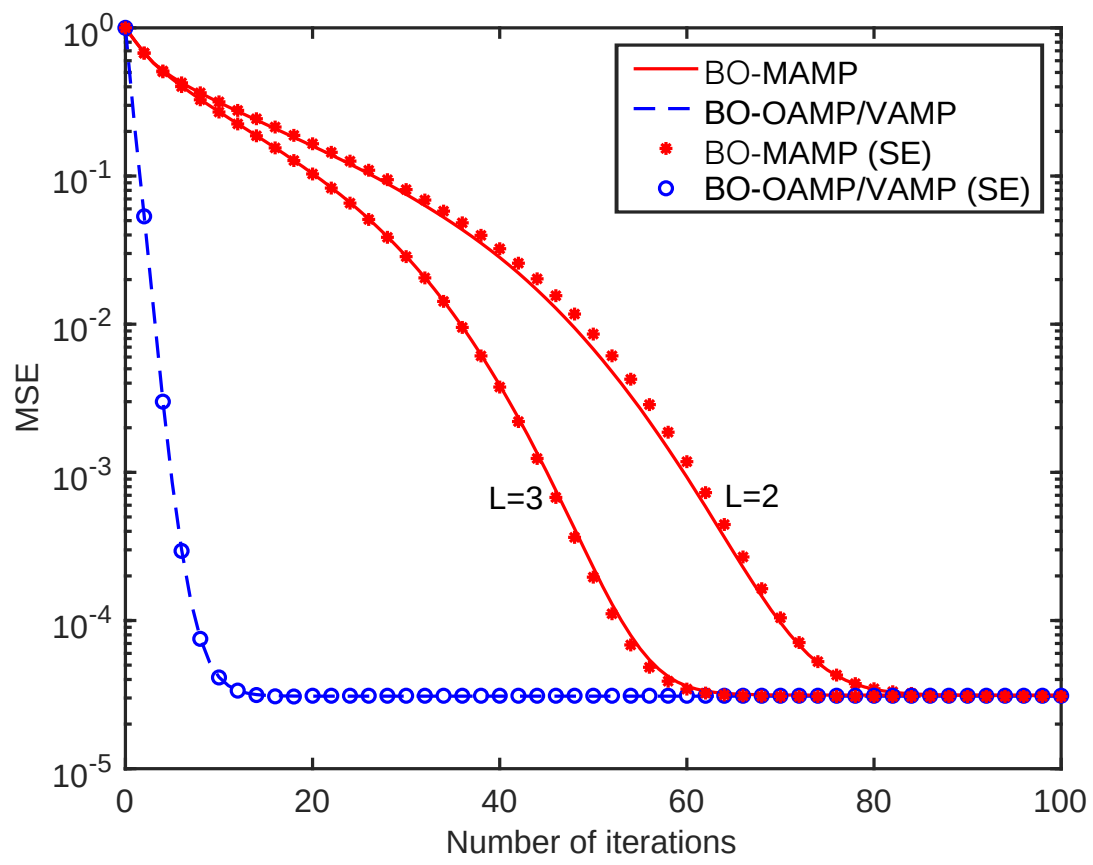

(b) $\kappa = 10^2$

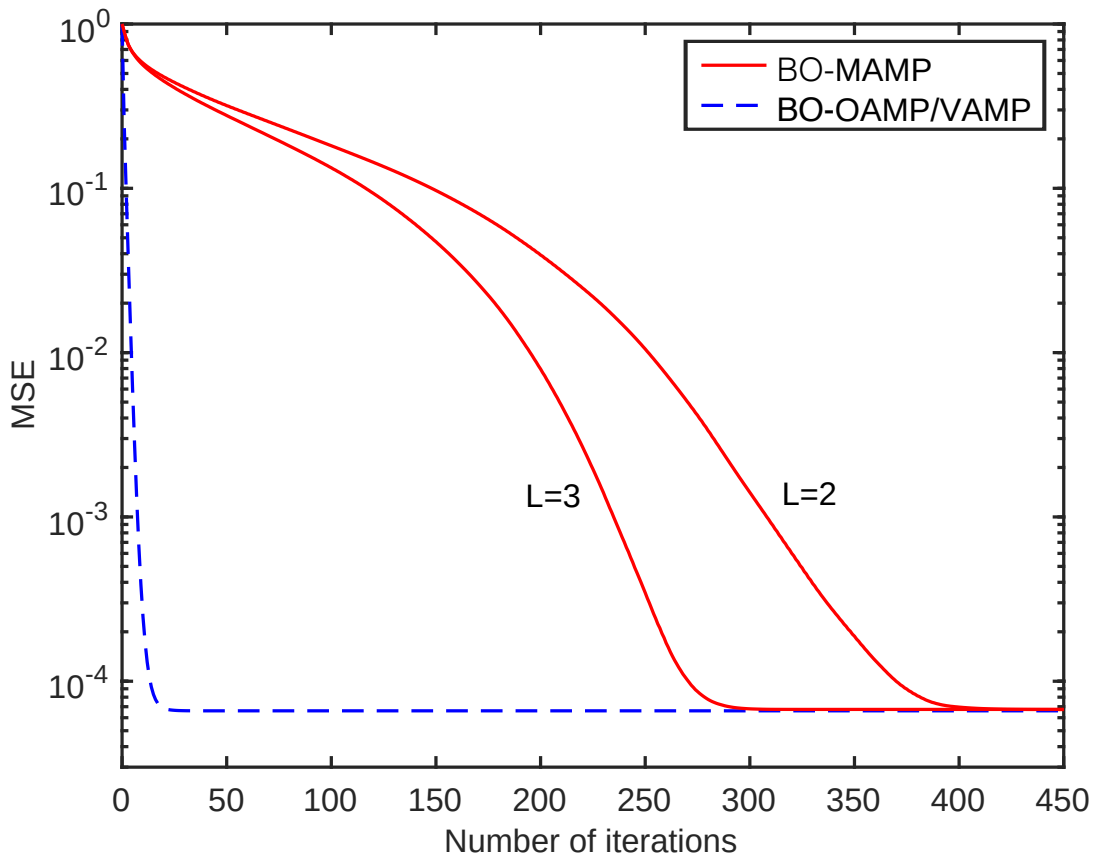

(c) $\kappa = 10^3$

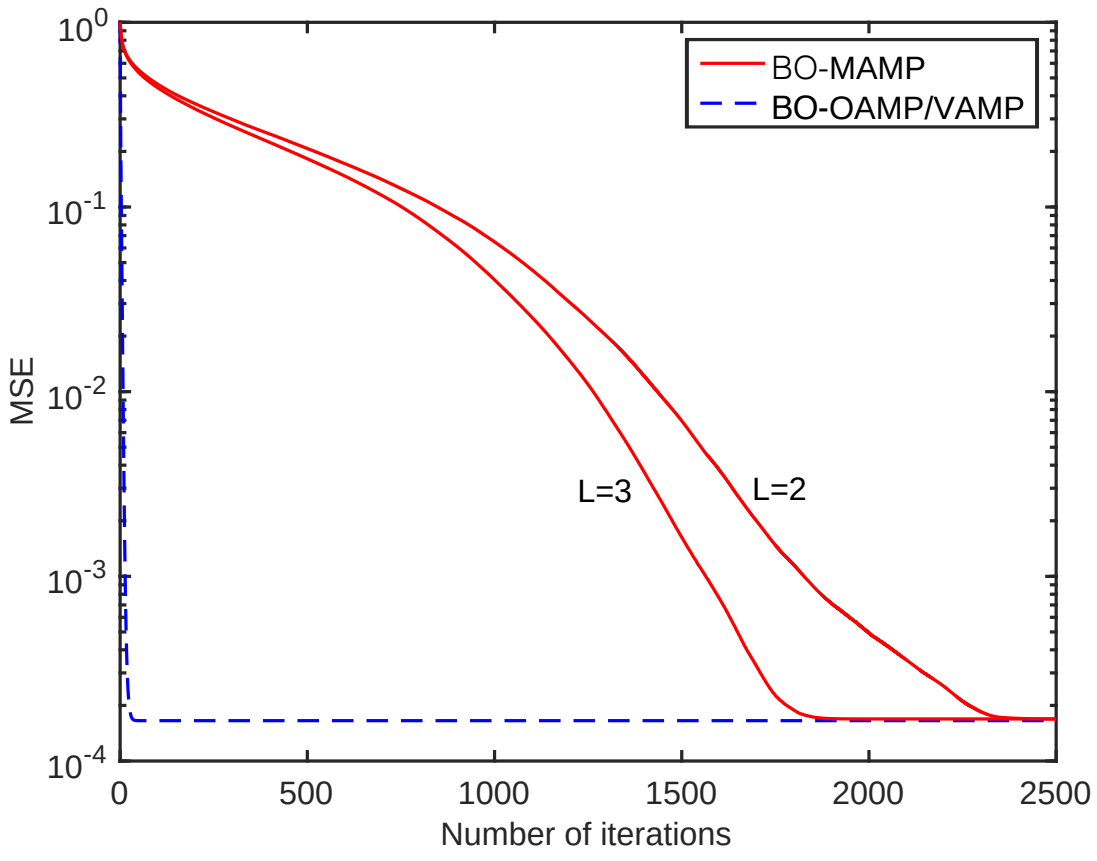

(d) $\kappa = 10^4$

Fig. 9. MSE versus the number of iterations for BO-MAMP and BO-OAMP/VAMP with different condition numbers of $\boldsymbol{A}$. $M = 8192$, $N = 16384$, $\mu = 0.1$, SNR = 40 dB and $\kappa \in \{10, 10^2, 10^3, 10^4\}$. SE denotes state evolution. We did not show the SEs of $\kappa = 10^3$ and $\kappa = 10^4$ since Algorithm 2 becomes unstable in these cases.

### *C. Influence of Condition Number and Damping Length*

Fig. 9 shows MSE versus the number of iterations for BO-MAMP with different condition numbers and damping lengths. As can be seen, BO-MAMP converges to BO-OAMP/VAMP for all the condition number shown. However, for matrices with larger condition numbers, BO-MAMP needs more iterations to converge. Note that CAMP fails to converge when $\kappa > 15$ (see Fig. 4 in [31]). The state evolution of BO-MAMP matches well with the simulated MSE when $\kappa = 10$ and $\kappa = 10^2$. We did not show the SEs of $\kappa = 10^3$ and $\kappa = 10^4$ since Algorithm 2 becomes unstable in these cases. The reason is that for large $\kappa$, $\boldsymbol{V}_t^{\gamma}$ may become nearly singular, which results in unstable $\eta_t$. In detail, numerical calculations of $(\boldsymbol{V}_t^{\gamma})^{-1}$ in (87b) become inaccurate. We are still working on this issue. In addition, BO-MAMP with $L = 3$ (damping length) has significant improvement in convergence speed compared with $L = 2$ when the condition number of $\boldsymbol{A}$ is large (e.g., $\kappa \geq 50$). It should be mentioned that $L = 3$ is generally sufficient for BO-MAMP convergence, since the MSEs of BO-MAMP are almost the same when $L \geq 3$. Thus, we did not show the MSE curves of BO-MAMP with $L \geq 4$. This is an interesting observation, but we still have no mathematical explanation for this phenomenon.

### *D. Influence of Compression Ratio*

Fig. 10 shows MSE versus the number of iterations for BO-MAMP with $N = 16384$, $\mu = 0.1$, $\kappa = 10$, SNR = 30 dB and $\delta = M/N \in \{0.2, 0.25, 0.4, 0.7, 1\}$ (compression ratios). As can be seen, BO-MAMP converges to BO-OAMP/VAMP for all the matrices with small-to-large compression ratios, and the state evolution of BO-MAMP matches well with the simulated MSE. For the matrices with small compression ratios, BO-MAMP needs more iterations to converge. Damping length $L = 2$ is enough for compression ratios $\delta \in \{0.2, 0.7, 1\}$, and $L = 3$ is enough for compression ratio $\delta \in \{0.25, 0.4\}$.

*Under-loaded systems:* In under-loaded case (i.e., $N \leq M$), we let

$$\boldsymbol{A} = \begin{bmatrix} \boldsymbol{\Sigma}_{N\times N} \\ \boldsymbol{0}_{(M-N)\times N} \end{bmatrix} \boldsymbol{\Pi} \boldsymbol{F}. \tag{77}$$

Hence, $\lambda_{\min} = 0$ when $N < M$. Here, we let $\sum_{i=1}^{N} d_i^2 = M$, i.e., the measurement energy increases with $M$. Fig. 11 shows MSE versus the number of iterations for BO-MAMP with $\delta \in \{1, 24, 32, 64\}$. As can be seen, BO-MAMP converges fast to BO-OAMP/VAMP in under-loaded case with small-to-large compression ratios. In addition, the state evolution of BO-MAMP matches well with the simulated MSE.

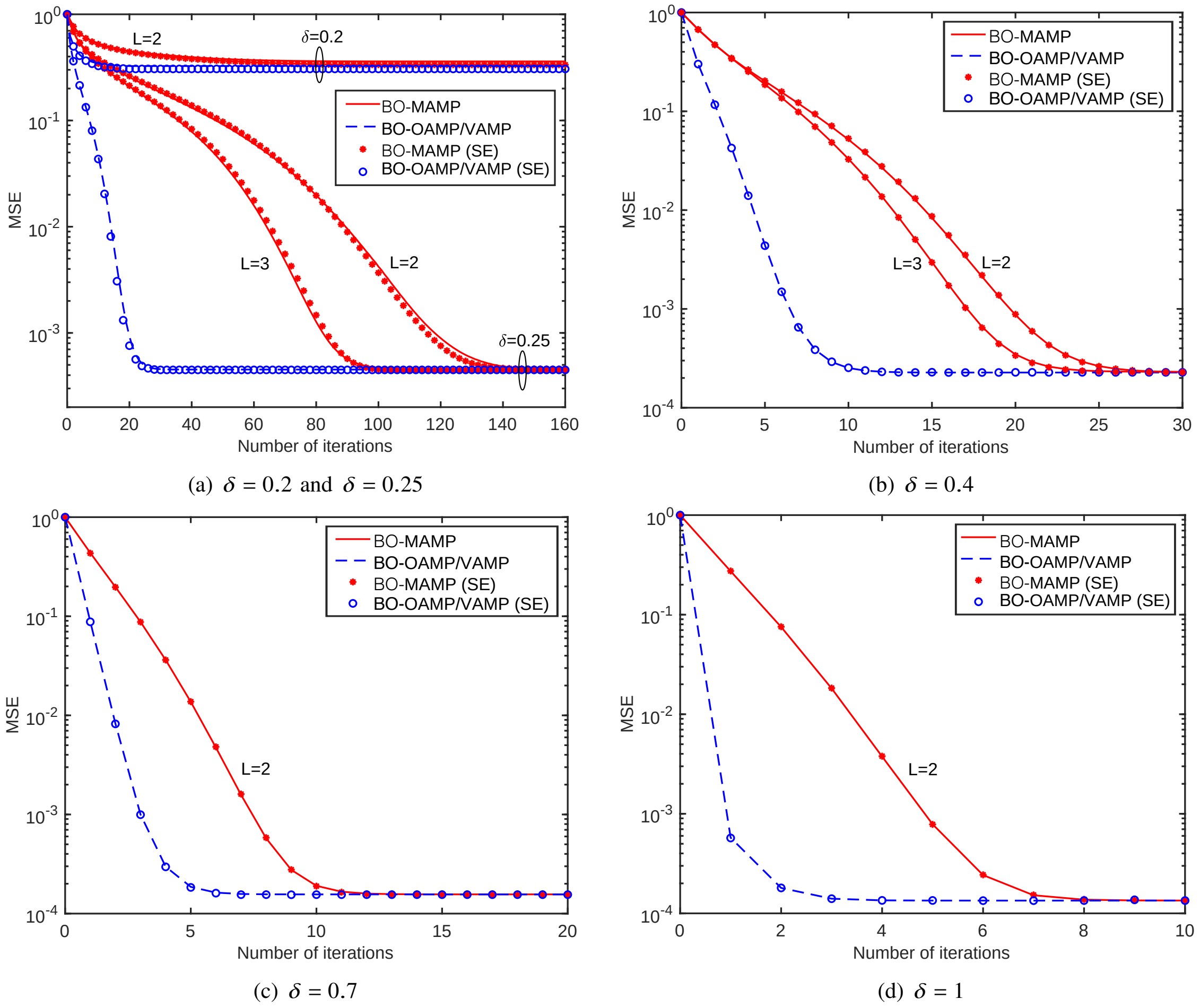

(a) $\delta = 0.2$ and $\delta = 0.25$

(b) $\delta = 0.4$

(c) $\delta = 0.7$

(d) $\delta = 1$

Fig. 10. MSE versus the number of iterations for BO-MAMP and BO-OAMP/VAMP with different compression ratios. $N = 16384$, $\mu = 0.1$, $\kappa = 10$, SNR = 30 dB and $\delta = M/N \in \{0.2, 0.25, 0.4, 0.7, 1\}$. SE denotes state evolution. Damping length $L = 2$ is good enough for $\delta = M/N \in \{0.2, 0.7, 1\}$, and $L = 3$ is good enough for $\delta = M/N \in \{0.25, 0.4\}$.

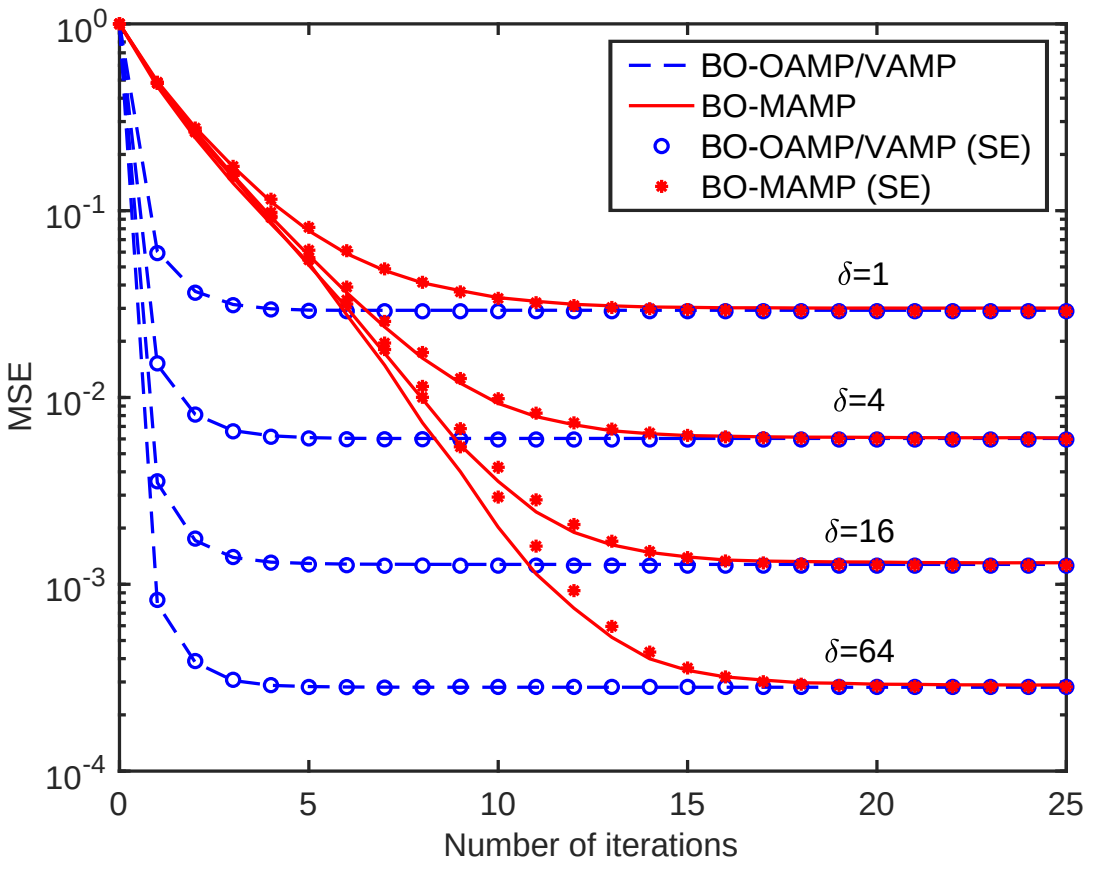

Fig. 11. MSE versus the number of iterations for BO-MAMP with different compression ratios $\delta \in \{1, 24, 32, 64\}$. $N = 512$, $\mu = 0.3$, $\kappa = 10$, $L = 3$ and SNR = 15 dB. SE denotes state evolution.

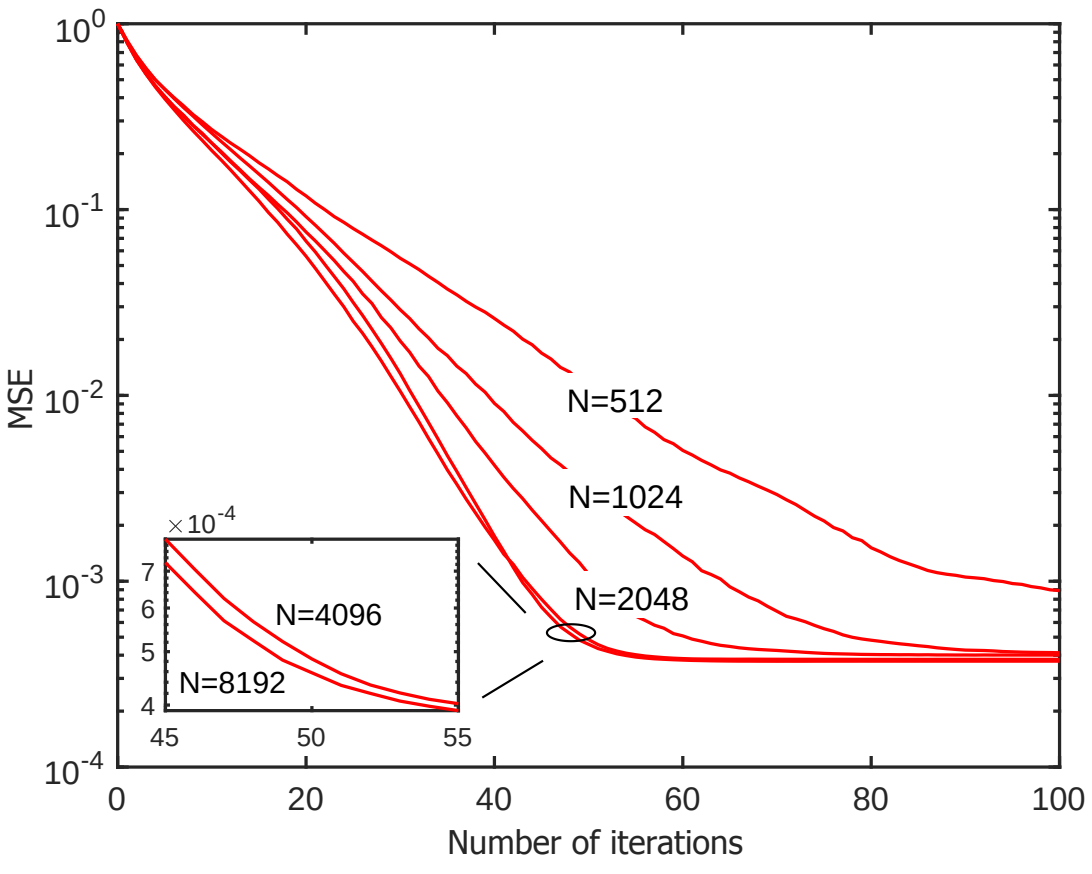

Fig. 12. MSE versus the number of iterations for BO-MAMP with different system sizes $N \in \{512, 1024, 2048, 4096, 8192\}$. $\delta = 0.5$, $\mu = 0.1$, $\kappa = 80$, $L = 3$ and SNR = 30 dB.

## *E. Influence of System Size*

Fig. 12 shows MSE versus the number of iterations for BO-MAMP with different system sizes $N \in \{512, 1024, 2048, 4096, 8192\}$ and $\delta = 0.5$. As can be seen, BO-MAMP works well for $N$ as small as 512 when $\kappa = 80$. However, for very small system size (e.g., $N < 100$) and very large condition number (e.g., $\kappa > 10^3$), the state evolution and the parameters of BO-MAMP may be inaccurate. In this case, BO-MAMP becomes unstable, i.e., fails to converge.

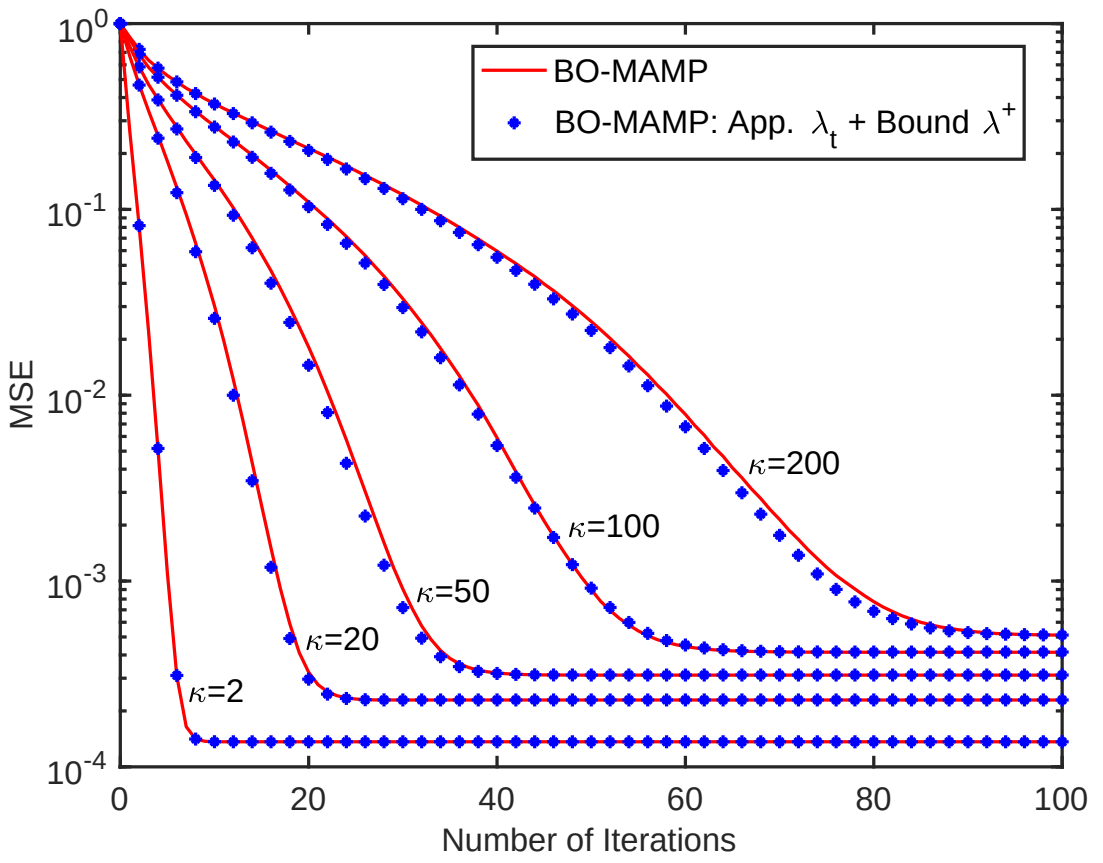

Fig. 13. MSE versus the number of iterations for (1) BO-MAMP with exact $\{\lambda_t, \lambda_{\min}, \lambda_{\max}\}$ and (2) BO-MAMP with the approximate $\{\lambda_t\}$ in (4) in Proposition 1 and the bounds $\lambda_{\min}^{\rm low} = 0$ and $\lambda_{\max}^{\rm up} = (N\lambda_{2\mathcal{T}})^{\frac{1}{2\mathcal{T}}}$, where $\mathcal{T} \in \{14, 30, 50, 80, 100\}$ correspond to $\kappa \in \{2, 20, 50, 100, 200\}$, respectively. $N = 16384$, $\delta = 0.5$, $\mu = 0.1$, $L = 3$ and SNR = 30 dB.

### *F. Influence of Approximate* $\{\lambda_t\}$ *and Bounds of* $\{\lambda_{\min}, \lambda_{\max}\}$

Fig. 13 compares the BO-MAMP with exact $\{\lambda_{\min}, \lambda_{\max}, \lambda_t\}$ and the BO-MAMP with approximate $\{\lambda_t\}$ in (4) in Proposition 1 and bounds $\lambda_{\min}^{\rm low} = 0$ and $\lambda_{\max}^{\rm up} = (N\lambda_{2\mathcal{T}})^{\frac{1}{2\mathcal{T}}}$, where $\mathcal{T} \in \{14, 30, 50, 80, 100\}$ correspond to $\kappa \in \{2, 20, 50, 100, 200\}$, respectively. As can be seen, the MSEs of BO-MAMP with the approximate $\{\lambda_t\}$ in (4) and the bounds $\{\lambda_{\min}^{\rm low}, \lambda_{\max}^{\rm up}\}$ are almost the same as that with the exact values of $\{\lambda_t, \lambda_{\min}, \lambda_{\max}\}$.

## VII. Conclusions

This paper proposes an MAMP for high-dimensional linear systems with right-unitarily transform matrices under the orthogonality principle, which guarantees asymptotic IID Gaussianity of estimation errors in MAMP. A kind of orthogonal memory LE and orthogonal memory NLE are developed to realize the required orthogonality for MAMP. In addition, we propose a low-cost BO-MAMP. The proposed BO-MAMP achieves the Bayes-optimal MSE predicted by the replica method for right-unitarily-invariant matrices (under the unique fixed point assumption), and has comparable complexity to AMP. Specifically, the techniques of long memory and orthogonalization are used to achieve the potentially Bayes-optimal solution of the problem with a low-complexity matched filter. The convergence of BO-MAMP is optimized with relaxation parameters at LE and a damping vector at NLE. It is proved that the optimized BO-MAMP converges to the high-complexity BO-OAMP/VAMP for all right-unitarily-invariant matrices.

Recently, the MAMP was extended to solve the generalized linear problem in [51]. Apart from that, iterative LMMSE and AMP/OAMP/VAMP have been extended to handle the coded linear system [29], [30], [52], the intersymbol interference (ISI) channels [28], [52], a more general class of matrices in the linear layers [53], [54], and MMV problems [55], [56]. MAMP can be naturally extended to such problems, which is an interesting future work.

## Acknowledgment

The authors would like to thank Keigo Takeuchi, Ramji Venkataramanan (Associate Editor), Yao Ge, Qinghua Guo, Junjie Ma and the anonymous reviewers for their comments or discussions that have improved the quality of the manuscript greatly. The authors also appreciate Keigo Takeuchi's source code and simulation data for CAMP.

## Appendix A
## Proof of Theorem 1

We first prove the orthogonality in (20) for BO-MAMP. Then, following Lemma 4, we get the asymptotic IID Gaussianity in (24) for BO-MAMP.

First, following Proposition 3, we have

$$\boldsymbol{r}_t = \tfrac{1}{\varepsilon_t^\gamma}\boldsymbol{Q}_t\boldsymbol{y} + \tfrac{1}{\varepsilon_t^\gamma}\textstyle\sum_{i=1}^t \boldsymbol{H}_{t,i}\boldsymbol{x}_i, \tag{78a}$$

with

$$\begin{aligned}
\tfrac{1}{N}{\rm tr}\{\tfrac{1}{\varepsilon_t^\gamma}\boldsymbol{Q}_t\boldsymbol{A}\} &= \tfrac{1}{\varepsilon_t^\gamma}\textstyle\sum_{i=1}^t \vartheta_{t,i}\tfrac{1}{N}{\rm tr}\{\boldsymbol{A}^{\rm H}\boldsymbol{B}^{t-i}\boldsymbol{A}\} && (78b)\\
&= \frac{1}{\sum_{i=1}^t\vartheta_{t,i}w_{t-i}}\textstyle\sum_{i=1}^t \vartheta_{t,i}\tfrac{1}{N}{\rm tr}\{\boldsymbol{W}_{t-i}\} && (78c)\\
&= 1, && (78d)\\
{\rm tr}\{\tfrac{1}{\varepsilon_t^\gamma}\boldsymbol{H}_{t,i}\} &= \tfrac{1}{\varepsilon_t^\gamma}\vartheta_{t,i}({\rm tr}\{w_{t-i}\boldsymbol{I}-\boldsymbol{W}_{t-i}\}) && (78e)\\
&= 0, \qquad i=1,\dots,t, && (78f)
\end{aligned}$$

where (78b) follows $\boldsymbol{Q}_t = \sum_{i=1}^t\vartheta_{t,i}\boldsymbol{A}^{\rm H}\boldsymbol{B}^{t-i}$ (see (41c)), (78e) follows $\boldsymbol{H}_{t,i} = \vartheta_{t,i}(w_{t-i}\boldsymbol{I}-\boldsymbol{W}_{t-i})$ (see (41d)), (78c) follows $\varepsilon_t^\gamma = \sum_{i=1}^t\vartheta_{t,i}w_{t-i}$ (see (40b) and (40c)) and $\boldsymbol{W}_{t-i} = \boldsymbol{A}^{\rm H}\boldsymbol{B}^{t-i}\boldsymbol{A}$ (see (39a)), and (78d) and (78f) follow $w_{t-i} = \frac{1}{N}{\rm tr}\{\boldsymbol{W}_{t-i}\}$ (see (39c)). From (78), we know that the memory LE in BO-MAMP is an instance of orthogonal memory LE (see Definition 5). Therefore, from (26), the following orthogonality holds: For $1 \le t' \le t$,

$$\lim_{N\to\infty} \tfrac{1}{N}\boldsymbol{x}^{\rm H}\boldsymbol{g}_t \overset{\rm a.s.}{=} 0, \tag{79a}$$

$$\lim_{N\to\infty} \tfrac{1}{N}\boldsymbol{f}_{t'}^{\rm H}\boldsymbol{g}_t \overset{\rm a.s.}{=} 0. \tag{79b}$$

Second, since $\phi_t(\cdot)$ is the same as the NLE in (15b) of BO-OAMP/VAMP, the following orthogonality holds for the orthogonal $\phi_t(\cdot)$ [24]: $\forall t \ge 1$,

$$\lim_{N\to\infty} \tfrac{1}{N}\boldsymbol{g}_t^{\rm H}(\phi_t(\boldsymbol{r}_t)-\boldsymbol{x}) \overset{\rm a.s.}{=} 0. \tag{80}$$

In addition, following (79), we have

$$\lim_{N\to\infty} \tfrac{1}{N}\boldsymbol{g}_t^{\rm H}\boldsymbol{f}_{t'} \overset{\rm a.s.}{=} 0,\ 1 \le t' \le t. \tag{81}$$

Therefore,

$$\begin{aligned}
&\lim_{N\to\infty} \tfrac{1}{N}\boldsymbol{g}_t^{\rm H}\boldsymbol{f}_{t+1}\\
&= \lim_{N\to\infty} \tfrac{1}{N}\boldsymbol{g}_t^{\rm H}(\bar{\phi}_t(\boldsymbol{r}_t)-\boldsymbol{x}) && (82a)\\
&= \lim_{N\to\infty} \tfrac{1}{N}\boldsymbol{g}_t^{\rm H}([\boldsymbol{x}_1,\dots,\boldsymbol{x}_t,\phi_t(\boldsymbol{r}_t)]\cdot\boldsymbol{\zeta}_{t+1}-\boldsymbol{x}) && (82b)\\
&= \lim_{N\to\infty} \tfrac{1}{N}\boldsymbol{g}_t^{\rm H}[\boldsymbol{f}_1,\dots,\boldsymbol{f}_t,\phi_t(\boldsymbol{r}_t)-\boldsymbol{x}]\cdot\boldsymbol{\zeta}_{t+1} && (82c)\\
&\overset{\rm a.s.}{=} 0, && (82d)
\end{aligned}$$

where (82b) follows (38b), (82c) follows the damping constraint $\sum_{i=1}^{t}\zeta_{t,i}=1$, and (82d) follows (80) and (81). Furthermore, since $[\boldsymbol{g}_1\cdots\boldsymbol{g}_t]$ are column-wise IID and row-wise joint Gaussian with zero mean (which is guaranteed by the orthogonality in (78) for the memory LE in (38a) by induction [31], see also III-C), following the Stein's lemma (see Lemma 2), we have: $\forall 1\le t'\le t$,

$$\lim_{N\to\infty}\tfrac{1}{N}\boldsymbol{g}_{t'}^{\rm H}\boldsymbol{f}_{t+1}\overset{\text{a.s.}}{=}\lim_{N\to\infty}\tfrac{1}{N}\tfrac{v_{t',t}^{\gamma}}{v_{t,t}^{\gamma}}\boldsymbol{g}_{t}^{\rm H}\boldsymbol{f}_{t+1}\overset{\text{a.s.}}{=}0. \tag{83}$$

Finally, combining (79) and (83), we obtain the orthogonality in (20) for BO-MAMP. Following (20) and Lemma 4, we then get the asymptotic IID Gaussianity in (24) for BO-MAMP. Thus, we complete the proof of Theorem 1.

## Appendix B
## Poof of (46)

Following the definition in (23a), we have

$$\begin{aligned}
v_{t,t'}^{\gamma} &\equiv \tfrac{1}{N}{\rm E}\{\boldsymbol{g}_t^{\rm H}\boldsymbol{g}_{t'}\} && (84a)\\
&\overset{\text{a.s.}}{=}\lim_{N\to\infty}\tfrac{1}{N\varepsilon_t^{\gamma}\varepsilon_{t'}^{\gamma}}\big(\boldsymbol{Q}_t\boldsymbol{n}+\textstyle\sum_{i=1}^{t}\boldsymbol{H}_{t,i}\boldsymbol{f}_i\big)^{\rm H}\big(\boldsymbol{Q}_{t'}\boldsymbol{n}+\textstyle\sum_{i=1}^{t'}\boldsymbol{H}_{t',i}\boldsymbol{f}_i\big) && (84b)\\
&\overset{\text{a.s.}}{=}\tfrac{1}{N\varepsilon_t^{\gamma}\varepsilon_{t'}^{\gamma}}\big[\sigma^2{\rm tr}\{\boldsymbol{Q}_t^{\rm H}\boldsymbol{Q}_{t'}\}+\textstyle\sum_{i=1}^{t}\sum_{j=1}^{t'}v_{i,j}^{\phi}{\rm tr}\{\boldsymbol{H}_{t,i}^{\rm H}\boldsymbol{H}_{t',j}\}\big] && (84c)\\
&=\tfrac{1}{\varepsilon_t^{\gamma}\varepsilon_{t'}^{\gamma}}\textstyle\sum_{i=1}^{t}\sum_{j=1}^{t'}\vartheta_{t,i}\vartheta_{t',j}[\sigma^2 w_{t+t'-i-j}+v_{i,j}^{\bar\phi}\bar{w}_{t-i,t'-j}], && (84d)
\end{aligned}$$

where (84b) follows (41b) in Proposition 3, (84c) follows that $\boldsymbol{n}$ is independent with $\{\boldsymbol{f}_i\}$ and $\{\boldsymbol{n},\boldsymbol{f}_i\}$ are column-wise IID (see Theorem 1), and (84d) follows

$$\begin{aligned}
&\tfrac{1}{N}{\rm tr}\{\boldsymbol{Q}_t^{\rm H}\boldsymbol{Q}_{t'}\}\\
&=\textstyle\sum_{i=1}^{t}\sum_{j=1}^{t'}\vartheta_{t,i}\vartheta_{t',j}\tfrac{1}{N}{\rm tr}\{\boldsymbol{B}^{t-i}\boldsymbol{A}\boldsymbol{A}^{\rm H}\boldsymbol{B}^{t'-j}\} && (85a)\\
&=\textstyle\sum_{i=1}^{t}\sum_{j=1}^{t'}\vartheta_{t,i}\vartheta_{t',j}\tfrac{1}{N}{\rm tr}\{\boldsymbol{A}^{\rm H}\boldsymbol{B}^{t+t'-i-j}\boldsymbol{A}\} && (85b)\\
&=\textstyle\sum_{i=1}^{t}\sum_{j=1}^{t'}\vartheta_{t,i}\vartheta_{t',j}w_{t+t'-i-j}, && (85c)
\end{aligned}$$

and

$$\begin{aligned}
&\tfrac{1}{N}{\rm tr}\{\boldsymbol{H}_{t,i}^{\rm H}\boldsymbol{H}_{t',j}\}\\
&=\vartheta_{t,i}\vartheta_{t',j}\tfrac{1}{N}{\rm tr}\{(\boldsymbol{W}_{t-i}-w_{t-i}\boldsymbol{I})^{\rm H}(\boldsymbol{W}_{t'-j}-w_{t'-j}\boldsymbol{I})\} && (85d)\\
&=\vartheta_{t,i}\vartheta_{t',j}\tfrac{1}{N}{\rm tr}\{\boldsymbol{W}_{t-i}\boldsymbol{W}_{t'-j}-w_{t-i}\boldsymbol{W}_{t'-j}\\
&\qquad -w_{t'-j}\boldsymbol{W}_{t-i}+w_{t-i}w_{t'-j}\boldsymbol{I}\} && (85e)\\
&=\vartheta_{t,i}\vartheta_{t',j}\big(\tfrac{1}{N}{\rm tr}\{\boldsymbol{W}_{t-i}\boldsymbol{W}_{t'-j}\}-w_{t-i}w_{t'-j}\\
&\qquad -w_{t-i}w_{t'-j}+w_{t-i}w_{t'-j}\big) && (85f)\\
&=\vartheta_{t,i}\vartheta_{t',j}\big(\lambda^{\dagger}w_{t+t'-i-j}-w_{t+t'-i-j+1}-w_{t-i}w_{t'-j}\big) && (85g)\\
&=\vartheta_{t,i}\vartheta_{t',j}\bar{w}_{t-i,t'-j}, && (85h)
\end{aligned}$$

where (85a) follows (41c), (85c) follows (39a) and (39c), (85d) follows (41d), (85f) follows (39c), (85h) follows (39d), and (85g) follows

$$\begin{aligned}
&\tfrac{1}{N}{\rm tr}\{\boldsymbol{W}_{t-i}\boldsymbol{W}_{t'-j}\} && (86a)\\
&=\tfrac{1}{N}{\rm tr}\{\boldsymbol{A}^{\rm H}\boldsymbol{B}^{t-i}\boldsymbol{A}\boldsymbol{A}^{\rm H}\boldsymbol{B}^{t'-j}\boldsymbol{A}\} && (86b)\\
&=\tfrac{1}{N}{\rm tr}\{\boldsymbol{A}^{\rm H}\boldsymbol{B}^{t-i}(\lambda^{\dagger}\boldsymbol{I}-\boldsymbol{B})\boldsymbol{B}^{t'-j}\boldsymbol{A}\} && (86c)\\
&=\tfrac{1}{N}{\rm tr}\{\lambda^{\dagger}\boldsymbol{A}^{\rm H}\boldsymbol{B}^{t+t'-i-j}\boldsymbol{A}-\boldsymbol{A}^{\rm H}\boldsymbol{B}^{t+t'-i-j+1}\boldsymbol{A}\} && (86d)\\
&=\lambda^{\dagger}w_{t+t'-i-j}-w_{t+t'-i-j+1}, && (86e)
\end{aligned}$$

where (86b) follows (39a), (86c) follows $\boldsymbol{B}=\lambda^{\dagger}\boldsymbol{I}-\boldsymbol{A}^{\rm H}\boldsymbol{A}$, and (86e) follows (39a) and (39c). Therefore, we complete the proof of (46).

## Appendix C
## Correlated Noise Vector Generation for NLE

In the state evolution of BO-MAMP (see Algorithm 2), it is slightly unstable if we generate $\boldsymbol{\eta}_t=[\eta_1\cdots\eta_t]^{\rm T}$ based on $\mathcal{CN}(\boldsymbol{0},\boldsymbol{V}_t^{\gamma})$ in each iteration. For example, it may result in a non-positive semidefinite $\boldsymbol{\mathcal{V}}_t$. It is more stable to generate $\eta_t$ one by one in each iteration as follows.

In the first iteration, we generate $\eta_1$ based on $\mathcal{CN}(0,v_{1,1}^{\gamma})$. In the next iterations, we generate $\eta_t$ given $\boldsymbol{\eta}_{t-1}$ by induction as follows.

$$\eta_t=\boldsymbol{\alpha}_t^{\rm T}\boldsymbol{\eta}_{t-1}+g_t, \tag{87a}$$

where

$$\begin{aligned}
\boldsymbol{\alpha}_t&=[\boldsymbol{V}_{t-1}^{\gamma}]^{-1}[v_{1,t}^{\gamma}\cdots v_{t-1,t}^{\gamma}]^{\rm T}, && (87b)\\
g_t&\sim\mathcal{CN}\big(0,v_{g_t}\big), && (87c)\\
v_{g_t}&=v_{t,t}^{\gamma}-[v_{t,1}^{\gamma}\cdots v_{t,t-1}^{\gamma}]\boldsymbol{\alpha}_t. && (87d)
\end{aligned}$$

Then, we can verify that

$$\begin{aligned}
{\rm E}\{\|\eta_t\|^2\}&=\boldsymbol{\alpha}_t^{\rm H}{\rm E}\{\boldsymbol{\eta}_{t-1}^{*}\boldsymbol{\eta}_{t-1}^{\rm T}\}\boldsymbol{\alpha}_t+{\rm E}\{\|g_t\|^2\} && (88a)\\
&=[v_{t,1}^{\gamma}\cdots v_{t,t-1}^{\gamma}][\boldsymbol{V}_{t-1}^{\gamma}]^{-1}\boldsymbol{V}_{t-1}^{\gamma}\boldsymbol{\alpha}_t+v_{g_t} && (88b)\\
&=v_{t,t}^{\gamma}, && (88c)
\end{aligned}$$

and

$$\begin{aligned}
{\rm E}\{\eta_t^{*}\boldsymbol{\eta}_{t-1}^{\rm T}\}&=\boldsymbol{\alpha}_t^{\rm H}{\rm E}\{\boldsymbol{\eta}_{t-1}^{*}\boldsymbol{\eta}_{t-1}^{\rm T}\} && (88d)\\
&=[v_{1,t}^{\gamma}\cdots v_{t-1,t}^{\gamma}]^{*}[\boldsymbol{V}_{t-1}^{\gamma}]^{-1}\boldsymbol{V}_{t-1}^{\gamma} && (88e)\\
&=[v_{t,1}^{\gamma}\cdots v_{t,t-1}^{\gamma}]. && (88f)
\end{aligned}$$

That is, $\eta_t$ satisfies the required covariance constraint of $\boldsymbol{\eta}_{t-1}$.

## Appendix D
## Proof of Proposition 5

The following asymptotic IID Gaussianity proposition can be proved via [31], [46].

*Proposition 9 (Asymptotically IID Gaussianity):* Under the orthogonality in (20), we have

$$\begin{aligned}
&\lim_{N\to\infty}\tfrac{1}{N}\|\boldsymbol{A}(\phi_{t-1}(\boldsymbol{r}_{t-1})-\boldsymbol{x})-\boldsymbol{n}\|^2\\
&\qquad\overset{\text{a.s.}}{=}\tfrac{1}{N}{\rm E}\{\|\boldsymbol{A}\boldsymbol{\eta}_t-\boldsymbol{n}\|^2\}, && (89a)\\
&\lim_{N\to\infty}\tfrac{1}{N}[\boldsymbol{A}(\phi_{t-1}(\boldsymbol{r}_{t-1})-\boldsymbol{x})-\boldsymbol{n}]^{\rm H}[\boldsymbol{A}(\boldsymbol{x}_{t'}-\boldsymbol{x})-\boldsymbol{n}]\\
&\qquad\overset{\text{a.s.}}{=}\tfrac{1}{N}{\rm E}\{(\boldsymbol{A}\boldsymbol{\eta}_t-\boldsymbol{n})^{\rm H}(\boldsymbol{A}\bar{\boldsymbol{\eta}}_{t'}-\boldsymbol{n})\}, && (89b)
\end{aligned}$$

where $t'<t$, and $\boldsymbol{\eta}_t$ and $\bar{\boldsymbol{\eta}}_{t'}$ are Gaussian random vectors with ${\rm E}\{\boldsymbol{\eta}_t\boldsymbol{\eta}_t^{\rm H}\}=v_{t,t}^{\phi}\boldsymbol{I}$ and ${\rm E}\{\boldsymbol{\eta}_t\bar{\boldsymbol{\eta}}_{t'}^{\rm H}\}=v_{t',t}^{\phi}\boldsymbol{I}$. Besides, $\boldsymbol{\eta}_t$ and $\bar{\boldsymbol{\eta}}_{t'}$ are independent of $\boldsymbol{A}$ and $\boldsymbol{n}$.

Following Proposition 9, we have

$$\begin{aligned}
\lim_{N\to\infty}\tfrac{1}{N}\boldsymbol{z}_t^{\rm H}\boldsymbol{z}_t&=\lim_{N\to\infty}\tfrac{1}{N}\|\boldsymbol{A}(\phi_{t-1}(\boldsymbol{r}_{t-1})-\boldsymbol{x})-\boldsymbol{n}\|^2 && (90a)\\
&\overset{\text{a.s.}}{=}\tfrac{1}{N}{\rm E}\{\|\boldsymbol{A}\boldsymbol{\eta}_t-\boldsymbol{n}\|^2\} && (90b)\\
&=\tfrac{1}{N}{\rm tr}\{\boldsymbol{A}^{\rm H}\boldsymbol{A}\}v_{t,t}^{\phi}+\delta\sigma^2, && (90c)
\end{aligned}$$

and

$$\begin{aligned}
&\lim_{N\to\infty} \tfrac{1}{N} \boldsymbol{z}_t^{\mathrm{H}} \bar{\boldsymbol{z}}_{t'} \\
&= \lim_{N\to\infty} \tfrac{1}{N} [\boldsymbol{A}(\phi_{t-1}(\boldsymbol{r}_{t-1}) - \boldsymbol{x}) - \boldsymbol{n}]^{\mathrm{H}} [\boldsymbol{A}(\boldsymbol{x}_{t'} - \boldsymbol{x}) - \boldsymbol{n}] && \text{(91a)}\\
&\overset{\text{a.s.}}{=} \tfrac{1}{N} \mathrm{E}\{(\boldsymbol{A}\boldsymbol{\eta}_t - \boldsymbol{n})^{\mathrm{H}} (\boldsymbol{A}\bar{\boldsymbol{\eta}}_{t'} - \boldsymbol{n})\} && \text{(91b)}\\
&= \tfrac{1}{N} \mathrm{tr}\{\boldsymbol{A}^{\mathrm{H}}\boldsymbol{A}\} v^{\phi}_{t,t'} + \delta\sigma^2. && \text{(91c)}
\end{aligned}$$

Therefore, we complete the proof of Proposition 5.

## Appendix E
## Proof of Proposition 8

Note that $v^{\gamma}_{1,1}$ is independent of $\xi_1$. For simplicity, we can set $\xi_1 = 1$. Thus, it only needs to consider the case $t \geq 2$. Since $v^{\gamma}_{t,t}(\xi_t) = \frac{1}{N}\mathrm{E}\{\|\boldsymbol{r}_t - \boldsymbol{x}\|^2\} > 0$, we have $v^{\gamma}_{t,t}(-c_{t,0}) > 0$, i.e.,

$$c_{t,1}c^2_{t,0} + 2c_{t,2}c_{t,0} + c_{t,3} \geq 0. \tag{92}$$

In addition, we can see that $v^{\gamma}_{t,t}$ is differentiable with respect to $\xi_t$ except at the point $\xi_t = -c_{t,0}$, but $v^{\gamma}_{t,t}(-c_{t,0}) = +\infty$. Therefore, the optimal $\xi_t$ is either $\pm\infty$ or $\partial v^{\gamma}_{t,t}/\partial \xi_t = 0$, i.e., $(c_{t,1}c_{t,0} + c_{t,2})\xi_t - (c_{t,2}c_{t,0} + c_{t,3}) = 0$. Fig. 14 gives a graphic illustration of $v^{\gamma}_{t,t}(\xi_t)$. Then, the optimal $\xi^{\mathrm{opt}}_t$ can be solved as follows.

- As shown in Fig. 14(a), when $c_{t,1}c_{t,0} + c_{t,2} = 0$, there is no solution of $\partial v^{\gamma}_{t,t}/\partial\xi_t = 0$. In this case, we can set

$$\xi^{\mathrm{opt}}_t = +\infty. \tag{93}$$

- As shown in Fig. 14(b), when $c_{t,1}c_{t,0} + c_{t,2} \neq 0$, there is one solution of $\partial v^{\gamma}_{t,t}/\partial\xi_t = 0$, i.e.,

$$\xi_t = \frac{c_{t,2}c_{t,0} + c_{t,3}}{c_{t,1}c_{t,0} + c_{t,2}}. \tag{94}$$

Furthermore,

$$\begin{aligned}
&v^{\gamma}_{t,t}(\pm\infty) - v^{\gamma}_{t,t}(\xi^{\mathrm{opt}}_t) \\
&= \frac{1}{w_0^2}\left[c_{t,1} - \frac{c_{t,1}(\xi^{\mathrm{opt}}_t)^2 - 2c_{t,2}\xi^{\mathrm{opt}}_t + c_{t,3}}{(\xi^{\mathrm{opt}}_t + c_{t,0})^2}\right] && \text{(95a)}\\
&= \frac{c_{t,1}c^2_{t,0} + 2c_{t,2}c_{t,0} + c_{t,3}}{w_0^2(\xi^{\mathrm{opt}}_t + c_{t,0})^2} && \text{(95b)}\\
&\geq 0, && \text{(95c)}
\end{aligned}$$

where (95a) follows (57), (95b) follows (94), and (95c) follows (92). Therefore, the optimal $\xi^{\mathrm{opt}}_t$ is given by (94).

Thus, we complete the proof of Proposition 8.

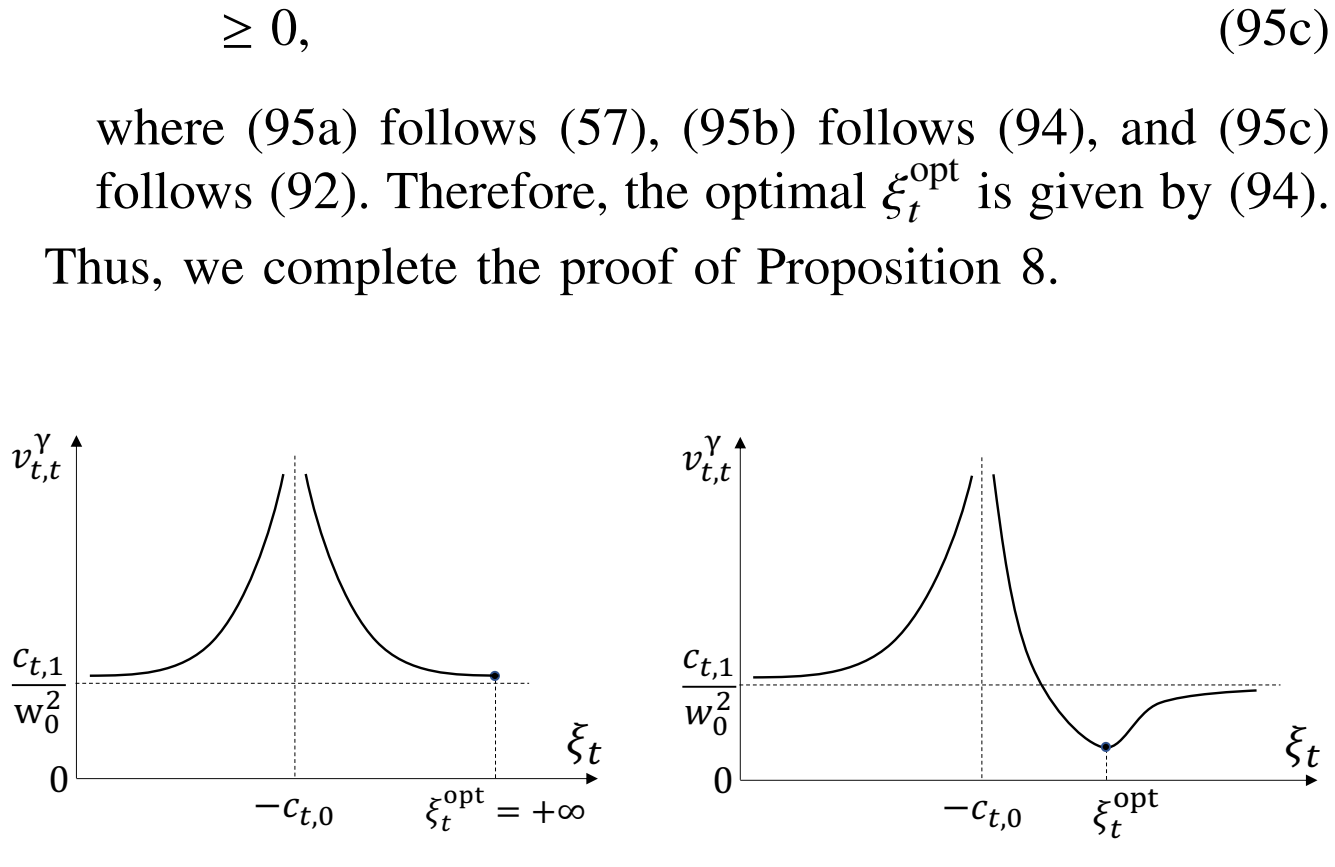

(a) $c_{t,1}c_{t,0} + c_{t,2} = 0$ (b) $c_{t,1}c_{t,0} + c_{t,2} \neq 0$

Fig. 14. Graphic illustration of $v^{\gamma}_{t,t}$ with respect to $\xi_t$.

## Appendix F
## Proof of Lemma 8

From (47), we have the following quadratic programming problem.

$$\begin{aligned}
\min_{\boldsymbol{\zeta}^{\mathcal{I}}_{t+1}} \quad & \tfrac{1}{2}[\boldsymbol{\zeta}^{\mathcal{I}}_{t+1}]^{\mathrm{H}} \boldsymbol{\mathcal{V}}^{\mathcal{I}}_{t+1} \boldsymbol{\zeta}^{\mathcal{I}}_{t+1}, && \text{(96a)}\\
\text{s.t.} \quad & \boldsymbol{1}^{\mathrm{T}}\boldsymbol{\zeta}^{\mathcal{I}}_{t+1} = 1. && \text{(96b)}
\end{aligned}$$

Since $\boldsymbol{\mathcal{V}}^{\mathcal{I}}_{t+1}$ is positive semi-definite, $v^{\bar{\phi}}_{t+1,t+1}$ is a convex function with respect to $\boldsymbol{\zeta}^{\mathcal{I}}_{t+1}$. We write the Lagrangian function as

$$\mathcal{L}(\boldsymbol{\zeta}^{\mathcal{I}}_{t+1}, \beta) = \tfrac{1}{2}[\boldsymbol{\zeta}^{\mathcal{I}}_{t+1}]^{\mathrm{H}} \boldsymbol{\mathcal{V}}^{\mathcal{I}}_{t+1} \boldsymbol{\zeta}^{\mathcal{I}}_{t+1} + \beta(1 - \boldsymbol{1}^{\mathrm{T}}\boldsymbol{\zeta}^{\mathcal{I}}_{t+1}). \tag{97}$$

We solve the problem by

$$\begin{aligned}
\nabla_{\boldsymbol{\zeta}^{\mathcal{I}}_{t+1}} \mathcal{L}(\boldsymbol{\zeta}^{\mathcal{I}}_{t+1}, \beta) &= \boldsymbol{0}, && \text{(98a)}\\
\partial\mathcal{L}(\boldsymbol{\zeta}^{\mathcal{I}}_{t+1}, \beta)/\partial\beta &= 0. && \text{(98b)}
\end{aligned}$$

That is,

$$\begin{bmatrix} \boldsymbol{\mathcal{V}}^{\mathcal{I}}_{t+1} \\ \boldsymbol{1}^{\mathrm{T}} \end{bmatrix} \boldsymbol{\zeta}^{\mathcal{I}}_{t+1} = \begin{bmatrix} \beta\boldsymbol{1} \\ 1 \end{bmatrix}. \tag{99}$$

Since $\boldsymbol{\mathcal{V}}^{\mathcal{I}}_{t+1}$ is positive semi-definite, $[\boldsymbol{\zeta}^{\mathcal{I}}_{t+1}]^{\mathrm{H}} \boldsymbol{\mathcal{V}}^{\mathcal{I}}_{t+1} \boldsymbol{\zeta}^{\mathcal{I}}_{t+1} \geq 0$, i.e., the problem in (96) is bounded. Therefore, (99) must be solvable, otherwise the problem in (96) is unbounded [57, Chapter 10.1], which contradicts $[\boldsymbol{\zeta}^{\mathcal{I}}_{t+1}]^{\mathrm{H}} \boldsymbol{\mathcal{V}}^{\mathcal{I}}_{t+1} \boldsymbol{\zeta}^{\mathcal{I}}_{t+1} \geq 0$. In this case, any solution of (99) is optimal, i.e., minimizes objective function $[\boldsymbol{\zeta}^{\mathcal{I}}_{t+1}]^{\mathrm{H}} \boldsymbol{\mathcal{V}}^{\mathcal{I}}_{t+1} \boldsymbol{\zeta}^{\mathcal{I}}_{t+1}$.

Eq. (99) can be solved in two cases.

1) When $\boldsymbol{\mathcal{V}}^{\mathcal{I}}_{t+1}$ is invertible, the problem has a unique optimal solution given by

$$\boldsymbol{\zeta}^{\mathcal{I}}_{t+1} = [\boldsymbol{\mathcal{V}}^{\mathcal{I}}_{t+1}]^{-1}\boldsymbol{1} / \big[\boldsymbol{1}^{\mathrm{T}}[\boldsymbol{\mathcal{V}}^{\mathcal{I}}_{t+1}]^{-1}\boldsymbol{1}\big], \tag{100}$$

where $[\boldsymbol{\mathcal{V}}^{\mathcal{I}}_{t+1}]^{-1}$ is the inverse of $\boldsymbol{\mathcal{V}}^{\mathcal{I}}_{t+1}$. In this case, since the optimal solution is unique, we have $v^{\bar{\phi}}_{t+1,t+1} < v^{\bar{\phi}}_{t,t}$.

2) When $\boldsymbol{\mathcal{V}}^{\mathcal{I}}_{t+1}$ is singular, there are infinite optimal solutions, and we now prove that $\boldsymbol{\zeta}^{\mathcal{I}}_{t+1} = [0 \ldots 1 \; 0]^{\mathrm{T}}$ is one of the optimal solutions. First, $\beta = 0$ is impossible otherwise we obtain a error-free recovery by damping (i.e., $v^{\bar{\phi}}_{t+1,t+1} = 0$), which is not true. If it happens, we can stop the iteration since we obtain a perfect estimate of $\boldsymbol{x}$. In other words, we only need to consider that

$$\begin{bmatrix} \boldsymbol{\mathcal{V}}^{\mathcal{I}}_{t+1} \\ \boldsymbol{1}^{\mathrm{T}} \end{bmatrix} \boldsymbol{\zeta}^{\mathcal{I}}_{t+1} = \begin{bmatrix} \boldsymbol{0} \\ 1 \end{bmatrix}. \tag{101}$$

is unsolvable. Furthermore, without loss of generality, we assume that $\boldsymbol{V}^{\bar{\phi}_{\mathcal{I}}}_t$ is invertible (see Remark 3). Since $\boldsymbol{\mathcal{V}}^{\mathcal{I}}_{t+1}$ is singular, and (101) is unsolvable, we have

$$\boldsymbol{1}^{\mathrm{T}}(\boldsymbol{V}^{\bar{\phi}_{\mathcal{I}}}_t)^{-1}\boldsymbol{v}^{\phi_{\mathcal{I}_t}}_{t+1} = 1, \tag{102}$$

where $\boldsymbol{v}^{\phi_{\mathcal{I}_t}}_{t+1} = [v^{\phi}_{i,t+1}, i \in \mathcal{I}_t]^{\mathrm{T}}$. The proof of (102) is provided in Appendix F-A. Following Lemma 10, we have $\boldsymbol{1}^{\mathrm{T}}(\boldsymbol{V}^{\bar{\phi}_{\mathcal{I}}}_t)^{-1} = [0 \ldots 0 \; (v^{\bar{\phi}}_{t,t})^{-1}]$. Then, Eq. (102) is simplified to

$$v^{\phi}_{t+1,t} = v^{\phi}_{t,t+1} = v^{\bar{\phi}}_{t,t}. \tag{103}$$

Following (63) in Lemma 9, we further have

$$v^{\bar{\phi}}_{i,t} = v^{\bar{\phi}}_{t,t}, \quad i = 1,\dots,t. \tag{104}$$

From (103) and (104), we know that

$$\zeta^{\mathcal{I}}_{t+1} = [0\dots1\ 0]^{\mathrm{T}} \tag{105}$$

is an solution of (99), i.e., minimizes objective function $[\zeta^{\mathcal{I}}_{t+1}]^{\mathrm{H}}\boldsymbol{\mathcal{V}}^{\mathcal{I}}_{t+1}\zeta^{\mathcal{I}}_{t+1}$. As a by product, we have $v^{\phi}_{t+1,t+1} \geq v^{\bar{\phi}}_{t,t}$ when $\boldsymbol{\mathcal{V}}^{\mathcal{I}}_{t+1}$ is singular, otherwise $\zeta^{\mathcal{I}}_{t+1} = [0\dots0\ 1]^{\mathrm{T}}$ is better than $\zeta^{\mathcal{I}}_{t+1} = [0\dots1\ 0]^{\mathrm{T}}$, which contradicts (105).

Thus, we complete the proof of Lemma 8.

### A. *Proof of* (102)

Since $\boldsymbol{\mathcal{V}}^{\mathcal{I}}_{t+1}$ is singular, we have $\mathrm{rank}(\boldsymbol{\mathcal{V}}^{\mathcal{I}}_{t+1}) = t$. In addition, since (101) is unsolvable, we obtain

$$t \leq \mathrm{rank}\left(\begin{bmatrix}\boldsymbol{\mathcal{V}}^{\mathcal{I}}_{t+1}\\ \mathbf{1}^{\mathrm{T}}\end{bmatrix}\right) < \mathrm{rank}\left(\begin{bmatrix}\boldsymbol{\mathcal{V}}^{\mathcal{I}}_{t+1} & \mathbf{0}\\ \mathbf{1}^{\mathrm{T}} & 1\end{bmatrix}\right) \leq t+1. \tag{106}$$

In other words, there exists non-all-zero $\{\alpha_1,...,\alpha_t\}$ such that $\sum_{i=1}^{t}\alpha_i\tilde{\boldsymbol{v}}_i = \tilde{\boldsymbol{v}}_{t+1}$, where $\tilde{\boldsymbol{v}}_i$ represents the $i$-th column of $\tilde{\boldsymbol{V}} = [\boldsymbol{\mathcal{V}}^{\mathcal{I}}_{t+1}, \mathbf{1}]^{\mathrm{T}}$. It is easy to verify that $\sum_{i=1}^{t}\alpha_i = 1$ (see last row of $\tilde{\boldsymbol{V}}$) and $[\alpha_1,...,\alpha_t]^{\mathrm{T}} = (\boldsymbol{V}^{\bar{\phi}_{\mathcal{I}}}_t)^{-1}\boldsymbol{v}^{\phi_{\mathcal{I}_t}}_{t+1}$ (see the first $t$ rows of $\tilde{\boldsymbol{V}}$), where $\boldsymbol{v}^{\phi_{\mathcal{I}_t}}_{t+1} = [v^{\phi}_{i,t+1}, i\in\mathcal{I}_t]^{\mathrm{T}}$. Thus, we can obtain $\mathbf{1}^{\mathrm{T}}(\boldsymbol{V}^{\bar{\phi}_{\mathcal{I}}}_t)^{-1}\boldsymbol{v}^{\phi_{\mathcal{I}_t}}_{t+1} = 1$.

## Appendix G
## Proof of Theorem 2

We prove Theorem 2 by two steps. First, we prove Lemma 11 that the state-evolution fixed points of BO-OAMP/VAMP and BO-MAMP are the same. Then, following Lemma 3 and Lemma 9, we can say that the optimized BO-MAMP achieves the Bayes optimal MSE predicted by the replica method for all right-unitarily-invariant transform matrices if its state evolution has a unique fixed point. Thus, we complete the proof of Theorem 2.

*Lemma 11 (Consistency of Fixed Points):* The state-evolution fixed points of BO-OAMP/VAMP and BO-MAMP with optimized $\{\theta_t, \zeta_t\}$ are the same. It is given by

$$v^{\gamma}_* = \gamma_{\mathrm{SE}}(v^{\phi}_*) = \tfrac{1}{(\varepsilon^{\gamma}_*)^2}\sum_{j=0}^{\infty}\sum_{i=0}^{\infty}\theta^{i+j+2}\big(\sigma^2 w_{i+j} + v^{\phi}_*\bar{w}_{i,j}\big), \tag{107a}$$

$$v^{\phi}_* = \phi_{\mathrm{SE}}(v^{\gamma}_*), \tag{107b}$$

where $\phi_{\mathrm{SE}}(\cdot)$ given in (15f), and

$$\theta = (\lambda^{\dagger} + \rho_*)^{-1}, \tag{107c}$$

$$\varepsilon^{\gamma}_* = \textstyle\sum_{i=0}^{\infty}\theta^{i+1}(\lambda^{\dagger}b_i - b_{i+1}). \tag{107d}$$

*Proof:* In Appendix G-A, we show that the state-evolution fixed point of BO-OAMP/VAMP is given by (107). In Appendix G-B, we show that the state-evolution fixed point of BO-MAMP is given by (107). Hence, we complete the proof of Lemma 11. ∎

### A. *State Evolution Fixed Point of BO-OAMP/VAMP*

The convergence of SE for BO-OAMP/VAMP has been rigorously proved in [40], [41], [58], [59] (see Theorem 3 in [40], [41] and Corollary 3 in [58], [59]). Define $v^{\phi}_* \equiv \lim_{t\to\infty} v^{\phi}_t$, $v^{\gamma}_* \equiv \lim_{t\to\infty} v^{\gamma}_t$, $\rho_* \equiv \lim_{t\to\infty}\rho_t = \sigma^2/v^{\phi}_*$ and $\varepsilon^{\gamma}_* \equiv \lim_{t\to\infty}\varepsilon^{\gamma}_t$. We give a Taylor series expansion for the state evolution function of LE in BO-OAMP/VAMP as follows. Note that $\{\boldsymbol{r}_t, \boldsymbol{x}_t, \boldsymbol{g}_t, \boldsymbol{f}_t\}$ are not required to converge.

Let $\boldsymbol{C} = \theta\boldsymbol{B} = \theta(\lambda^{\dagger}\boldsymbol{I} - \boldsymbol{A}\boldsymbol{A}^{\mathrm{H}})$, where $\theta = (\lambda^{\dagger}+\rho_*)^{-1}$. Since $\rho_* > 0$ and $\boldsymbol{A}\boldsymbol{A}^{\mathrm{H}}$ is non-negative, the matrix $\rho_*\boldsymbol{I} + \boldsymbol{A}\boldsymbol{A}^{\mathrm{H}}$ is invertible. From (107c), we then have

$$\big(\rho_*\boldsymbol{I} + \boldsymbol{A}\boldsymbol{A}^{\mathrm{H}}\big)^{-1} = \theta(\boldsymbol{I} - \boldsymbol{C})^{-1}. \tag{108}$$

Furthermore, the spectral radius of $\boldsymbol{C}$ is given by

$$\rho(\boldsymbol{C}) = \frac{\lambda_{\max} - \lambda_{\min}}{\lambda_{\max} + \lambda_{\min} + 2\rho_*} < 1. \tag{109}$$

Following Proposition 7, we have

$$(\boldsymbol{I} - \boldsymbol{C})^{-1} = \textstyle\sum_{i=0}^{\infty}\boldsymbol{C}^i = \sum_{i=0}^{\infty}\theta^i\boldsymbol{B}^i. \tag{110}$$

Following $\boldsymbol{B} = \lambda^{\dagger}\boldsymbol{I} - \boldsymbol{A}\boldsymbol{A}^{\mathrm{H}}$, (108) and (110), we have

$$\begin{aligned}
\varepsilon^{\gamma}_* &= \tfrac{1}{N}\mathrm{tr}\big\{\boldsymbol{A}^{\mathrm{H}}\big(\rho_*\boldsymbol{I} + \boldsymbol{A}\boldsymbol{A}^{\mathrm{H}}\big)^{-1}\boldsymbol{A}\big\} && \text{(111a)}\\
&= \textstyle\sum_{i=0}^{\infty}\frac{\theta^{i+1}}{N}\mathrm{tr}\{\lambda^{\dagger}\boldsymbol{B}^i - \boldsymbol{B}^{i+1}\} && \text{(111b)}\\
&= \textstyle\sum_{i=0}^{\infty}\theta^{i+1}(\lambda^{\dagger}b_i - b_{i+1}) && \text{(111c)}\\
&= \textstyle\sum_{i=0}^{\infty}\theta^{i+1}w_i. && \text{(111d)}
\end{aligned}$$

Therefore, we have

$$\begin{aligned}
v^{\gamma}_* &\overset{\text{a.s.}}{=} \lim_{t\to\infty}\lim_{N\to\infty}\tfrac{1}{N}\|\boldsymbol{r}_t - \boldsymbol{x}\|^2 && \text{(112a)}\\
&= \lim_{t\to\infty}\lim_{N\to\infty}\tfrac{1}{N}\big\|\tfrac{1}{\varepsilon^{\gamma}_t}\boldsymbol{A}^{\mathrm{H}}\big(\rho_t\boldsymbol{I}+\boldsymbol{A}\boldsymbol{A}^{\mathrm{H}}\big)^{-1}(\boldsymbol{n}-\boldsymbol{A}\boldsymbol{f}_t)+\boldsymbol{f}_t\big\|^2 && \text{(112b)}\\
&= \lim_{N\to\infty}\tfrac{1}{N}\big\|\tfrac{1}{\varepsilon^{\gamma}_*}\boldsymbol{A}^{\mathrm{H}}\big(\rho_*\boldsymbol{I} + \boldsymbol{A}\boldsymbol{A}^{\mathrm{H}}\big)^{-1}\boldsymbol{n}\\
&\qquad + \lim_{t\to\infty}[\boldsymbol{I} - \tfrac{1}{\varepsilon^{\gamma}_*}\boldsymbol{A}^{\mathrm{H}}\big(\rho_*\boldsymbol{I} + \boldsymbol{A}\boldsymbol{A}^{\mathrm{H}}\big)^{-1}\boldsymbol{A}]\boldsymbol{f}_t\big\|^2 && \text{(112c)}\\
&= \lim_{N\to\infty}\tfrac{1}{N}\big\|\textstyle\sum_{i=0}^{\infty}\frac{\theta^{i+1}}{\varepsilon^{\gamma}_*}\boldsymbol{A}^{\mathrm{H}}\boldsymbol{B}^i\boldsymbol{n}\\
&\qquad + \lim_{t\to\infty}[\boldsymbol{I} - \textstyle\sum_{i=0}^{\infty}\frac{\theta^{i+1}}{\varepsilon^{\gamma}_*}(\lambda^{\dagger}\boldsymbol{B}^i - \boldsymbol{B}^{i+1})]\boldsymbol{f}_t\big\|^2 && \text{(112d)}\\
&= \tfrac{\sigma^2}{(\varepsilon^{\gamma}_*)^2}\sum_{j=0}^{\infty}\sum_{i=0}^{\infty}\theta^{i+j+2}\underbrace{(\lambda^{\dagger}b_{i+j} - b_{i+j+1})}_{w_{i+j}}\\
&\quad + \tfrac{v^{\phi}_*}{(\varepsilon^{\gamma}_*)^2}\Big[\sum_{j=0}^{\infty}\sum_{i=0}^{\infty}\theta^{i+j+2}\underbrace{\big[(\lambda^{\dagger})^2 b_{i+j} - 2\lambda^{\dagger}b_{i+j+1} + b_{i+j+2}\big]}_{\lambda^{\dagger}w_{i+j} - w_{i+j+1}}\\
&\quad - \underbrace{(\varepsilon^{\gamma}_*)^2}_{\sum_{j=0}^{\infty}\sum_{i=0}^{\infty}\theta^{i+j+2}w_i w_j}\Big] && \text{(112e)}\\
&= \tfrac{\sigma^2}{(\varepsilon^{\gamma}_*)^2}\sum_{j=0}^{\infty}\sum_{i=0}^{\infty}\theta^{i+j+2}w_{i+j}\\
&\quad + \tfrac{v^{\phi}_*}{(\varepsilon^{\gamma}_*)^2}\sum_{j=0}^{\infty}\sum_{i=0}^{\infty}\theta^{i+j+2}\underbrace{(\lambda^{\dagger}w_{i+j} - w_{i+j+1} - w_i w_j)}_{\bar{w}_{i,j}}\Big] && \text{(112f)}\\
&= \tfrac{1}{(\varepsilon^{\gamma}_*)^2}\sum_{j=0}^{\infty}\sum_{i=0}^{\infty}\theta^{i+j+2}\big(\sigma^2 w_{i+j} + v^{\phi}_*\bar{w}_{i,j}\big), && \text{(112g)}
\end{aligned}$$

where (112a) follows $v^{\gamma}_* \equiv \lim_{t\to\infty} v^{\gamma}_t$ and $v^{\gamma}_t = \frac{1}{N}\|\boldsymbol{r}_t - \boldsymbol{x}\|^2$, (112b) follows (14) and (15a), (112c) follows $\rho_* \equiv \lim_{t\to\infty}\rho_t$

and $\varepsilon_*^\gamma \equiv \lim_{t\to\infty} \varepsilon_t^\gamma$, (112d) follows (108), (110) and $\boldsymbol{A}\boldsymbol{A}^{\rm H} = \lambda^\dagger \boldsymbol{I} - \boldsymbol{B}$, (112e) follows (39b) and that $\boldsymbol{n}$ is independent with $\{\boldsymbol{f}_i\}$ and $\{\boldsymbol{n}, \boldsymbol{f}_i\}$ are column-wise IID (see Theorem 1), (112f) follows (39c) and (111), and (112g) follows (39d). Thus, the transfer function of LE in BO-OAMP/VAMP converges to

$$v_*^\gamma = \gamma_{\rm SE}(v_*^\phi) \overset{\rm a.s.}{=} \tfrac{1}{(\varepsilon_*^\gamma)^2} \sum_{j=0}^{\infty}\sum_{i=0}^{\infty} \theta^{i+j+2}\big(\sigma^2 w_{i+j} + v_*^\phi \bar{w}_{i,j}\big). \quad (113)$$

From (17b) and (113), we get the state-evolution fixed point of BO-OAMP/VAMP in (107).

### B. State Evolution Fixed Point of BO-MAMP

For the BO-MAMP with optimized $\{\theta_t, \zeta_t\}$, from Lemma 9, we have

$$\lim_{t\to\infty} v_{t,i}^{\bar\phi} = v_*^{\bar\phi}, \quad \forall i \le t. \quad (114)$$

Since $\{\xi_t = \xi_*, \forall t\}$, we then have

$$\lim_{t\to\infty} \rho_t = \rho_* = \sigma^2 / v_*^{\bar\phi}, \quad (115a)$$

$$\lim_{t\to\infty} \theta_t = \theta = (\lambda^\dagger + \rho_*)^{-1}, \quad (115b)$$

$$\lim_{i\to\infty} \vartheta_{t,i} = \xi_* \theta^{t-i}, \quad \text{for } i \le t. \quad (115c)$$

In addition, for finite $i$ and $j$,

$$\lim_{t\to\infty} \vartheta_{t,i} w_{t-i} = \lim_{t\to\infty} \tfrac{1}{N} {\rm tr}\{\vartheta_{t,i} \boldsymbol{W}_{t-i}\} \to 0, \quad (116a)$$

$$\lim_{t\to\infty} \theta^{t-i} w_{t-i} = \lim_{t\to\infty} \tfrac{1}{N} {\rm tr}\{\theta^{t-i} \boldsymbol{W}_{t-i}\} \to 0, \quad (116b)$$

$$\lim_{t\to\infty} \vartheta_{t,i}\vartheta_{t,j} w_{2t-i-j} = \lim_{t\to\infty} \tfrac{1}{N} {\rm tr}\{\vartheta_{t,i}\vartheta_{t,j} \boldsymbol{W}_{2t-i-j}\} \to 0, \quad (116c)$$

$$\lim_{t\to\infty} \vartheta_{t,i}\vartheta_{t,j} \bar{w}_{t-i,t-j} = \lim_{t\to\infty} \tfrac{1}{N} {\rm tr}\{\vartheta_{t,i}\vartheta_{t,j} \boldsymbol{H}_{t,i}\boldsymbol{H}_{t,j}\} \to 0, \quad (116d)$$

$$\lim_{t\to\infty} \theta^{2t-i-j} w_{2t-i-j} = \lim_{t\to\infty} \tfrac{1}{N} {\rm tr}\{\theta^{2t-i-j} \boldsymbol{W}_{2t-i-j}\} \to 0, \quad (116e)$$

$$\lim_{t\to\infty} \theta^{2t-i-j} \bar{w}_{t-i,t-j} = \lim_{t\to\infty} \tfrac{1}{N} {\rm tr}\{\theta^{2t-i-j} \boldsymbol{H}_{t,i}\boldsymbol{H}_{t,j}\} \to 0, \quad (116f)$$

since $\vartheta_{t,i}\boldsymbol{W}_{t-i}, \vartheta_{t,i}\vartheta_{t,j}\boldsymbol{W}_{2t-i-j}, \vartheta_{t,i}\vartheta_{t,j}\boldsymbol{H}_{t,i}\boldsymbol{H}_{t,j}, \theta^{t-i}\boldsymbol{W}_{t-i}, \theta^{2t-i-j}\boldsymbol{W}_{2t-i-j}$ and $\theta^{2t-i-j}\boldsymbol{H}_{t,i}\boldsymbol{H}_{t,j}$ are all products of infinite matrices whose spectral radiuses are less than 1 (see (54)). Therefore, we can obtain the following limits.

$$\lim_{t\to\infty} \varepsilon_t^\gamma \overset{\rm a.s.}{=} \lim_{t\to\infty} \textstyle\sum_{i=1}^{t} \xi_* \theta^{t-i} w_{t-i} \quad (117a)$$

$$= \xi_* \textstyle\sum_{i'=0}^{\infty} \theta^{i'} w_{i'} \quad (117b)$$

$$= \xi_* \varepsilon_*^\gamma / \theta, \quad (117c)$$

and

$$v_*^\gamma \equiv \lim_{t\to\infty} v_{t,t}^\gamma \quad (117d)$$

$$\overset{\rm a.s.}{=} \lim_{t\to\infty} \tfrac{\theta^2}{(\xi_*\varepsilon_*^\gamma)^2} \sum_{i=1}^{t}\sum_{j=1}^{t} \xi_*^2 \theta^{2t-i-j}\big(\sigma^2 w_{2t-i-j} + v_*^{\bar\phi} \bar{w}_{t-i,t+j}\big) \quad (117e)$$

$$= \tfrac{1}{(\varepsilon_*^\gamma)^2} \textstyle\sum_{i'=0}^{\infty}\sum_{j'=0}^{\infty} \theta^{i'+j'+2}\big(\sigma^2 w_{i'+j'} + v_*^{\bar\phi} \bar{w}_{i',j'}\big), \quad (117f)$$

where (117a) follows (40), (115c), (116a) and (116b), (117b) follows $i' = t - i$, (117c) follows (111), (117e) follows (46) and (114)-(116), and (117f) follows $i' = t - i$ and $j' = t - j$.

Thus, the state evolution function of LE in BO-MAMP converges to:

$$v_*^\gamma = \gamma_{\rm SE}(v_*^{\bar\phi}) \overset{\rm a.s.}{=} \tfrac{1}{(\varepsilon_*^\gamma)^2} \sum_{j=0}^{\infty}\sum_{i=0}^{\infty} \theta^{i+j+2}\big(\sigma^2 w_{i+j} + v_*^{\bar\phi} \bar{w}_{i,j}\big). \quad (118)$$

In addition, $\phi_t(\cdot)$ in BO-OAMP/VAMP and BO-MAMP are the same. From (38b) and $\sum_{i=1}^{t+1} \zeta_{t+1,i} = 1$, the state evolution function of NLE in BO-MAMP converges to

$$v_*^{\bar\phi} = \phi_{\rm SE}(v_*^\gamma), \quad (119)$$

where $\phi_{\rm SE}(\cdot)$ is the same as that of BO-OAMP/VAMP in (15f) and (17b). Replacing $v_*^{\bar\phi}$ by $v_*^{\phi}$ in (118) and (119), we obtain the state-evolution fixed point of BO-MAMP in (107).

## Appendix H
## Algorithm Summary

We summarize BO-MAMP and the state evolution of BO-MAMP in the following algorithms. Note that the subscript $t$ of some variables (e.g., $\theta, \vartheta_i, \xi, \varepsilon, c_i \cdots$) is omitted in Algorithms 1 and 2 to reduce the storage in the program.

**Algorithm 1** BO-MAMP

**Input:** $\{\hat\phi_t(\cdot)\}$, $\boldsymbol{A}$, $\boldsymbol{y}$, $P_x$, $\{\lambda_{\min}, \lambda_{\max}, \lambda_t\}$, $\sigma^2$, $\mathcal{T}$, $L$.

**Initialization:** $\boldsymbol{z}_1 = \boldsymbol{y}$, $\hat{\boldsymbol{r}} = \boldsymbol{0}$, $\vartheta_1 = \xi = 1$, $\lambda^\dagger = \frac{\lambda_{\max}+\lambda_{\min}}{2}$, $\{w_i, 0 \le i < 2\mathcal{T}\}$ by (39c), $\{\bar{w}_{i,j}, 0 \le i,j \le \mathcal{T}\}$ by (39d), $v_{1,1}^\phi = (\frac{1}{N}\boldsymbol{y}^{\rm H}\boldsymbol{y} - \delta\sigma^2)/w_0$.

**for** $t = 1$ to $\mathcal{T}$ **do**

% Memory LE

$\theta = \big(\lambda^\dagger + \sigma^2/v_{t,t}^\phi\big)^{-1}$, $\big\{\vartheta_i = \theta\vartheta_i,\ p_i = -\vartheta_i w_{t-i}, 1 \le i < t\big\}$

$\{c_i, 0 \le i \le 3\}$ by (58), $\big\{t \ge 2 : \xi = \vartheta_t = \frac{c_2 c_0 + c_3}{c_1 c_0 + c_2}\big\}$ % $\xi$

$\{p_t = -\xi w_0,\ \varepsilon = w_0 c_0 - p_t\}$, $v_\gamma = \frac{1}{\varepsilon^2}(c_1\xi^2 - 2c_2\xi + c_3)$

$\hat{\boldsymbol{r}} = \xi \boldsymbol{z}_t + \theta(\lambda^\dagger \hat{\boldsymbol{r}} - \boldsymbol{A}\boldsymbol{A}^{\rm H}\hat{\boldsymbol{r}})$, $\boldsymbol{r} = \frac{1}{\varepsilon}\big(\boldsymbol{A}^{\rm H}\hat{\boldsymbol{r}} - \sum_{i=1}^{t} p_i \boldsymbol{x}_i\big)$

% NLE

$(\hat{\boldsymbol{x}}_{t+1}, \hat{v}_{t+1}) = \hat\phi_t(\boldsymbol{r}, v_\gamma)$, $\boldsymbol{x}_{t+1} = \big(\frac{1}{\hat{v}_{t+1}} - \frac{1}{v_\gamma}\big)^{-1}\big(\frac{\hat{\boldsymbol{x}}_{t+1}}{\hat{v}_{t+1}} - \frac{\boldsymbol{r}_t}{v_\gamma}\big)$

$\boldsymbol{z}_{t+1} = \boldsymbol{y} - \boldsymbol{A}\boldsymbol{x}_{t+1}$

$\big\{v_{t+1,t'}^\phi = (\{v_{t',t+1}^\phi)^* = \frac{1}{w_0}\big[\frac{1}{N}\boldsymbol{z}_{t+1}^{\rm H}\boldsymbol{z}_{t'} - \delta\sigma^2\big], 1 \le t' \le t+1\big\}$

% Damping

$l_{t+1} = \min\{L, t+1\}$, $\big\{\boldsymbol{\zeta} = \boldsymbol{\mathcal{V}}_{t+1}^{-1}\boldsymbol{1}, v_\phi = \frac{1}{\boldsymbol{1}^{\rm T}\boldsymbol{\zeta}}, \boldsymbol{\zeta} = v_\phi\boldsymbol{\zeta}\big\}$ % $\boldsymbol{\zeta}$

$\boldsymbol{x}_{t+1} = \sum_{i=1}^{l_{t+1}} \zeta_i \boldsymbol{x}_{t-l_{t+1}+1+i}$, $\boldsymbol{z}_{t+1} = \sum_{i=1}^{l_{t+1}} \zeta_i \boldsymbol{z}_{t-l_{t+1}+1+i}$

$v_{t+1,t+1}^\phi = v_\phi$

$\big\{v_{t+1,t'}^\phi = (v_{t',t+1}^\phi)^* = \sum_{i=1}^{l_{t+1}} \zeta_i v_{t-l_{t+1}+1+i,t'}^\phi, 1 \le t' \le t\big\}$

**end for**

**Output:** $\{\hat{\boldsymbol{x}}_{t+1}, \hat{v}_{t+1}\}$.

**Algorithm 2** State Evolution (SE) of BO-MAMP

**Input:** $\boldsymbol{A}$, $\{\hat{\phi}_t(\cdot)\}, x \sim P_x, \{\lambda_{\min}, \lambda_{\max}, \lambda_t\}$, $\sigma^2$, $\mathcal{T}$, $L$.

**Initialization:** $\vartheta_{1,1} = \xi = 1$, $\lambda^\dagger = \frac{\lambda_{\max}+\lambda_{\min}}{2}$, $\{w_i, 0 \le i < 2\mathcal{T}\}$ by (39c), $\{\bar{w}_{i,j}, 0 \le i, j < \mathcal{T}\}$ by (39d), $v^{\phi}_{1,1} = \mathrm{E}\{x^2\}$, $\tilde{\boldsymbol{x}}_1 = 0$.

1: **for** $t = 1$ to $\mathcal{T}$ **do**

2: % Memory LE

$\big\{\vartheta_{t,i} = \vartheta_{t-1,i}/\big(\lambda^\dagger + \sigma^2/v^{\phi}_{t,t}\big),\ p_i = -\vartheta_{t,i} w_{t-i}, 1 \le i < t\big\}$

$\{c_i, 0 \le i \le 3\}$ by (58), $\quad \varepsilon_t = w_0(\xi + c_0)$

$\{\textbf{If } t \ge 2 : \xi = \vartheta_{t,t} = (c_2 c_0 + c_3)(c_1 c_0 + c_2)\}$

$\big\{v^{\gamma}_{t,t'} = (v^{\gamma}_{t',t})^*, 1 \le t' \le t-1\big\}$ by (46)

$v^{\gamma}_{t,t} = (c_1\xi^2 - 2c_2\xi + c_3)/\varepsilon_t^2, \quad \big\{\eta_t \text{ by } (87),\ r = x + \eta_t\big\}$

3: % NLE

$(\hat{x}, \hat{v}) = \hat{\phi}_t(r, v^{\gamma}_{t,t}),\ x_{t+1} = (\frac{1}{\hat{v}} - \frac{1}{v^{\gamma}_{t,t}})^{-1}(\frac{\hat{x}}{\hat{v}} - \frac{r}{v^{\gamma}_{t,t}})$

$\boldsymbol{v}^{\phi}_{t+1} = \mathrm{E}\big\{(x_{t+1} - x)^*([\tilde{\boldsymbol{x}}_t\ x_{t+1}]^{\mathrm{T}} - x\mathbf{1})\big\}$

4: % Damping

$l_{t+1} = \min\{L, t+1\},\ \big\{\boldsymbol{\zeta} = \boldsymbol{\mathcal{V}}^{-1}_{t+1}\mathbf{1}, v^{\phi}_{t+1,t+1} = \frac{1}{\mathbf{1}^{\mathrm{T}}\boldsymbol{\zeta}}, \boldsymbol{\zeta} = v^{\phi}_{t+1,t+1}\boldsymbol{\zeta}\big\}$

$\big\{v^{\phi}_{t+1,t'} = (v^{\phi}_{t',t+1})^* = \sum_{i=1}^{l_{t+1}} \zeta_i v^{\phi}_{t-l_{t+1}+1+i,t'}, 1 \le t' \le t\big\}$

$\tilde{\boldsymbol{x}}_{t+1} = \big[\tilde{\boldsymbol{x}}_t\ \ \sum_{i=1}^{l_{t+1}} \zeta_i x_{t-l_{t+1}+1+i}\big]$

5: **end for**

**Output:** $\hat{\boldsymbol{v}}$.